\documentclass[10pt,prd,aps,twocolumn,a4paper,
floatfix,nofootinbib,showpacs,superscriptaddress,longbibliography]{revtex4-2}

\usepackage[utf8]{inputenc}
\usepackage{graphicx,psfrag}
\usepackage{graphics}

\usepackage{physics}
\usepackage{mathrsfs}          %
\usepackage{mathtools,amsfonts,amssymb}
\usepackage{multirow}
\usepackage{comment}
\usepackage[colorlinks=true]{hyperref}
\hypersetup{
  linkcolor=blue,
  citecolor=blue,
  urlcolor=blue
}
\usepackage{bbm}
\usepackage{bm}
\usepackage{placeins}
\usepackage[normalem]{ulem}
\usepackage{tensor} %
\usepackage{verbatim}     %

\usepackage[dvipsnames]{xcolor}
\usepackage{orcidlink}

\newcommand{\bdryequal}{\overset{\scriptscriptstyle\wedge}{=}}

\newcommand{\eg}{e.\,g.\@}
\newcommand{\ie}{i.\,e.\@}

\definecolor{cyan}{rgb}{0,0.9,0.9}
\definecolor{orange}{rgb}{0.9,0.5,0}
\definecolor{purple}{rgb}{0.8,0.4,0.8}
\definecolor{gray}{rgb}{0.8242,0.8242,0.8242}
\definecolor{grey}{rgb}{0.5,0.5,0.5}
\definecolor{pink}{rgb}{1.0, 0.0, 0.5}
\definecolor{dgreen}{rgb}{0.35, 0.8, 0.35}
\begin{document}

\title{Worldtube excision method for intermediate-mass-ratio inspirals:
  scalar-field toy model}

\newcommand{\affilaei}{Max Planck Institute for Gravitational Physics (Albert Einstein Institute), Am Mühlenberg 1, Potsdam 14476, Germany}

\newcommand{\affilSouthampton}{School of Mathematical Sciences and STAG Research Centre, University of Southampton, Southampton, SO17 1BJ, United Kingdom}

\author{Mekhi \surname{Dhesi}\,\orcidlink{0000-0003-0017-4302}}
\affiliation{\affilSouthampton}
\author{Hannes R. \surname{Rüter}\,\orcidlink{0000-0002-3442-5360}}
\affiliation{\affilaei}
\author{Adam \surname{Pound}\,\orcidlink{0000-0001-9446-0638}}
\author{Leor \surname{Barack}\,\orcidlink{0000-0003-4742-9413}}
\affiliation{\affilSouthampton}
\author{Harald P. \surname{Pfeiffer}\,\orcidlink{0000-0001-9288-519X}}
\affiliation{\affilaei}

\date{\today}

\begin{abstract}
  The computational cost of inspiral and merger simulations for black-hole
  binaries increases in inverse proportion to the square of the
  mass ratio $q:=m_2/m_1\leq 1$. One factor of $q$ comes from the number
  of orbital cycles, which is proportional to $1/q$, and another is associated
  with the required number of time steps per orbit, constrained (via the
  Courant-Friedrich-Lewy condition) by the need to resolve the two disparate length scales.
  This problematic scaling makes simulations progressively less tractable
  at smaller $q$. Here we propose and explore a method for alleviating
  the scale disparity in simulations with mass ratios in the
  intermediate astrophysical range ($10^{-4} \lesssim q\lesssim 10^{-2}$), where
  purely perturbative methods may not be adequate. A region of
  radius much larger than $m_2$ around the smaller object is excised from
  the numerical domain, and replaced with an analytical model
  approximating a tidally deformed black hole. The analytical model
  involves certain \emph{a priori} unknown parameters,  associated with
  unknown bits of physics together with gauge-adjustment terms; these
  are dynamically determined by matching to the numerical solution outside
  the excision region.  In this paper we develop the basic idea and
  apply it to a toy model of a scalar charge in a circular geodesic orbit
  around a Schwarzschild black hole, solving for the massless Klein-Gordon field
  in a 1+1D framework. Our main goal here is to explore
  the utility and properties of different matching strategies, and to this
  end we develop two independent implementations, a finite-difference one
  and a spectral one.  We discuss the extension of our method to a full
  3D numerical evolution and to gravity.
\end{abstract}

\maketitle

\section{Introduction}\label{Section:Introduction}

All gravitational-wave signals
reported so far
by the LIGO-Virgo Collaboration~\cite{LIGOScientific:2018mvr,Abbott:2020niy}
originated from compact-object binaries in which the two components had fairly
comparable masses. The most extreme mass disparity to date was observed in
GW190814, whose likely source was the coalescence of a
$2.50$--$2.67 M_\odot$ object (either an
exceptionally heavy neutron star or an exceptionally light black hole) with a
$22.2$--$24.3 M_\odot$ black hole
\cite{Abbott:2020khf}. Upgrades and future generations of ground-based
detectors \cite{KAGRA:2020npa, Hall:2019xmm}, and especially the planned space-based
detector LISA \cite{LISA:2017pwj}, will open up a
new window of observation in the low-frequency band of the gravitational-wave
spectrum, enabling the detection of signals from ever heavier binary systems,
including ones containing intermediate-mass and supermassive black holes. In
consequence, it is expected that the detection of high mass ratio events will
become routine, and that the catalogue of detected binary sources will extend
to include
a broad range of mass ratios---potentially down to
$\sim 1$:$10^6$ with LISA
\cite{Jani:2019ffg, Salcido:2016oor, Volonteri:2020wkx}.

In anticipation of this remarkable expansion in observational reach, it is
important to develop accurate theoretical waveform templates that reliably
cover the entire relevant range of mass ratios. Standard Numerical Relativity
(NR) methods~\cite{BauSha10} work well for mass ratios in the range
$ 0.1 \lesssim q:=m_2/m_1 \leq 1$ (see \eg{}~\cite{Boyle:2019kee}).
However, simulations become progressively less tractable at smaller
$q$, and few numerical simulations have been performed at
$q < 0.1$ so far. The root cause is a problematic scaling of the
required simulation time with $q$. Fundamentally, one
expects the required simulation time to grow in proportion to
$q^{-2}$, where one factor of $q^{-1}$ is
associated with the number of in-band orbital cycles, and the second factor
$q^{-1}$ comes from the Courant-Friedrich-Lewy (CFL) stability
limit on the time step
of the numerical simulation, constrained by the need to spatially resolve the
small object. The state of the art in small-$q$ NR is
represented by the recent simulations performed at RIT of the last 13 orbital
cycles prior to merger of a black-hole binary system with
$q=1/128$~\cite{Lousto:2020tnb, Rosato:2021jsq}. Such simulations remain
extremely computationally expensive.

For \emph{extreme} mass ratios (say, $q \lesssim 10^{-4}$), it
is more natural to apply an alternative treatment based on black-hole
perturbation theory.  Here, the field equations are formally expanded in powers
of $q$, and the orbital dynamics are described in terms of
a point-particle inspiral trajectory on the fixed geometry of the large black
hole. In the limit $q\to 0$, the trajectory is geodesic. Back
reaction from the small object's self-field, which drives the slow inspiral, is
accounted for order-by-order in $q$, in what is known as
the gravitational self-force (GSF) approach \cite{Barack:2018yvs, Pound:2021qin}. GSF is
currently the only viable method for modelling astrophysical extreme-mass-ratio
inspirals (EMRIs), in which a compact object orbits a massive black hole in a
galactic
nucleus. Development continues towards an accurate model of EMRI waveforms
suitable for signal identification and interpretation with
LISA~\cite{vandeMeent:2017bcc,Chua:2020stf,Hughes:2021exa,Pound:2019lzj,Warburton:2021kwk,WardellMG}.

The \emph{intermediate} range of mass ratios, say
$10^{-4}\lesssim q\lesssim 10^{-1}$, poses a unique modelling challenge.
\emph{A priori}, it is
hard to ascertain whether GSF calculations can in practice cover with
sufficient accuracy the entire range of $q$ where the
computational cost of full NR simulations is prohibitive. An initial
study~\cite{vandeMeent:2020xgc} suggested that this may well be the case
for sufficiently simple binary systems (of nonspinning black holes in a
quasicircular inspiral), and recent computations of so-called
``post-adiabatic'' GSF waveforms~\cite{WardellMG} have borne out that
prediction. However, it remains unclear whether the
two
methods, separately applied, can provide us with a reliable model
of intermediate-mass-ratio inspirals (IMRIs) over the full parameter space of
astrophysically plausible sources. The relevance and pressing nature of this
question became self-evident with the first conclusive observation of
an intermediate-mass black hole ($M\sim 142^{+28}_{-16}M_{\odot}$)
as the merger product
in GW190521~\cite{Abbott:2020tfl}.

\begin{figure}
  \includegraphics[width=\columnwidth]
  {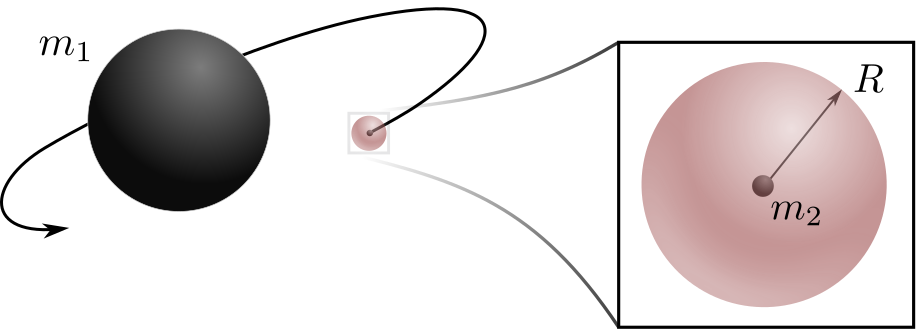}
  \caption{
    \label{fig:excision_cartoon} Our basic strategy:
    A region of radial extent $m_2\ll R\ll \mathcal{ R}$ is excised
    from the computational domain ($\mathcal{ R}$ being the characteristic lengthscale
    associated with the tidal field of the large mass $m_1$ at the location of the small mass $m_2$).
    An approximate analytical solution is used inside the excised region.
    The analytical and numerical solutions are dynamically
    matched at each step of the numerical evolution.}
\end{figure}
In this work we explore a new, synergistic approach to IMRI modelling,
featuring a direct synthesis of black-hole perturbation
and NR techniques. The central idea is simple, and illustrated in
Fig.~\ref{fig:excision_cartoon} as applied to a compact-object binary with
masses $m_1\gg m_2$. An excision region is introduced around the
small object,
of radial extent $R$ chosen such that
$m_2\ll R\ll \mathcal{ R}$, where $\cal R$
is the characteristic lengthscale associated with the tidal field of
$m_1$ at the
location of $m_2$ (such that $\mathcal{ R}\sim m_1$ near the
end of the inspiral).
Inside this region---a ``worldtube'' in spacetime---an approximate analytical
solution is prescribed for the spacetime metric, arising from the perturbation
theory of compact objects
in a tidal environment. An NR simulation is set up for the binary, in which the
worldtube's interior is excised from the numerical domain, and replaced with
the analytical solution. At each time step of the numerical evolution, the
numerical solution (outside the tube) and analytical solution (inside the tube)
are matched across the tube's boundary, in a process that fixes
\emph{a priori}
unknown tidal coefficients in the analytical solution, as well as gauge degrees
of freedom.
The intended effect of this construction is to partially alleviate the scale
disparity that thwarts the efficiency of the numerical evolution at small
$q$.  An outline of such a strategy was first (to our
knowledge) put forward by B. Schutz in a conference talk a few years ago
\cite{Schutz2017}.

To begin thinking about how such a strategy may work in practice, we restrict
attention to the simplest scenario, where the smaller object is a black hole.
The appropriate analytical solution inside the worldtube is then that of a
tidally perturbed Kerr black hole, where the tidal perturbation arises from the
presence of the larger body. Such geometries are examples of a broader class of
spacetimes studied extensively in recent
literature~\cite{Poisson:2005pi, Poisson:2009qj, Taylor:2008xy, Poisson:2018qqd, LeTiec:2020bos, Damour:2009vw, Raposo:2020yjy, Yagi:2016ejg, Hinderer:2009ca},
where the tidal response of a compact object to an external perturbative tidal
field is derived analytically order by order in
$s/\mathcal{R}$---the ratio of distance $s$ from
the smaller object and the characteristic lengthscale
$\mathcal{R}$ of the tidal field---under the assumption
$s\ll \mathcal{R}$. For a nonrotating black hole, the perturbed metric has
so far been constructed through order
$(s/\mathcal{R})^{4}$~\cite{Poisson:2018qqd}.
For our excision method we would need the perturbed metric near the worldtube's
boundary,
where it takes the form of an expansion in $R/\mathcal{ R}\ll 1$.

Since the analytically prescribed metric on the worldtube's boundary is only
an approximation, the spacetime constructed in the numerical simulation is
also approximate, even if numerical error could be reduced to zero. If our
analytical solution is correct only up to $\mathcal{O} \left((R/\mathcal{ R})^n\right)$---what we
later call an ``$n$th-order model''---then, in general, an
error of $\mathcal{O}\left((R/\mathcal{ R})^{n+1}\right)$ is fed from the tube's boundary to the
numerical solution, and propagates to the bulk of the numerical domain. One
could then only hope to construct the binary's spacetime up to an error of
$\mathcal{O}\left((R/\mathcal{ R})^{n+1}\right)$, even in the continuum limit. This worldtube error can
be reduced either by increasing the order $n$ of the
analytical model, or by decreasing the tube's radius $R$.
Of course, decreasing $R$ restores the original
scale disparity and thus diminishes the gain from the introduction of a tube.
There is hence a fundamental trade-off in our method between precision and
computational cost, with $R$ serving as a control
parameter.
At the end of this introduction we give a rough
estimation of the potential computational savings for an ``optimal'' choice of
$R$.

Our primary purpose in this initial study is to develop and test a matching
methodology for the field across the worldtube's boundary.  For that purpose we
employ a linear scalar-field toy model, in which the small black hole is
replaced with a pointlike scalar charge, and the large object is a
Schwarzschild black hole. Instead of tackling the full Einstein's equations, we
thus solve the massless linear Klein-Gordon equation for a scalar field on a
fixed
Schwarzschild background. Furthermore, we decompose the
field equation into multipole modes on the Schwarzschild geometry, and solve
for each mode of the field individually as an evolution problem in 1+1
dimensions (radius+time). Our worldtube is then a 2-dimensional ``strip''
confined between two parallel timelike curves (cf.~Fig.~\ref{fig:matching_method1}
below). As a final simplification, we set the scalar charge to move on a fixed
circular geodesic orbit around the large black hole (ignoring radiation
reaction), meaning we can fix our worldtube in advance of the evolution, and it
has a simple geometry. All of these simplifications take us very far, of
course, from the actual physical problem in question. However, our toy problem
retains enough relevant features to make it useful as a development platform
for
worldtube matching procedures.

We develop and explore two such procedures. The first is based on matching
the analytical and numerical solutions in an open ``buffer'' region around the
tube's boundaries. At each step of the time evolution, the matching determines
a set of unknown coefficients in the analytical solution. Once the
analytical solution has been fixed inside the tube, the evolution can proceed
to the next time step. This approach is close in spirit to the standard method
of
matched asymptotic expansions, which underlies most of the literature on
tidally perturbed black hole spacetimes (as well as GSF theory). But
whereas in standard matched expansions one matches together two asymptotic
expansions, here one matches an asymptotic expansion (the approximate
analytical solution in the tube) to an ``exact'' numerical solution. The second
matching approach we explore is conceptually different, reminiscent more of
the standard treatment of interfaces between media in hyperbolic systems using
a junction condition. In this approach we regard the worltube boundary as a
strict interface, where boundary conditions are set for the numerical evolution
outside the tube. These boundary conditions are obtained (at each time step)
from solutions of a certain set of first-order ordinary differential equations
(ODEs) along the boundary, formulated in a way that ensures well-posedness of
the evolution scheme.

We formulate each of the two matching approaches quite independently of any
implementation details; indeed, each approach can in principle be implemented
using whichever one's favorite numerical evolution method happens to be
(finite difference or spectral, Cauchy or characteristic, etc.).  Here, to
illustrate the applicability of our two approaches and test their performance,
we present two independent numerical implementations, one for each approach.
For the first approach (matching in a buffer region) we present a
finite-difference implementation in characteristic coordinates. For the second
approach (matching on the boundary) we present a spectral implementation with
Cauchy evolution. For each approach we demonstrate the stability and
convergence of the numerical evolution, compare with analytical solutions where
possible, and explore the dependence of the solutions on the worldtube radius
$R$.

The paper is organised as follows. We begin in
Sec.~\ref{Section:Scalar Toy Model} by
setting up our scalar-field toy model, with
a point scalar charge on a circular geodesic sourcing a linear scalar field on
a Schwarzschild background. We introduce a multipole-mode decomposition to
reduce to problem to 1+1 dimensions, and (at the single-mode level) prescribe a
suitable approximate analytical solution for the scalar field near the scalar
charge, later to populate the interior of the worldtube around the charge.

Section \ref{Section: Matching Methods} explains the general principles of our two
matching approaches, in a language that is divorced from any implementation
details.
Our two particular numerical
implementations---henceforth referred to as ``scheme~I'' and
``scheme~II''---are
described and explored in Secs.~\ref{Section:Scheme I}
to~\ref{section:Scheme2Results}. Section~\ref{Section:Scheme I}
begins
with a detailed description of our numerical method in scheme~I, based on a
finite-difference formula in null coordinates and characteristic evolution.
Particular attention is paid to the development of matching architectures in a
buffer region around the worldtube boundaries.
In Sec.~\ref{section:Results} we  present various validation
tests to demonstrate the stability and numerical convergence of our code and
the correctness of the numerical results, and then focus on exploring the
dependence of the numerical solution on $R$. Sections
\ref{section:Scheme2Method} and \ref{section:Scheme2Results} do the same
for scheme~II, beginning with a detailed
description of our spectral method and detailing the way boundary conditions
are imposed on the worldtube.
Section \ref{Section: Conclusion} contains a recap of our results, and a
discussion of the next steps in the development of our approach to IMRI
modelling.

First, however, let us conclude this introduction with a rough
estimate
of the runtime savings one might hope to achieve with our method.

\subsection{Potential Runtime Savings}
\label{Subsection:Runtime}
As already mentioned, the
approximation error of the
perturbative solution on the worldtube boundary is expected to be
\begin{equation}\label{eq:err_worldtube}
  \varepsilon_\mathrm{WT}~\sim \left(\frac{R}{\mathcal{ R}}\right)^{n+1},
\end{equation}
where $n$ is the order of the analytic solution, and
$\cal R$ is the characteristic length scale
associated with the tidal field of $m_1$ at
$m_2$.
Optimally, the approximation error $\varepsilon_\mathrm{WT}$ should be
comparable to the error $\varepsilon_\mathrm{NR}$ of the NR simulation,
\ie{}
$\varepsilon_\mathrm{WT}\sim \varepsilon_\mathrm{NR}$.  This gives an ``optimal'' worldtube radius
\begin{equation}
  R\sim \varepsilon_\mathrm{NR}^{1/(n+1)}\;\mathcal{ R}.
\end{equation}
For $n=4$ (as presently available for a tidally perturbed
Schwarzschild black hole~\cite{Poisson:2009qj}) the
dependence on $\varepsilon_\mathrm{NR}$ is quite weak.  As an example,
$\varepsilon_\mathrm{NR}=10^{-5}$ and $n=4$ yield
$R\sim 0.1 \mathcal{ R}$.
As a measure of $\cal R$ we may use the Kretschmann
scalar $K=R_{\alpha\beta\gamma\delta}R^{\alpha\beta\gamma\delta}$ associated with the Schwarzschild field of
$m_1$, with Riemann tensor $R_{\alpha\beta\gamma\delta}$. This
gives
$\mathcal{ R}\sim K^{-1/4}\sim 0.4 (D^3/m_1)^{1/2}$, where $D$ is the separation
between the
two black holes. For example, near the end of the inspiral
($D\sim 6m_1$) we have $\mathcal{ R}\sim 6 m_1$,
and an optimal choice of $R\sim 0.6 m_1$.

The efficiency gain of the worldtube method arises from the weakened
CFL condition: The smallest scale on the numerical
grid with a worldtube is $\sim R$ (as long as the worldtube is
smaller than the
more massive BH), while the smallest scale for the
traditional simulation is $\sim m_2$.  Therefore, the CFL
condition
allows a time-step
larger by a factor $\sim R/m_2$.
Assuming a comparable computational cost per time-step between
worldtube and traditional methods, the speed-up will be
\begin{equation}\label{eq:speedup}
  \text{speedup}
  \sim \frac{R}{m_2}
  \sim \varepsilon_\mathrm{NR}^{1/(n+1)}\,\frac{\mathcal{ R}}{m_2}=
  \varepsilon_\mathrm{NR}^{1/(n+1)}\frac{\mathcal{ R}}{m_1}\;q^{-1}.
\end{equation}

Equation~\eqref{eq:speedup} suggests  a potential speed-up proportional to
$q^{-1}\gg 1$,
with the constant of proportionality depending on the target error
$\varepsilon_\mathrm{NR}$, the order of the analytical approximation
$n$ and the length scale $\cal R$, itself
depending on the orbital radius $D$.
For $D$ in the relevant strong-field range between
$\sim 6m_1$ and $\sim 10m_1$,
and with our sample values $n=4$ and $\varepsilon_\mathrm{NR}=10^{-5}$,
the constant of
proportionality is around unity.  Therefore, for
example, a speed-up by a factor 100 seems feasible for mass-ratio
$1:100$.
To phrase this differently, the computational cost of evolving for one
orbit with the worldtube approach could be similar to evolving one
orbit of a comparable-mass BBH with traditional NR methods at the same
numerical error $\varepsilon_\mathrm{NR}$.

We caution that our estimate here is extremely crude.
For one, the error scaling in Eq.~\eqref{eq:err_worldtube}
turns out not to hold in that precise form in our actual numerical
implementations, as
described below.  Moreover, Eq.~\eqref{eq:speedup} assumes that
time-stepping error is always subdominant, which may only hold for high-order
time-stepping schemes like those employed by the SpEC
code~\cite{Boyle:2019kee}.
And even if the substantial speed-up of Eq.~\eqref{eq:speedup} can be
realized, high-mass-ratio simulations will remain more challenging
than comparable mass simulations, because the duration of the inspiral
increases with more extreme mass-ratios.  A tighter
$\varepsilon_\mathrm{NR}$ might also be required at more extreme mass ratios, to
resolve the smaller amplitude of the
gravitational waves and to preserve phase accuracy over the longer
inspiral.

\section{Scalar-field Toy Model}\label{Section:Scalar Toy Model}

Our toy model replaces the smaller black hole with a pointlike test particle
endowed with a scalar charge $e$. The particle is in a
circular geodesic orbit around the larger object, taken to be a Schwarzschild
black hole of mass $M$.  The orbiting charge sources a
linear scalar field $\Phi$, which satisfies the Klein-Gordon
equation
\begin{equation}  \label{eqn:1}
  g^{\alpha\beta}\nabla_{\alpha}\nabla_{\beta}\Phi(x)=-4 \pi \rho(x)\, .
\end{equation}
Here $\nabla_\alpha$ is the covariant derivative compatible with the
background Schwarzschild metric $g_{\alpha\beta}$, and
$\rho(x)$ is the scalar charge density, represented by the
distribution
\begin{equation}\label{eq:delta-source}
  \rho(x)=e \int_{-\infty}^{\infty}
  \frac{\delta^4[x^\mu-x^\mu_p(\tau)]}{\sqrt{-g}} d\tau\, ,
\end{equation}
in which $x^{\mu}_p$ denotes the coordinates of the particle's
worldline, parametrised with proper time
$\tau$, and $g$ is the determinant of
$g_{\alpha\beta}$. In a Schwarzschild coordinate system attached to the
background Schwarzschild  geometry we have, for our circular orbit,
$r_p:= x_p^r=\text{const}$, and, without loss of generality, we set
$\theta_p:=x_p^\theta\equiv\pi/2$. The particle's geodesic orbit then has a tangent
four-velocity given (in Schwarzschild coordinates $t,r,\theta,\phi$) by
\begin{equation}
  u^\alpha := dx^\alpha_p/d\tau =  \gamma \left(1,\, 0,\, 0,\, \Omega\right)\, ,
\end{equation}
where $\Omega:=(d\phi_p/d\tau)/(dt_p/d\tau) = (M/r_p^3)^{1/2}$ is the orbital angular velocity with respect to
time $t$, and $\gamma:=(1-3M/r_p)^{-1/2}$ is a gravitational
redshift factor. In terms of time $t$, the particle's
Schwarzschild coordinates are
\begin{equation}\label{eq:coords}
  x_p^\alpha = \left(t,r_p,\frac{\pi}{2},\Omega t\right)\, ,
\end{equation}
where, again without loss of generality, we have set $\phi_p=0$ at
$t=0$.

Our toy model makes a further simplification: rather than tackling the field
equation~\eqref{eqn:1} in the 3+1D spacetime, we separate it into
spherical-harmonic multipole modes (taking advantage of the background's
spherical symmetry), and solve for each multipole of the field in 1+1D
(time+radius). To achieve this, we write
\begin{equation}
  \Phi=\frac{e}{r}\sum_{\ell=0}^{\infty}\sum_{m=-\ell}^{\ell}\Psi_{\ell m}(r,t)Y_{\ell m}(\theta,\phi)\, ,
  \label{eqn:psi}
\end{equation}
where $Y_{\ell m}(\theta,\phi)$ are standard spherical harmonics, defined on
2-spheres $r=\text{const}$ around the large black hole, and the factor
$\frac{1}{r}$ is introduced for later convenience. We insert the
expansion~\eqref{eqn:psi} into Eq.~\eqref{eqn:1}, and on
the right-hand side of the latter we substitute  the completeness
relation
$\delta(\theta-\theta_p)\delta(\phi-\phi_p)/\sin\theta= \sum_{\ell m}Y_{\ell m}(\theta,\phi)\bar{Y}_{\ell m}(\theta_p,\phi_p)$, where an overbar denotes complex conjugation. By
virtue of the orthogonality of the $Y_{\ell m}$ functions, one
immediately obtains a separate equation for each of the time-radial functions
$\Psi_{\ell m}(r,t)$. The equation reads
\begin{equation}
  \frac{\partial^2 \Psi_{\ell m}}{\partial t^2}-\frac{\partial^2\Psi_{\ell m}}{\partial {r^*}^2}+V_{\ell}(r)\Psi_{\ell m}= \\
  S_{\ell m}(t)\delta(r^*-r_p^*) \, ,
  \label{eqn:overall}
\end{equation}
where
\begin{equation}
  V_\ell(r)= \left(1-\frac{2M}{r}\right)\left(\frac{\ell(\ell+1)}{r^{2}}+\frac{2M}{r^{3}}\right),
\end{equation}
and
\begin{equation}\label{eq:S}
  S_{\ell m}(t)= \frac{4\pi}{\gamma r_p}\, \bar{Y}_{\ell m}\left(\frac{\pi}{2},\Omega t\right)\, .
\end{equation}
Here we have introduced the tortoise radial coordinate $r^*= r+2M\ln[r/(2M)-1]$,
with $r^*_p:=r^*(r_p)$.

\begin{figure}
  \includegraphics[width=\columnwidth]
  {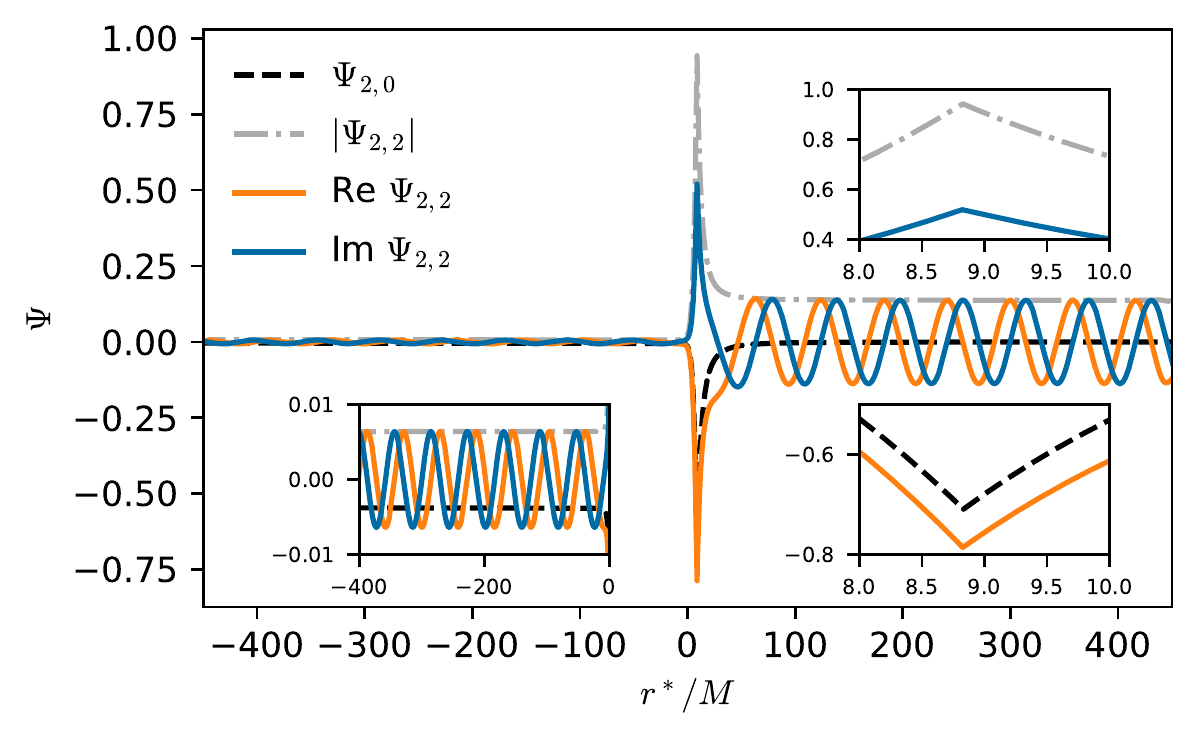}
  \caption{
    \label{fig:FDS_waveform_inset}
    Illustration of the form of physical solutions to the modal
    scalar-field equation~\eqref{eqn:overall}, with a source
    corresponding to a scalar charge on a circular geodesic orbit
    of radius $r_p=7M$ ($r_p^*\approx 8.83M$). Shown here are $\Psi_{22}$
    and $\Psi_{20}$ as functions of the radial coordinate $r^*$ at some
    constant value of the time $t$. The field $\Psi_{22}$, which is
    radiative, is computed using the code presented in
    Subsec.~\ref{subsec:algorithm}; it displays scalar waves emanating from
    the particle down towards the event horizon ($r^*\to-\infty$) and
    out towards infinity ($r^*\to\infty$). The field $\Psi_{20}$, which
    is static, is given analytically in Eq.~\eqref{eqn:analytical}.
    For all values of $\ell,m$, the field is continuous at the location
    of the particle but has a finite jump discontinuity in its first radial
    derivative there.
  }
\end{figure}

Equation~\eqref{eqn:overall} is the basic field equation of our
toy model, and in the rest of this work we apply our excision method
to it in order to develop our strategy and test its performance. We
aim to construct a solution of~\eqref{eqn:overall} subject to
``physical'' boundary conditions, namely that there is no radiation coming in
from past null infinity or out of the past event horizon; we refer to
that solution as the ``physical'' one. For benchmarking, it is useful to have
at hand the actual physical solution of Eq.~\eqref{eqn:overall} (without
a worldtube), and for
that purpose we have developed a simple time-domain
numerical code capable of accurately computing $\Phi_{\ell m}$ for
given mode numbers $\ell,m$ and orbital radius
$r_p$. The algorithm of our code, to be described in
Sec.~\ref{subsec:algorithm}, is based on characteristic evolution with a
second-order-convergent finite-difference formula, with the
$\delta$-function source term incorporated by way of imposing
suitable jump conditions along the particle's worldline (see
Sec.~\ref{subsec:algorithm} for details). The typical form of the solution is
illustrated in Fig.~\ref{fig:FDS_waveform_inset}, showing a
$t = \mathrm{constant}$ snapshot of the field $\Psi_{22}$.
Notable features of the solution are (i) scalar-field waves (of frequency
$m\Omega$) that emanate from the particle and show in the outer
``wave zone'', $r^*\gg M$; (ii) scalar-field waves (again of
frequency $m\Omega$ but typically of a lower amplitude) going
into the black hole, visible at $r^*\ll -M$; and (iii) the cusp in
the scalar field at the particle's location, where $\Psi_{\ell m}$ is
continuous but has a finite jump in its first radial derivative. Similar
features characterise the solution for other $\ell$ and
$m\ne 0$ modes.

For $m=0$ (axially symmetric) modes of the scalar-field
perturbation, the source $S_{\ell m}$ becomes time-independent, and
the physical solution is static. The field equation~\eqref{eqn:overall}
then reduces to an ordinary differential equation, and admits
simple \emph{analytical} solutions. Such solutions are particularly
useful for benchmarking purposes, and they will serve us well in that capacity
later in our analysis. For a ``physical'' $m=0$ field we look
for a static solution of Eq.~\eqref{eqn:overall} for which the modal
Klein-Gordon field $\Phi_{\ell 0}(r):= \Psi_{\ell 0}(r)/r$ is bounded on the event horizon and
falls off at infinity. It is not hard to see that these conditions define a
unique solution for each $\ell$. The solution is given by
\begin{equation}
  \label{eqn:analytical}
  \begin{split}
    \Psi_{\ell 0}(r)= \frac{r r_p}{M} S_{\ell 0}\Big(& Q_{\ell}(z_p)P_{\ell}(z)\Theta(r_{p}-r)  \\
    & + Q_{\ell}(z)P_{\ell}(z_p)\Theta(r-r_{p})\Big)\, ,
  \end{split}
\end{equation}
where $P_{\ell}$ and $Q_{\ell}$ are Legendre
functions of the first and second kind, respectively, with the arguments
$z:=r/M-1$ and $z_p:=r_p/M-1$, and
$\Theta(\cdot)$ is the Heaviside step function. An example of such a
static solution, with $\ell=2$, is also shown in
Fig.~\ref{fig:FDS_waveform_inset}. The static modes, too, are continuous at the
location of particle, and display a finite jump discontinuity in the first
radial derivative there.

\subsection{Local Approximate Solution}
\label{Subsection:Local Approximate Solution}

In our toy model we replace the actual solution in a worldtube surrounding
the particle's worldline with an analytical approximation
$\Psi^A_{\ell m}$.  The analytical solution consists of two terms: a
``puncture'' field $\Psi^\mathcal{P}_{\ell m}$, which captures the local
irregularity in (\ie{}, discontinuous derivatives of) the
field at the particle, and
a ``regular'' field $\Psi^\mathcal{R}_{\ell m}$, which accounts for the remaining,
smooth part of the local field. Both these terms are expressed as a power
series in the distance to the worldline, truncated at a certain order (to be
referred to as ``the order'' of the analytical model). The expansion
coefficients of $\Psi^\mathcal{P}_{\ell m}$ can be determined analytically from the
field equation~\eqref{eqn:overall} using a local asymptotic analysis, as
we explain below, and are fixed in advance in our model. The expansion
coefficients of $\Psi^\mathcal{R}_{\ell m}$, on the other hand, can only be
determined by matching to the external field outside the worldtube; these
coefficients remain \emph{a priori} unknown, and they are to be
determined dynamically during the numerical evolution as described in later
sections. In the rest of this section we describe the construction of a
suitable local analytical model $\Psi^A_{\ell m}$ for the scaler field.
Preliminary considerations regarding the construction of such a model in the
3+1D gravity problem of our ultimate interest are discussed in Sec.\
\ref{Section: Conclusion}.

We begin with the construction of a suitable puncture field
$\Psi^\mathcal{P}_{\ell m}$. Recalling our observation that the physical solution
is continuous but has a finite jump discontinuity in its first radial
derivative at the particle, we introduce the ansatz
\begin{equation}
  \label{eqn:PunctureAnsatz}
  \Psi^\mathcal{P}_{\ell m}(r,t)=\lvert \Delta r \rvert \sum_{j=1}^{n}a_{j \ell m}(\Delta r)^{j-1} S_{\ell m}(t)\, ,
\end{equation}
where $\Delta r := r-r_p$. Our choice of time dependence here makes sense,
because the source function $S_{\ell m}(t)$ depends on
$t$ harmonically, via the factor $e^{-im\Omega t}$
implicit in $\bar{Y}_{\ell m}\left(\frac{\pi}{2},\Omega t\right)$ in Eq.~\eqref{eq:S}, and the
retarded solution inherits this harmonic time dependence.
We terminate the expansion at order $(\Delta r)^n$ for some
$n\geq 1$, referring to the resulting field as an
``$n$th-order puncture'', denoted $\Psi_{\ell m}^{\mathcal{P}(n)}$.

The constant coefficients $a_{j \ell m}$ in Eq.~\eqref{eqn:PunctureAnsatz}
are determined by
substituting~\eqref{eqn:PunctureAnsatz} in the field
equation~\eqref{eqn:overall}, re-expanding in powers of
$\Delta r$,
and then demanding that the resulting equation is satisfied at the particle as
a distributional equality. This produces a hierarchy of algebraic equations for
$a_{j \ell m}$, which we can solve recursively order by order in
$\Delta r$. More specifically, once~\eqref{eqn:PunctureAnsatz} is
substituted in~\eqref{eqn:overall}, the requirement that the
delta-function terms balance in the equation immediately determines
$a_{1 \ell m}$. Then, the requirement that the remaining discontinuity
vanishes at $\mathcal{O}(\Delta r^0)$ determines $a_{2 \ell m}$ in terms
of $a_{1 \ell m}$, the requirement that it  vanishes at
$\mathcal{O}(\Delta r^1)$ determines $a_{3 \ell m}$ in terms of
$a_{1 \ell m}$ and $a_{2 \ell m}$, and so on. For the first
five coefficients one obtains, in this fashion,
\begin{subequations}
  \begin{align}
    a_{1 \ell m} & =  -\frac{1}{2f_p}\,  ,
    \\
    a_{2 \ell m} & =  \frac{M}{2f_p^2 r_p^2}\, ,
    \\
    a_{3 \ell m} & =  \frac{r_p^4m^2\Omega^2-\lambda r_p^2 f_p -2M(3r_p-2M)}{12f_p^3 r_p^4}\, ,
    \\
    a_{4 \ell m} & =  \frac{\lambda r_p^3 f_p -3Mr_p^4m^2\Omega^2 +2M(3r_p^2-4M r_p+2M^2)}{12f_p^4 r_p^6},
    \\
    a_{5 \ell m} & = \frac{1}{240 f_p^5 r_p^8}
    \Big[2r_p^4 m^2\Omega^2\left(\lambda r_p^2 f_p +2M(11r_p+13 M)\right)
    \nonumber                                                                                              \\
                 & \quad - r_p^8 m^4 \Omega^4
    -2\lambda r_p^2 f_p(9r_p^2 +2M r_p-4M^2)
    \nonumber                                                                                              \\
                 & \quad  -24M(5r_p^3-10 M r_p^2+10M^2 r_p-4M^3)\nonumber                                  \\
                 & \quad - \lambda^2 r_p^4 f_p^2 \Big]\, ,
  \end{align}
\end{subequations}
where $f_p:=f(r_p)=1-2M/r_p$ and $\lambda:=\ell(\ell+1)$.
With this, we have all that we need to construct puncture fields through fifth
order.

Next, consider the remaining piece of the local field, $\Psi_{\ell m}^{\mathcal{R}(n)}$,
which we now \emph{define} as the difference
$\Psi_{\ell m}-\Psi_{\ell m}^{\mathcal{P}(n)}$ between the full physical field and the
$n$th-order puncture field, expanded in
$\Delta r$, with the expansion truncated at
$\mathcal{O}(\Delta r^n)$.
Since, by construction, $\Psi_{\ell m}^{\mathcal{P}(n)}$ has
the same singular structure as $\Psi_{\ell m}$ through
$\mathcal{O}(\Delta r^n)$, the so-defined field $\Psi_{\ell m}^{\mathcal{R}(n)}$ is
smooth, and takes the form of a polynomial:
\begin{equation}
  \Psi_{\ell m}^{\mathcal{R}(n)}(t,r)=\sum_{j=0}^n \psi^\mathcal{R}_j(t)(\Delta r)^j
  \, .
  \label{eqn:regular}
\end{equation}
The $n+1$ coefficients $\psi^\mathcal{R}_k(t)$ (their
$\ell,m$ indices suppressed for brevity)
are \emph{a priori} unknown; they are to be determined by matching
to the numerical field outside the worldtube at each time step in the numerical
evolution, as we describe in the next section.

Our full $n$th-order analytical approximate field inside
the worldtube is given by
\begin{equation}\label{eq:psiA def}
  \Psi_{\ell m}^{A(n)}(t,r;\psi^\mathcal{R}_k) = \Psi_{\ell m}^{\mathcal{P}(n)}(t,r) +\Psi_{\ell m}^{\mathcal{R}(n)}(t,r;\psi^\mathcal{R}_k)\, ,
\end{equation}
where our notation reminds the reader that $\Psi_{\ell m}^{A(n)}$ inherits
from
$\Psi_{\ell m}^{\mathcal{R}(n)}$ a parametric dependence on the $n+1$
time-dependent coefficients $\psi^\mathcal{R}_k:=\{\psi^\mathcal{R}_0(t),\ldots,\psi^\mathcal{R}_n(t)\}$.
We use the field
$\Psi_{\ell m}^{A(n)}$ to populate the interior of the excision worldtube in
our numerical simulations, with $\psi^\mathcal{R}_k$ determined by matching
at each time step. The ``approximate'' nature of $\Psi_{\ell m}^{A(n)}$ comes
from the finite truncation of the expansion in $\Delta r$ at order
$n$. Note that, due to the finite truncation, our
definition of $\Psi_{\ell m}^{A(n)}$ is attached to our particular choice of a
distance expansion parameter: using \eg{}
$\Delta r^*$ instead of
$\Delta r$ would yield a slightly different (but equally valid)
analytic approximation. Note also that $\Psi_{\ell m}^{A(n)}$ cannot be ``made
exact'' (even in principle) with a fine-tuned choice of the parameters
$\psi^\mathcal{R}_k$, since these parameters control only the smooth piece
of the field and cannot correct the error in the non-smooth piece caused by the
finite truncation of $\Psi_{\ell m}^{\mathcal{P}(n)}$. The error in
$\Psi_{\ell m}^{A(n)}$ is inherent, and can only be controlled by varying the
model order $n$ (or the worldtube radius).

This concludes the formulation of our 1+1D scalar-field toy model. In the next
section we formulate two (alternative) matching strategies for the field
in and outside of the worldtube, and in later sections we use our toy model to
test the implementation of each of these strategies. In the rest of
the paper the $(n)$ superscript and $\ell m$
indices are mostly suppressed, for brevity.

\section{Two matching approaches}
\label{Section: Matching Methods}
Using our toy model, we now develop our two matching approaches: one based on
matching
the
numerical field to the analytical approximation $\Psi_{\ell m}^{A}$ in an
open region around the particle, and another based on junction conditions
imposed on the
surface of the excision region. In this section we describe the principles
behind each approach, keeping the descriptions independent of any particular
choice of discretization. %
\subsection{First approach: matching in a buffer region}
\label{Subsection:Matching Method I}
\begin{figure}
  \centering
  \includegraphics[width=.75\columnwidth]
  {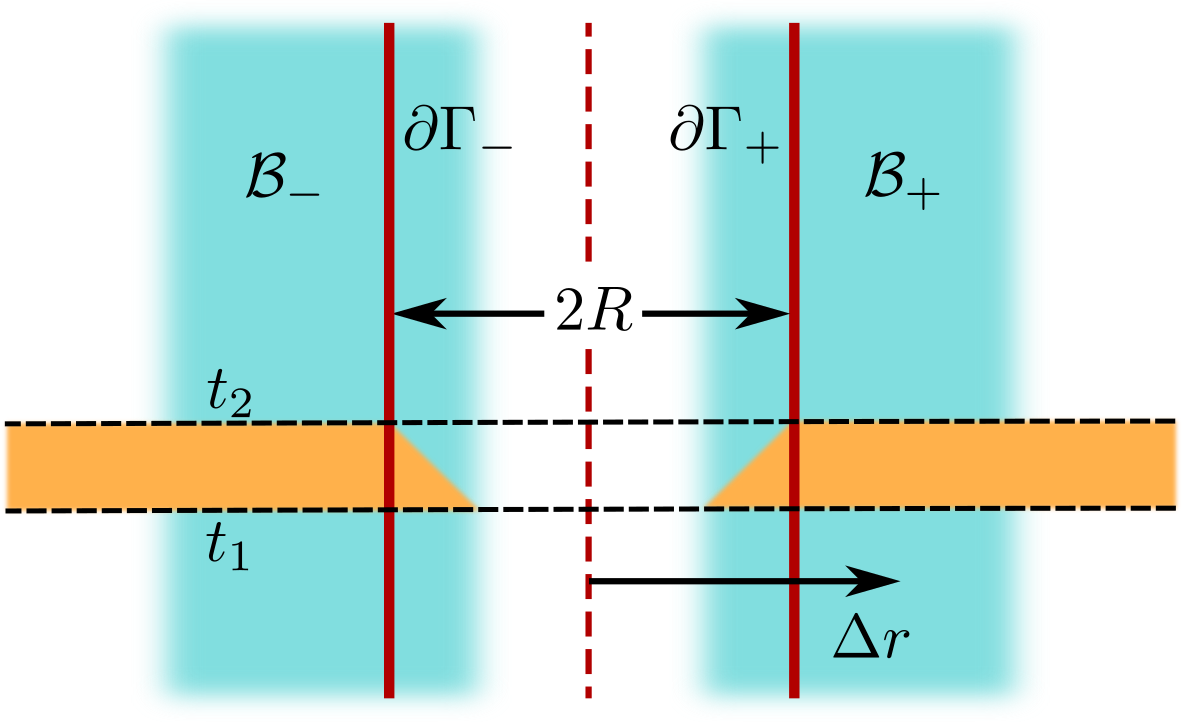}
  \caption{
    \label{fig:matching_method1} Spacetime diagram
    (in $t,r$ coordinates) illustrating
    the various regions involved in our first matching approach.
    The vertical dashed line indicates the particle's
    worldline at constant orbital radius.
    The disjoint shaded regions $\mathcal{B}_\pm$ make up the buffer
    region $\mathcal{B}=\mathcal{B}_-\cup \mathcal{B}_+$ where the matching occurs.
    The excision region $\Gamma$ has a disjoint boundary
    $\partial\Gamma = \partial\Gamma_+\cup\partial\Gamma_-$
    that lies within $\mathcal{B}$. We carry out the matching by
    expanding the numerical field $\Psi^N$ in powers of $\Delta r$
    and equating the coefficients in the expansion to the coefficients
    in $\Psi^A$. The orange shaded region shows the domain of dependence
    of the field outside $\Gamma$ at time $t_2$, given data at time $t_1$.}
\end{figure}

Our first approach is inspired by the method of matched
asymptotic expansions.
Let us recall how that method is traditionally applied to the small-mass-ratio
limit
of the binary problem~\cite{Poisson:2011nh}, with reference to
Fig.~\ref{fig:excision_cartoon}. The binary is treated as a one-parameter system,
using $m_1$ as an overall length scale and the small mass
ratio $q=m_2/m_1$ as a small parameter (in this treatment, the
length $\mathcal{R}$ utilized in the introduction is replaced with
$m_1$ rather than being treated as an independent scale). In
the
bulk of the binary spacetime, one expands the metric in powers of
$q$. Sufficiently near $m_2$, at
distances $\sim m_2$, such an expansion breaks down because the
gravity of $m_2$ dominates over that of
$m_1$. One then constructs a complementary local
approximation using an expansion in powers of $q$ while
holding $s/m_2$ fixed, where
$s$ is a suitable measure of
spatial distance from the companion's representative worldline. By holding
$s/m_2$ fixed, this expansion zooms in on the region
$s\sim m_2\ll m_1$, such that $s/m_1\sim q$. In a buffer
region $m_2\ll s\ll m_1$, $s/m_1$ and
$m_2/m_1$ are both small, and the
exterior and interior approximations must agree. This requirement translates
into a precise matching condition: if the local
approximation inside the worldtube is re-expanded in powers of
$q$ at fixed $s$ (no longer holding
$s/m_2$ fixed), and the external solution is re-expanded in
powers of
$s / m_1$, then in both cases one arrives at a double series in
$q$ and $s / m_1$, which should be a good
approximation in the buffer region. The matching condition states that because
they are expansions of the same metric, the two double expansions must agree
term by term.

Now consider the translation of these notions into our toy problem. For
simplicity we use $t$--$r$ coordinates
in our description rather than the double-null coordinates we ultimately use in
Scheme I. The setup  in the $t$--$r$
plane is illustrated in Fig.~\ref{fig:matching_method1}. We define a buffer region
$\mathcal{B}$, made up of disjoint regions $\mathcal{B}_\pm$,
in which $\Delta r$ is small compared to $M$.
We then define an excision region
$\Gamma=[-R\leq \Delta r\leq R]$ around the particle's orbital radius, with boundaries
$\partial\Gamma_\pm$ lying in $\mathcal{B}_\pm$. We loosely imagine
that outside of $\Gamma$, we solve for $\Psi$
using  the homogeneous field equation, Eq.~\eqref{eqn:overall}
with the right-hand side set to zero, and that inside,
we use the analytical approximation $\Psi_{\ell m}^A$. For convenience,
we label the numerically evolved field outside $\Gamma$ as
$\Psi_{\ell m}^N$. However, following the dictates of matched expansions,
we operate under the principle that in $\mathcal{B}$,
$\Psi_{\ell m}^N$ and $\Psi_{\ell m}^A$ can be used
interchangeably.

As in the method of matched expansions, to match the two fields we expand
$\Psi^N$ to put it in the same form as $\Psi^A$
(we hereafter omit the indices $\ell m$ for brevity).
In each of the regions $\mathcal{B}_\pm$, $\Psi^N$ can be
approximated by a power series in $\Delta r$,
\begin{equation}
  \label{psiN local approx}
  \Psi^{N\pm}(t,r)
  = \sum_{j=0}^{n} \psi^\pm_j(t)(\Delta r)^j  + \mathcal{O}(\Delta r^{n+1})\, ,
\end{equation}
where $\Psi^{N\pm}$ denotes the restriction of
$\Psi^N$ to $\mathcal{B}_\pm$. The coefficients
$\psi^\pm_j$ can be found by projecting $\Psi^N$ onto
the basis of functions $\{(\Delta r)^0,\ldots,(\Delta r)^n\}$ using
a suitable inner product
\begin{equation}\label{inner product}
  \langle x,y\rangle = \int_{\mathcal{B}'}x(r)y(r)dr.
\end{equation}
Here $\mathcal{B}'$ is some open interval (with fixed
$t$) in $\mathcal{B}$, or a collection of
multiple such intervals; we consider the choice of integration domain below.
Our matching condition is then that the coefficients in
Eq.~\eqref{psiN local approx} are identical to the coefficients in
$\Psi^A$:
\begin{equation}\label{matching condition}
  \psi^\pm_j(t) = \psi^\mathcal{R}_{j}(t) + \psi^{\mathcal{P}\pm}_j(t)\, ,
\end{equation}
where $\psi^{\mathcal{P}\pm}$ are the puncture coefficients, which can be read
off Eq.~\eqref{eqn:PunctureAnsatz}:
$\psi^{\mathcal{P}\pm}_j(t) = \pm a_{j \ell m}S_{\ell m}(t)$.

To satisfy Eq.~\eqref{matching condition}, we must ensure that
$\psi^\pm_j$ satisfies the same jump conditions as
$\psi^{\mathcal{P}\pm}_j$, meaning $\psi^+_j - \psi^-_j = \psi^{\mathcal{P}+}_j - \psi^{\mathcal{P}-}_j$.
If we were to
construct the approximations~\eqref{psiN local approx} separately in their
respective regions $\mathcal{B}^\pm$, with no regard to the relationship
between them, then these jump conditions would \emph{not} be
precisely satisfied. We enforce the correct jumps by demanding that the
difference $\Psi^N - \Psi^\mathcal{P}$ is approximated by the smooth field
$\Psi^\mathcal{R}$,
\begin{align}\label{matching condition 2}
  \Psi^N(t,r) - \Psi^\mathcal{P}(t,r) = \sum_{j=0}^n \psi^\mathcal{R}_j(t)(\Delta r)^j
  + \mathcal{O}(\Delta r^{n+1}) \, .
\end{align}
This requires choosing the integration domain in Eq.~\eqref{inner product} to
have support in both ${\mathcal B}_+$ and ${\mathcal B}_-$.
Taking the inner product of
Eq.~\eqref{matching condition 2} with $(\Delta r)^k$ and discarding
higher-order terms, we obtain a linear system for $\psi^\mathcal{R}_j$,
\begin{equation}\label{matching linear system}
  \sum_{j=0}^n A_{jk} \psi^\mathcal{R}_j(t) = b_k(t)~~~~~\text{for }k=0,\dots,n\, ,
\end{equation}
with $A_{jk} = \langle (\Delta r)^j,(\Delta r)^k\rangle$ and
$b_k = \langle \Psi^N - \Psi^\mathcal{P},(\Delta r)^k\rangle$.
We note that the
solution to Eq.~\eqref{matching linear system} yields the
$L^2$ best approximation
of $\Psi^N - \Psi^\mathcal{P}$. Since this equation must hold for all
$t$, it also implies an analogous equation for
$\partial_t\psi^\mathcal{R}_j$, which is required for a Cauchy evolution.

To enforce the matching condition in a numerical evolution, we can use the
following scheme:
\begin{enumerate}
  \item Suppose that at time $t_1$, we have data for
        $\Psi^N$ and $\partial_t\Psi^N$ everywhere outside
        $\Gamma$.
  \item Determine the approximate solution $\Psi^A(t_1)$ and
        $\partial_t\Psi^A(t_1)$ by solving
        Eq.~\eqref{matching linear system} and the analogous
        equation for $\partial_t\psi^\mathcal{R}_j$. We then have $\Psi$ and
        $\partial_t\Psi$ for all $r$ at time
        $t_1$, given by the field values from Step 1 outside
        $\Gamma$ and by $\Psi^A$ and $\partial_t\Psi^A$
        inside $\Gamma$.
  \item Use the homogeneous equation, Eq.~\eqref{eqn:overall}
        with the right-hand side set to zero, together with the data at
        $t_1$ to obtain $\Psi^N$ at a later time
        $t_2$ everywhere outside $\Gamma$, as
        illustrated in Fig.~\ref{fig:matching_method1}. This requires data
        from inside
        $\Gamma$ at $t_1$, which is provided by
        $\Psi^A(t_1,r)$ and $\partial_t\Psi^A(t_1,r)$.
\end{enumerate}
This can then be repeated indefinitely. Note that the time interval from one
slice to the next is tied to the length scale of the buffer region. The
evolution from $t_k$ to $t_{k+1}$ should only
draw upon data for $\Psi^A$ in the buffer region, implying that
the time intervals must be of order $R$ or shorter.
In principle, this division of spacetime into time intervals need not be
associated with one's numerical discretisation, and the spacetime region
between $t_k$ and $t_{k+1}$ can be spatially
discretised
in any convenient way.

Our description here refers to an evolution between slices of constant
$t$, but it extends straightforwardly to any choice of
slicing, including particularly the characteristic slicing we work with in
Sec.~\ref{Section:Scheme I}. In general, the one-dimensional series
approximation~\eqref{psiN local approx} is replaced by a two-dimensional series
in powers of coordinate distances ($\Delta t$ and
$\Delta r$ or appropriate null coordinates,
for example) from a reference point on the worldline. The inner
product~\eqref{inner product} is then replaced by an integral over a
two-dimensional region. We can also naturally extend the method to an evolution
in $3+1$ dimensions by matching to a local approximation in a
three- or four-dimensional region around the companion.

One additional aspect of this matching approach that should be noted is that it
does not inherently impose any degree of differentiability across
$\partial\Gamma_\pm$, except in the limit
$n\to\infty$. This contrasts with our second matching approach,
which
we describe next.

\subsection{Second approach: matching using junction conditions}
\label{Subsection:Matching Method II}

Our second approach consists of matching the fields and its derivatives
on the surface of the worldtube.
As in the first approach, the regular part $\Psi^\mathcal{R}$ is a
truncated Taylor
series in $\Delta r^*$.
However, here the coefficients are determined through a
Hermite interpolation using values of the field
and its derivatives up to a certain order $d$ on
$\partial \Gamma_-$
and $\partial \Gamma_+$,
\ie{} we solve the system
\begin{equation}
  \label{eq:approach2_matching}
  \begin{split}
    \left.
    \partial_{r^*}^k \left (
    \Psi^N - \Psi^\mathcal{P}
    - \sum_{j=0}^{2d+1} \psi^\mathcal{R}_j(\Delta r^*)^j
    \right )
    \right |_{\partial \Gamma_\pm} = 0 & \\
    \text{for }k=0,\dots,d & \, ,
  \end{split}
\end{equation}
which is a system of $(2d+2)$ linear equations for the
$(2d+2)$
coefficients $\psi^\mathcal{R}_j$.
Unlike in the previous scheme, here we take the expansion order of the puncture
field, $n_\mathcal{P}$,  and that of the
regular field, $n_\mathcal{R} = 2 d + 1$, as independent.
The overall convergence of the scheme with respect to $R$
is hence
limited by $n_\mathcal{P}$ and $n_\mathcal{R}$.
The same procedure is carried out for the time derivative of the
regular part $\dot \Psi^\mathcal{R}$ and potentially further
reduction variables.
The Taylor expansions are then used to construct the boundary data
that must be provided on the worldtube.

Boundary conditions can be interpreted and implemented as modifications
to the right-hand sides of the bulk partial differential equations (PDEs),
which
we assume to be strongly hyperbolic, as it is the case for our
wave equation toy model.
The system remains well-posed if the boundary conditions retain
strong hyperbolicity, which is the case when the coefficients of the
series expansion are constructed from non-principal derivatives of
$\Psi^N$ only.
At first sight this severely limits the achievable expansion order,
because we can use at most up to first derivatives of $\Psi^N$
and only field values of $\dot \Psi^N$.
However, this limitation can be overcome by introducing
an auxiliary system of ODEs evolving variables that represent
the derivatives $\partial_{r^*}^k \Psi^N|_{\partial\Gamma_\pm}$ and
$\partial_{r^*}^k \dot \Psi^N|_{\partial\Gamma_\pm}$
up to derivative order $d$.
The ODEs must be formulated compatibly with the
bulk PDEs, which can be done by taking
derivatives of the bulk equations.
The coefficients $\psi^\mathcal{R}_j$ are then computed using these
auxiliary
variables instead of data from the bulk PDEs.
The auxiliary ODE system is solved simultaneously with the PDE
system in a fashion not involving principal (or higher) derivatives of
the PDE variables.

The boundary regular field derivatives
$\partial_{r^*}^k \Psi^\mathcal{R}|_{\partial\Gamma_\pm}$
are related to the coefficients $\psi^\mathcal{R}_j$ through a simple
matrix transform that follows from Eq.~\eqref{eq:approach2_matching}:
\begin{equation}
  \label{eq:approach2_matrix_transform}
  \left.
  \frac{\partial^{k} \Psi^\mathcal{R}}{\partial {r^*}^{k}}
  \right |_{\partial \Gamma_\pm}
  =
  \sum_{j=0}^{2d+1} \psi^\mathcal{R}_j
  \left.
  \frac{\partial^k (\Delta r^*)^j}
  {\partial{r^*}^k}
  \right |_{\partial \Gamma_\pm} ~~~\text{for }k=0,\dots,d \, .
\end{equation}
Hence, for a linear system like the wave equation, this approach of using
auxiliary ODEs is equivalent to evolving the
regular part of the field inside the worldtube using a spectral method,
similar to puncture schemes in self-force calculations, as is discussed
in the next section.
For nonlinear systems however this split into a regular part might not
be possible and the scheme presented here could be a viable prototype
when dealing with such systems.
From the equivalence to a collocation-based spectral method it is possible
to derive a numerically stable scheme to couple the
bulk PDEs to the auxiliary system.
This equivalence also explains how to control non-local effects that
one might expect in an excision scheme. As long as the stability
criteria of the equivalent spectral method are satisfied, this
excision approach will satisfy them as well. These criteria entail
satisfaction of a CFL-like inequality and using
``energy preserving'' boundary conditions.
For a nonlinear field equation, this discretization
using boundary derivatives will no longer be equivalent to a collocation-based
method. However, both approaches converge to the same continuum limit and
all stability criteria should still apply.

Our second matching approach will be developed in full in
Sec.~\ref{section:Scheme2Method}
using a spectral method formulated on Cauchy slices, and its performance will
be explored in Sec.~\ref{section:Scheme2Results}.

\subsection{Error estimates and connection to standard puncture methods}
\label{Subsection:error estimates}

Our excision procedure is similar in some ways to the puncture schemes used in
numerous self-force calculations~\cite{Barack:2018yvs,Pound:2021qin}. However, there is a
crucial difference that we clarify (and motivate) here.

In a standard puncture scheme, one splits the exact field into two pieces,
$\Psi = \tilde\Psi^\mathcal{ P} + \tilde\Psi^\mathcal{ R}$. Here $\tilde\Psi^\mathcal{ P}$ captures the local
singularity at the particle but is attenuated to zero outside some neighborhood
of the particle. For example, it could be the field $\Psi^\mathcal{ P}$ we
work with in this paper but multiplied by a step function
$\theta(R-|\Delta r|)$ that vanishes outside $\Gamma$. Unlike
the field $\Psi^\mathcal{ R}$ that we work with, $\tilde\Psi^\mathcal{ R}$ is
the {\em exact} difference $\tilde\Psi^\mathcal{ R} := \Psi - \tilde\Psi^\mathcal{ P}$.

Using this split, one treats $\tilde\Psi^\mathcal{ R}$ as the field variable,
rearranging Eq.~\eqref{eqn:overall} to formulate a field equation with an
effective source,
\begin{equation}\label{eq:BoxPsiR}
  \Box \tilde\Psi^\mathcal{ R} = S(t)\delta(r^*-r^*_p) - \Box \tilde\Psi^\mathcal{ P} := S^{\rm eff} \, ,
\end{equation}
where for brevity we have defined $\Box := \partial_t^2-\partial^2_{r^*}+V$ and continued to omit
$\ell m$ labels. Equation~\eqref{eq:BoxPsiR} is solved over
the entire domain, without excising a region around the particle, and with the
same boundary conditions on $\tilde\Psi^\mathcal{ R}$ as on
$\Psi$.  In such a scheme, there is no approximation: outside
the support of $\tilde \Psi^\mathcal{ P}$, the solution for
$\tilde\Psi^\mathcal{ R}$ is identical to $\Psi$; inside, one
can add $\tilde \Psi^\mathcal{ P}$ to likewise obtain the exact
$\Psi$.

While this method is well suited to linear field equations, its applicability
to the fully nonlinear Einstein equations is unclear. Due to nonlinearities,
the metric of a tidally perturbed black hole is not a simple sum of singular
and regular pieces, and one cannot simply move a piece of the metric to the
right-hand side of the field equations. The excision methods we explore in this
paper represent an alternative that should extend to the nonlinear problem.
However, they do so at the cost of introducing an approximation: unlike a
traditional puncture scheme, our methods do not yield the exact field
$\Psi$.

First consider the error in our method inside $\Gamma$. In that
region we use the approximation $\Psi^A$, which differs from
$\Psi$ by an amount of order $(\Delta r)^{n+1}$ at best.
This is a best-case estimate because it assumes that our matching methods
enforce the exact values $\frac{1}{j!}\partial_r^j(\Psi-\Psi^\mathcal{ P})\rvert_{r=r_p}$ for the coefficients
$\psi^\mathcal{ R}_j$ in Eq.~\eqref{eqn:regular}. For simplicity, let us
assume this best case.

Now consider the field outside $\Gamma$. More concretely,
consider a bounded region $V$ with $\partial \Gamma$
as one of its boundaries; in a Cauchy evolution, the other boundaries might be
an initial-data surface (outside $\Gamma$) and timelike
boundaries far away, for example. Inside $V$, our field
$\Psi^N$ satisfies the same homogeneous field equation as
$\Psi$, $\Box\Psi^N=0$, but it inherits errors that
propagate out from $\Gamma$. Those errors can be understood by
writing $\Psi^N$ in a Kirchhoff integral
form~\cite{Poisson:2011nh}. We introduce a retarded Green's function
satisfying
\begin{equation}\label{eq:BoxG}
  \Box G(\bm{x},\bm{x}') = \Box' G(\bm{x},\bm{x}') = \delta^2(\bm{x},\bm{x}') \, ,
\end{equation}
where $\bm{x}=(t,r^*)$, $\Box':=\partial_{t'}^2-\partial^2_{r'^*}+V(r')$, and
$\delta^2(\bm{x},\bm{x}'):=\delta(t-t')\delta(r^*-r'^*)$. If we now take any point $\bm{x}\in V$, then
the equations~\eqref{eq:BoxG} and $\Box\Psi^N=0$ imply the
identity
\begin{multline}
  \Psi^N(\bm{x}')\Box' G(\bm{x},\bm{x}') \\
  - G(\bm{x},\bm{x}')\Box'\Psi^N(\bm{x}') = \Psi^N(\bm{x}') \delta^2(\bm{x},\bm{x}') \, .
\end{multline}
Integrating this equation over all $\bm{x}'\in V$ and then using
integration by parts, we obtain the Kirchhoff representation
\begin{align}
  \Psi^N(\bm{x}) & = \int_V \left[\Psi^N(\bm{x}')\Box' G(\bm{x},\bm{x}')\right.\nonumber                     \\
                 & \qquad\qquad \left. - G(\bm{x},\bm{x}')\Box'\Psi^N(\bm{x}')\right] d^2x' \, ,\nonumber    \\
                 & = \int_{\partial V} \left[\Psi^N(\bm{x}')\partial_{n'} G(\bm{x},\bm{x}') \right.\nonumber \\
                 & \qquad\qquad \left.- G(\bm{x},\bm{x}')\partial_{n'}\Psi^N(\bm{x}')\right]ds' \, .
  \label{eq:kirchhoff_representation}
\end{align}
Here the coordinate area element in $V$ is
$d^2x'=dt'dr'^*$. $\partial_{n'}$ is the partial derivative
normal to the boundary $\partial V$, and $ds'$ is
the coordinate line element on the boundary. For us the relevant portions of
$\partial V$ are the worldtube boundaries $\partial\Gamma_\pm$,
where $\partial_{n'}=\mp\partial_{r'^*}$ and $ds'=dt'$.

From the Kirchhoff form, we see that $\Psi^N$ inherits two
errors, respectively proportional to the errors in $\Psi^N\rvert_{\partial\Gamma}$ and
$\partial_{r^*}\Psi^N\rvert_{\partial\Gamma}$. Suppose that $\Psi^N=\Psi^A +\mathcal{O}(|\Delta r|^{n+1})$ in an open
neighbourhood of $\partial\Gamma$, as we seek to enforce in our first
matching approach. Then $\Psi^N\rvert_{\partial\Gamma}$ has an error of order
$R^{n+1}$, but $\partial_{r^*}\Psi^N\rvert_{\partial\Gamma}$ has an error of order
$R^{n}$. The field $\Psi^N$ therefore differs
from $\Psi$ by $\mathcal{O}(R^{n})$ throughout
$V$. This represents a loss of one order relative to the
$\mathcal{O}(R^{n+1})$ scaling that one might naively expect. Our numerical
analysis in Sec.~\ref{section:Results} confirms this $\mathcal{O}(R^{n})$
error estimate. However, our second matching approach more directly controls
derivatives at $\partial\Gamma$, and in Sec.~\ref{section:Scheme2Results} we
find that in certain cases this second approach yields the more rapid,
$\mathcal{O}(R^{n+1})$ convergence.

Analogous error estimates can be obtained for the
3+1D problem using a covariant Kirchhoff
representation of the form (138) in Ref.~\cite{Pound:2009sm}. A similar
estimation might also be possible in fully nonlinear general relativity using
Eq.~(39) of that reference (reproduced from Ref.~\cite{Sciama:1969vtz}).

\section{Scheme~I: Numerical Method}\label{Section:Scheme I}
\subsection{Setup}
\label{setup}
\begin{figure}
  \includegraphics[scale=0.2]
  {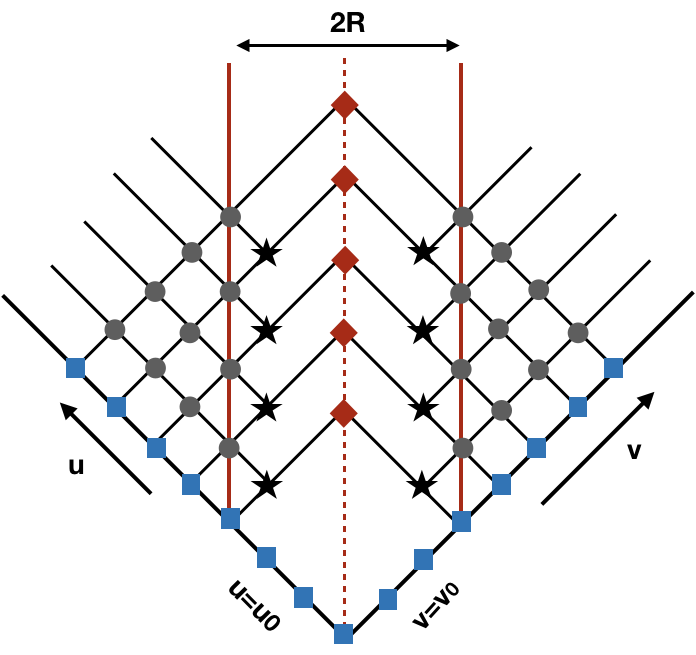}
  \caption{
    \label{fig:FDS_grid_new}
    The 1+1D characteristic mesh used in Scheme~I, with a fixed,
    uniform grid based on Eddington-Finkelstein coordinates $u,v$.
    The particle's circular orbit is represented by the dashed (red) line
    running vertically down the center. The solid (red) vertical lines
    mark the boundaries of the excision worldtube. Outside the worldtube
    we evolve the field equation numerically along characteristic rays
    (as 	described in the text) using a finite-difference formula
    detailed in Sec.~\ref{subsec:algorithm}. The evolution starts from
    characteristic
    initial data set on the two initial rays $v=v_0$ and $u=u_0$
    (blue square grid points) and proceeds to determine the data
    points in the bulk of spacetime outside the tube and on its
    boundaries (grey circle points). At each time step, a matching
    procedure, described in Sec.~\ref{subsec:matching}, is applied to
    determine the parameters of the approximate analytical solution on
    the remaining sections of the characteristic rays inside the tube,
    and in particular on the two ``ghost'' grid points (black, starred)
    needed at subsequent steps of the evolution.}
\end{figure}

In Scheme~I we use a finite-difference method based on a uniform, fixed
characteristic mesh in Eddington--Finkelstein
coordinates
\begin{equation}
  u=t-r^*, \quad\quad v=t+r^*.
\end{equation}
Figure \ref{fig:FDS_grid_new} illustrates the basic setup.
The numerical evolution starts from characteristic initial data specified on
two initial rays $v=v_0$ and $u=u_0$ (blue,
square grid points in the figure), chosen to intersect at the radius
$r^*=r^*_p$ of the scalar charge's circular orbit (dashed red line
running vertically at the centre of the grid); that is,
$v_0-u_0=2r_p^*$. An excision worldtube of width $\Delta r^*=2R$
is introduced around the orbit, with boundaries at $r^*=r^*_p\pm R$
(vertical red lines), chosen to intersect grid points. The numerical evolution
proceeds along successive characteristic rays using a finite-difference formula
to be described in Sec.~\ref{subsec:algorithm} below. At each ``time step'',
we
integrate first along a $u=\mathrm{const}$ ray starting at
$v=v_0$ and progressing outwards up until tube's left boundary,
and then along the corresponding incoming ray $v=u+2r_p^*$ starting
at $u=u_0$ and progressing inwards down to the tube's right
boundary. At each time step, a matching procedure then follows (described
in Sec.~\ref{subsec:matching} below), in which the value of the analytical
model $\Psi^A$ is determined along the remaining sections of the
two rays inside the tube. In particular, we assign analytical values to two
`ghost' grid points adjacent to the tube's boundary in its interior (black
starred points in the figure); these are needed for the subsequent time step of
the numerical evolution.

Our characteristic mesh has a fixed (pre-set) stepping interval
$h$ in both $u$ and
$v$.
The value of the field at a grid point with coordinates $(u,v)$
outside the tube (or on its boundary) is determined by our finite-difference
formula based only on previously obtained values at the three grid points with
coordinates ($u-h,v$), ($u,v-h$) and
$(u-h,v-h)$. This, as we show in Sec.~\ref{subsec:algorithm},
suffices for obtaining a quartic $\mathcal{O}(h^4)$ local convergence and
a
quadratic $\mathcal{O}(h^2)$ global convergence.

The physical initial data for the numerical evolution are, of course, unknown
to us except in the case of stationary, $m=0$ modes, where
the entire solution is known analytically, Eq.~\eqref{eqn:analytical}. We thus
resort to assigning fictitious
initial data, and rely on dissipation of the resulting junk radiation over
time. In post-processing we monitor the level of residual junk radiation,
discard the early, junk-contaminated portion of the evolution, and retain only
the remaining underlying ``physical'', approximately stationary solution. In
practice, we choose to set $\Psi=\Psi^{\mathcal P}$ on the portions of the rays
$v=v_0$ and $u=u_0$ inside the worldtube, and
attenuate smoothly to zero with a Gaussian across the tube's boundary. The
value of the field $\Psi$ on the two complete rays
$v=v_0$ and $u=u_0$ suffices, in principle, to
determine the solution anywhere in the domain of dependence
$u>u_0$ and $v>v_0$.

\subsection{Finite-difference formula}
\label{subsec:algorithm}

We now describe the algorithm used to integrate the wave equation numerically
outside the excision region.
Our method is a standard one, used extensively in self-force literature, legacy
of early work by Lousto
and Price~\cite{Lousto:1997wf}.
Since we evolve numerically only in the vacuum region outside the excision
tube, it suffices to consider
the scalar field equation~\eqref{eqn:overall} in vacuum. In terms of the
$u,v$ coordinates it reads
\begin{equation}
  \partial_{u}\partial_{v} \Psi + \frac{1}{4} V(r) \Psi=0\, ,
  \label{eqn:overalluv}
\end{equation}
where hereafter $\partial_u$ is taken with fixed
$v$ and $\partial_v$ is taken with fixed
$u$.
Our goal is to write a finite-difference version of this equation on the
characteristic grid described above.

To this end, consider a generic vacuum grid point with coordinates
$(u,v)$, and assume the field has been computed in previous
steps at all grid points within the past ``light cone'' of
$(u,v)$ (to the future of the initial surfaces). Consider the
grid `cell' with vertices $(u,v)$, ($u-h,v$),
($u,v-h$) and $(u-h,v-h)$, where, recall,
$h$ is our fixed step size in both $u$
and $v$. To obtain our finite-difference formula, it is
convenient to consider the formal integral of both sides of
Eq.~\eqref{eqn:overalluv} over the area of the grid cell. For the principal
part of the equation we obtain
\begin{equation}
  \begin{split}
    \iint_\mathrm{cell} \partial_{uv}\Psi  dudv &= \Psi(u,v)-\Psi(u,v-h)\\
    & -\Psi(u-h,v)+\Psi(u-h,v-h)\, ,
  \end{split}
  \label{eqn:fds_principal}
\end{equation}
which is \emph{exact}, and does not incur any
finite-differencing error.
For the potential term in Eq.~\eqref{eqn:overalluv} we obtain
\begin{align}\label{eqn:fds_potential}
  \frac{1}{4}\iint_\mathrm{cell} V(r) \Psi dudv = &
  \frac{h^{2}}{8} V(r_c) \Big[\Psi(u,v-h) + \Psi(u-h,v)\Big] \nonumber     \\
                                                  & + \mathcal{O}(h^{4}) ,
\end{align}
where $V(r_c)$ is the value of the potential at the centre of
cell, \ie{} at $r^*_c=(v-u-h)/2$.
Since the cell integral of the right-hand side of Eq.~\eqref{eqn:overalluv} is
zero, combining the above results gives
\begin{align}
  \Psi(u,v) = & \big[\Psi(u,v-h)+\Psi(u-h,v)\big]\bigg[1-\frac{h^2}{8}V(r_{c})\bigg] \nonumber \\
              & -\Psi_1(u-h,v-h)  +\mathcal{O}(h^{4}).
  \label{eqn:FDS+}
\end{align}
This simple finite-difference formula has a local error of
$\mathcal{O}(h^{4})$ at each vacuum grid point. Since the total number of
vacuum grid points scales as $1/h^2$ (for fixed physical grid
dimensions), we expect the global cumulative error to scale like
$h^2$.

We note that the above, quadratically convergent scheme requires only three
input data points to determine the field value at each vacuum point. These
three data points are always available from previous steps of the
characteristic evolution. To calculate points that are on the tube's boundary,
an input data point is required from inside the tube. For this internal point
we use the value of the approximate analytical field $\Psi^A$,
which will have been fitted for in the previous time step of the evolution,
using the procedure described in Sec.~\ref{subsec:matching} below.

\subsubsection{Test evolution with a point particle and no excision}
\label{subsec:TestEvolution}

For test and benchmarking, we have also developed a version of our code that
solves the full inhomogeneous field equation~\eqref{eqn:overall} as it
is, without an excision. In this case the vacuum regions extend to the exposed
scalar charge, and we must modify our finite-difference scheme to account for
the presence of the sourcing particle. Our vacuum
formula~\eqref{eqn:FDS+} still applies at all grid points, except points
sitting directly on the particle's worldline at $r^*=r_p^*$, for
which we need a modified formula.

In $u,v$ coordinates, the inhomoheneous field equation
\eqref{eqn:overall} becomes
\begin{equation}
  \label{uvsource}
  \partial_{u}\partial_{v}\Psi + \frac{1}{4}V(r)\Psi= \frac{1}{4}S(t) \delta(r^*-r^*_p)\, .
\end{equation}
Consider a generic worldline grid point at $(u,v)$, such that
$v-u=2r_p^*$. To write down a finite-difference expression for the
field at $(u,v)$, we again integrate both sides of the equation
over the cell with $(u,v)$ at its upper vertex. Recalling that
(in the continuous limit) solutions are continuous (albeit generally not
differentiable) on the particle's worldline, we find that
Eq.~\eqref{eqn:fds_principal} for the principal part still holds exactly, even
for
cells crossed by the particle. Equation~\eqref{eqn:fds_potential} for the
potential term also holds, but the error term is expected to be of
$\mathcal{O}(h^{3})$ in general, due to the discontinuous derivative. This,
however, would suffice for our purpose, since the number of worldline points
scale only as $1/h$, and so a local error of
$\mathcal{O}(h^{3})$ should lead to an cumulative global error of only
$\mathcal{O}(h^{2})$, consistent with our quadratic-convergence standard.

It remains only to evaluate the cell integral of the right-hand side
of~\eqref{uvsource}. To this end, we recall the form of the source
function $S(t)$, given in Eq.~\eqref{eq:S}; it
depends on $t$ only through the factor
$\bar{Y}_{\ell m}\left(\frac{\pi}{2},\Omega t\right)$, which itself depends on $t$ only
through the factor $e^{-im\Omega t}$. It is therefore convenient here to
write
\begin{equation}
  S(t)=A_{\ell m}e^{-im\Omega t},
\end{equation}
where, we obtain,
\begin{equation}
  A_{\ell m}=\frac{(-1)^{\frac{\ell+m}{2}}}{\gamma r_p}\bigg[\frac{4\pi(2\ell+1)(\ell+m-1)!!(\ell-m-1)!!}{(\ell+m)!!(\ell-m)!!}\bigg]^{1/2}
\end{equation}
when $\ell+m$ is even, or $A_{\ell m}=0$ when
$\ell+m$ is odd. The cell integral over the source can now be
readily evaluated in exact form, giving
\begin{equation}
  \begin{split}
    Z: &=  \frac{1}{4}\iint_\mathrm{cell} S(t) \delta(r^*-r^*_p)  dudv   \\
    &= \frac{1}{2}hA_{\ell m}\, \operatorname{sinc} \left(\frac{m\Omega h}{2}\right)e^{-im\Omega t_c},
  \end{split}
  \label{eqn:fds_source}
\end{equation}
where $\operatorname{sinc} x:=(\sin x)/x$ and $t_c$ is the value of
$t$ at the center of the cell in question,
\ie{}
$t_c=(v+u-h)/2$.

Collecting the above results, we arrive at the following finite-difference
formula, applicable at grid points traversed by the particle:
\begin{align}
  \Psi(u,v) = & \big[\Psi(u,v-h)+\Psi(u-h,v)\big]\bigg[1-\frac{h^2}{8}V(r_{c})\bigg] \nonumber \\
              & -\Psi_1(u-h,v-h) + Z +\mathcal{O}(h^{3}).
  \label{eqn:FDS_particle}
\end{align}
For our test evolution with a point particle, we use the vacuum
formula~\eqref{eqn:FDS+} at all grid points except those on the
particle's
worldline, for which we use~\eqref{eqn:FDS_particle}. With this, we expect
(and observe) a global quadratic convergence with $h$.

\subsection{Matching procedure}
\label{subsec:matching}

Scheme~I employs the matching approach described in
Sec.~\ref{Subsection:Matching Method I}, \ie{} matching in a buffer
region.
In practice, the implementation is a discretised version of the approach, and
the integral in Eq.~\eqref{inner product} reduces to a summation over
discrete
data points.
With $i$ labelling the discrete data points,
Eq.~\eqref{matching linear system} becomes
\begin{equation}
  \label{eq:discrete system}
  \sum_{j=0}^n A_{jk} \psi^\mathcal{R}_j(t) h = b_k(t) h~~~~~\text{for }k=0,\dots,n\, ,
\end{equation}
with $A_{jk} =\sum_{i=1}^{d} \Delta r_i^j \Delta r_i^k$ and $b_k =\sum_{i=1}^{d} (\Psi^N_i - \Psi^\mathcal{P}_i) \Delta r_i^k$.
The discretisation factor $h$, appears on both sides of
Eq.~\eqref{eq:discrete system}  and cancels. The number $d$ of
data points must be taken to be greater than or equal to the number of unknown coefficients $\psi^\mathcal{R}_j$, and the solution to Eq.~\eqref{eq:discrete system} then
yields the least-squares polynomial regression of
$\Psi^N_i - \Psi^\mathcal{P}_i$.
It should be noted that alternative matching methods could be used. However, we
adopt the standard least-squares polynomial regression for simplicity at this
trial
stage.

The above description assumes (for simplicity) a Cauchy-type evolution, and it
needs to be adapted for use in our characteristic evolution setup. In the
Cauchy evolution case, the regular field component of the analytical solution
is expanded in powers of $\Delta r$ about the point where the
current Cauchy slice intersects the particle's worldline (at the center of the
tube; refer again to Fig.~\ref{fig:matching_method1}). In our characteristic
implementation, we instead choose to expand $\Psi^\mathcal{R}(u,v)$ as a double
Taylor series
in $\Delta u:= u-u_p$ and $\Delta v:= v-v_p$ about the point of
intersection of the two current null slices $(u_p,v_p)$, which
in our setup is a point along the particle's worldline at the center of the
tube (refer again to Fig.~\ref{fig:FDS_grid_new}). The expansion takes the
form
\begin{equation}
  \label{eq:PsiR-scheme1}
  \Psi^\mathcal{R}(u,v) =
  \sum_{i=0}^n \sum_{j=0}^{n-i} \psi^\mathcal{R}_{ij} {\Delta u}^i {\Delta v}^j,
\end{equation}
where $n$ is the puncture order, and the coefficients
$\psi^\mathcal{R}_{ij}$ are \emph{a priori} unknown constant
coefficients at each time
step.
There are $N=(n+1)(n+2)/2$ such coefficients.
The matching conditions in Eq.~\eqref{eq:discrete system} are then replaced
with a suitable two-dimensional version, with solutions that are the
least-squares 2D polynomial regression model of
$\Psi^N_i - \Psi^\mathcal{P}_i$ over a 2D array of
data points in the $u,v$ plane. For this to work, one must
take $d\geq N$, \ie{} the number of data points
must be greater
than or equal to the number of coefficients $\psi^\mathcal{R}_{ij}$.

\begin{figure}
  \includegraphics[width=\columnwidth]
  {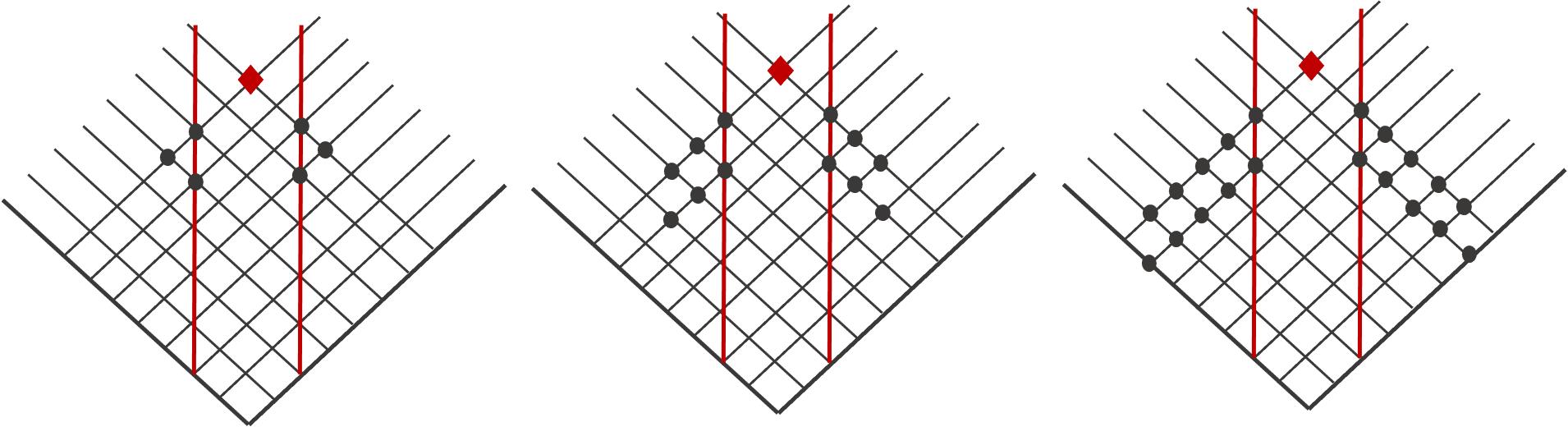}
  \caption{
    \label{fig:FDS_grid_2}
    Data used to fit for the unknown parameters $\psi^\mathcal{R}_{ij}$
    in the linear, quadratic and
    cubic-order approximate analytical models ($n=1,2$ and $3$, left
    to right respectively).
    Red vertical lines mark the worldtube's boundaries, and the red
    diamond is the reference point
    about which the regular field is expanded in a double Taylor series.
    Black circles represent the numerically determined field data points
    used to the fit the parameters
    $\psi^\mathcal{R}_{ij}$ of the analytical model inside the tube,
    using the procedure described in the text. }
\end{figure}
In  our particular implementation we choose to take $d=2N$,
\ie{} twice as many data points as unknown coefficients. This
choice
appeared to provide a good balance in the tradeoff between accuracy and
runtime. Since, with this choice, the number of data points is always even, it
allows us to distribute them evenly and symmetrically either side of the
worldtube. Our choice of data points for the matching, for model orders
$n=1,2$ and $3$, is shown in
Fig.~\ref{fig:FDS_grid_2}.
In all cases, we draw our matching points from the
current two null rays and from the two null rays in the preceding time step (it
is necessary to use data from more than a single time step in order to fit for
mix-derivative coefficients like $\psi^\mathcal{R}_{11}$).

With these choices, the matching procedure is as follows. As described already,
at each time step we evolve the initial data using successive applications of
our finite-difference formula along the corresponding two null rays
$u=\mathrm{const}$ and $v=\mathrm{const}$ (such that
$v-u=2r_p^*$) running from the initial null surfaces to the
boundaries of the tube. Once this step is completed, we record the
$d$ numerical data points $\Psi^N_{i}$ shown in
Fig.~\ref{fig:FDS_grid_2}, all of which are known to us from the current or
previous steps of the numerical evolution, and then construct the
$d$ values $\Psi^N_{i}-\Psi^\mathcal{P}_{i}$ by subtracting the
analytically known puncture values at the corresponding grid points. To these
$d$ values we now match the $n$th-order
2D polynomial given in Eq.~\eqref{eq:PsiR-scheme1} using a least-square
minimisation procedure to obtain the coefficients
$\psi^\mathcal{R}_{ij}$. This, in turn, determines the regular field
$\Psi^\mathcal{R}$, and thus also the complete analytical approximation
$\Psi^A=\Psi^\mathcal{R}+\Psi^\mathcal{P}$ inside the worldtube,
in the vicinity of the current
characteristic rays. We record the values of $\Psi^A$ at the two
``ghost'' grid points inside the tube adjacent to the boundaries on the current
ray (starred points in Fig.~\ref{fig:FDS_grid_new}); these two values will be
required when calculating the numerical field on the boundary in the next time
step. This concludes the computation for the current time step, and we can now
step
forward in (advance/retarded) time and repeat.

A few comments are in order. First, it may be noticed that in the first few
time steps of the evolution there may not be available sufficiently many data
points to fit all of the $N$ model parameters. In such
cases we simply set to zero the values of the ``missing'' data points. This
does not cause a problem, because the early evolution is in any case dominated
by non-physical junk radiation; all this does is modify the profile of the
initial junk.

Second, we note that in our procedure we choose not to impose that
$\Psi^A$ satisfies the field equation in the tube; if we did,
some of the coefficients $\psi^\mathcal{R}_{ij}$ would become mutually
dependent. For example, in the quadratic model with $n=2$,
imposing the field equation would determine the coefficient
$\psi^\mathcal{R}_{11}$ in terms $\psi^\mathcal{R}_{00}$,
$\psi^\mathcal{R}_{01}$ and $\psi^\mathcal{R}_{10}$. Such an alternative
approach is possible, but we find that it does not leads to any marked
improvement in either accuracy or speed. For simplicity, we thus opt to treat
all $N$ coefficients $\psi^\mathcal{R}_{ij}$ as independent
for the purpose of matching.

Finally, we comment on the degree of differentiability of our solution on the
tube's boundary. As already mentioned, since we are not explicitly imposing
continuity of the field or its derivatives at the tube's boundary, there is no
reason to expect that the field constructed via our matching procedure should
exhibit any level of differentiability there. In practice, for our specific
choice of matching data points, we find that the discrepancy between
$\Psi^N$ and $\Psi^A$, and between their radial
derivatives, are numerically small and seem to decrease to zero with
$h$, as expected on theoretical grounds.

\section{Scheme I: Tests and Analysis}\label{section:Results}

All of the results discussed below are for the fixed circular geodesic
orbit described in Eq.~\eqref{eq:coords}, with radius
$r_p=7M$.
We will consider two modes,
$(\ell,m)=(2,0)$ and $(2,2)$, as representative examples
of static
and radiating modes, respectively.
For the static mode we have the analytical solution~\eqref{eqn:analytical}
for comparison,
and for the radiating mode we compare the solutions obtained with
an excision worldtube to
numerical solutions produced by the `exposed' point-particle code described in
Sec.~\ref{subsec:TestEvolution}.

Our numerical solutions depend on three `control' parameters: the uniform grid
resolution $h$, the worldtube radius
$R$, and the order $n$ of the
analytical model inside the tube. For our numerical convergence tests we use
the sequence of values $h=\{0.02,0.01,0.005\}M$, fixing the resolution at
$h=0.005M$ for all other tests. The value of
$R$ for our various tests is chosen in the interval
$[0.0125M,0.8M]$. In scheme I we restrict to models with
$n=1,2,3$ (while scheme II extends this to
$n=4,5$). Convergence towards our benchmark solution is
observed, as expected, when decreasing $h$, or when
decreasing $R$, or when increasing $n$
(for a sufficiently small $R$). In what follows we
demonstrate, explore and better quantify this behavior using a range of
numerical experiments.

\subsection{Convergence with resolution}
\label{Subsection:Convergence with Resolution}

\begin{figure}
  \includegraphics[width=\columnwidth]
  {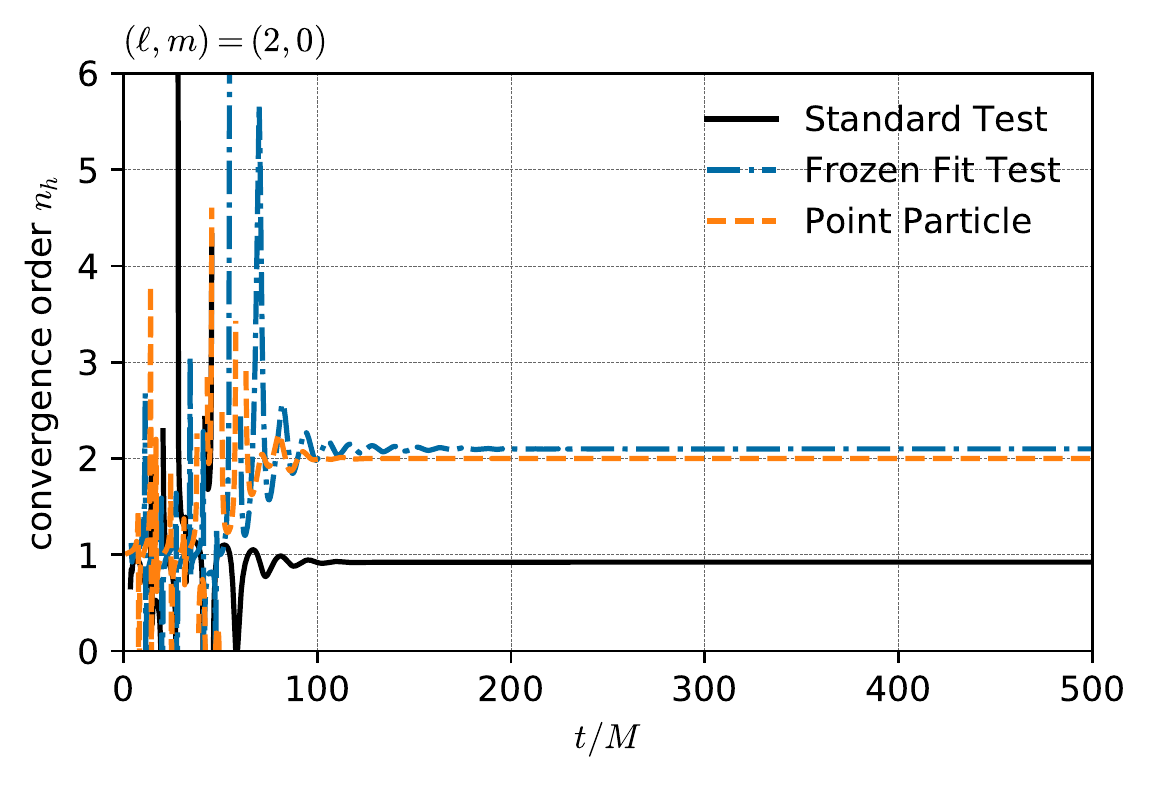}
  \caption{
    \label{fig:FDS_h}
    Convergence of the numerical solution with respect to grid resolution $h$.
    Plotted is the convergence index $n_h$, defined in Eq.~\eqref{eq:n_h},
    as a function of $t$ along a slice of constant radius $r^*=8.93258$
    (corresponding to the right boundary of the excision tube).
    In the ``standard'' test, the analytical solution in the tube is
    fitted for afresh for each choice of grid resolution, and the
    observed steady-state convergence is linear ($n_h\approx 1$). For
    comparison, when we fix the analytical solution in the tube as we
    vary $h$ (``frozen fit''), the observed convergence is quadratic
    ($n_h\approx 2$), as it is for a run with an exposed point particle without
    an excision tube (``point particle'').  The reduction in convergence
    rate evidently caused by the matching procedure is discussed in the text.
  }
\end{figure}

We start by examining the convergence of the finite difference scheme with
respect
to grid resolution $h$, using a local convergence test.
Three runs are performed, with fixed worldtube width
$R$ and model order $n$, and varying
$h$.  Denoting by $\Psi_{h}$ the field
computed with resolution $h$,
we construct the local convergence index
\begin{equation}\label{eq:n_h}
  n_h:=\log_2 \left |
  \frac{ \Psi_h-\Psi_{\frac{h}{2}} }
  { \Psi_{\frac{h}{2}}-\Psi_{\frac{h}{4}} }
  \right|\, ,
\end{equation}
which should yield approximately $2$ for a quadratic
convergence.
When applied to our exposed point-particle code, the test indeed yields
$n_h\approx 2$ after initial-junk
transients sufficiently subside; an example is shown in
Fig.~\ref{fig:FDS_h}. This confirms the quadratic
convergence of our basic finite-difference algorithm.

Next we perform our convergence test with a worldtube excision. Here we have a
choice in how the array of matching points outside the tube is modified as we
vary $h$. A sensible ``like-to-like'' comparison is one in
which the \emph{physical} position
and pattern of the data points around the worltube is held fixed as
$h$ is varied.
Proceeding in this way, our convergence test with a worldtube excision yields
$n_h\approx 1$,
indicating that the convergence is only linear---see again
Fig.~\ref{fig:FDS_h} for an example.
We find this deterioration in convergence rate affects all
$\ell,m$ modes examined (static as well
as radiative), and all model orders attempted ($n=1,2,3$).
Repeating the test with a sequence of smaller $h$
values does not improve the situation, and the convergence remains linear.
However, quadratic convergence is recovered if (for a static mode) we replace
the regression model in the tube with the known exact analytical solution. We
also
recover quadratic convergence if we ``freeze'' the matched analytical model in
the tube as we vary $h$ (\ie{}, fit the
model using one
value of $h$ and then apply the same polynomial regression
model when
running with the other two $h$ values participating in our
convergence test).

The apparent reduction in convergence rate may be explained as resulting from a
coupling between $h$-related and
$R$-related errors, expected
when the approximate analytical model in the tube is allowed to depend on
$h$, as in our convergence test. To understand this,
consider that the value of the numerical field at a point
$x$ outside the worldtube is a function
$\Psi(x;\tilde h,\tilde R)=\Psi_\mathrm{exact}(x)+\delta\Psi(x;\tilde h,\tilde R)$
depending parametrically on both
$\tilde h:= h/m_1$ and $\tilde R:=R/m_1$ (we ignore here the
dependence on $n$, assumed fixed for the rest of this
discussion). For small $\tilde h$ and $\tilde R$, the
error term may be expressed as a double Taylor expansion,
$\delta\Psi=\sum_{i,j}a_{ij}\tilde h^i \tilde R^j$.
The terms with $j=0$ describe the
usual discretization error for $R\to 0$ (exposed point
particle); we have $a_{00}=0=a_{10}$, with the leading term being
$a_{20}\tilde h^2$ for our quadratically convergence code. The terms with
$j\ne 0$ arise from the approximate nature of the analytical
solution in the tube. According to the argument in
Sec.~\ref{Subsection:error estimates}
(and as demonstrated in the next subsection),
the leading finite-$R$ error is of
$\mathcal{O}(R^n)$, and it is therefore expected to have the form
$\delta\Psi \simeq (a_{0n}+a_{1n}\tilde h+\cdots)\tilde R^n$ in general.
When we construct the index
$n_h$ in our convergence test, the contribution from the
$\propto a_{0n}$ term cancels out, and $n_h$ is
dominated by the $\propto a_{1n}$ error term, giving rise to the
observed linear convergence in $h$ at fixed
$R$. The crucial point here is that, in our convergence
test, we allow the value of the approximate analytical model on the tube's
boundary to depend on $h$ (in a complicated way, via a
matching procedure that involves numerical data points that themselves depend
on $h$), and as a result the
$R$-related error also becomes
$h$-dependent. When we freeze the analytical model (or use
the exact analytical solution for it) we decouple between the
$h$-related and $R$-related errors, and
quadratic convergence is recovered.

We note the occurrence of such linear-in-$h$ error terms is
not necessarily a weakness of our scheme: in practice, for a particular choice
of $h$ and $R$, the error term
$a_{1n} \tilde h R^n$ is not necessarily numerically larger than the term
$a_{02} \tilde h^2$. Rather, the occurrence of a linear term is a somewhat
artificial combined feature of the particular matching procedure applied and
the particular way the convergence test is designed. The lesson from the above
discussion is that one should exercise caution in designing and interpreting
convergence tests for a worldtube scheme, being mindful about the potential
effect of coupling between finite-difference and worldtube-related sources of
error.

\subsection{Convergence with Worldtube Size}

\begin{figure}
  \includegraphics[width=\columnwidth]{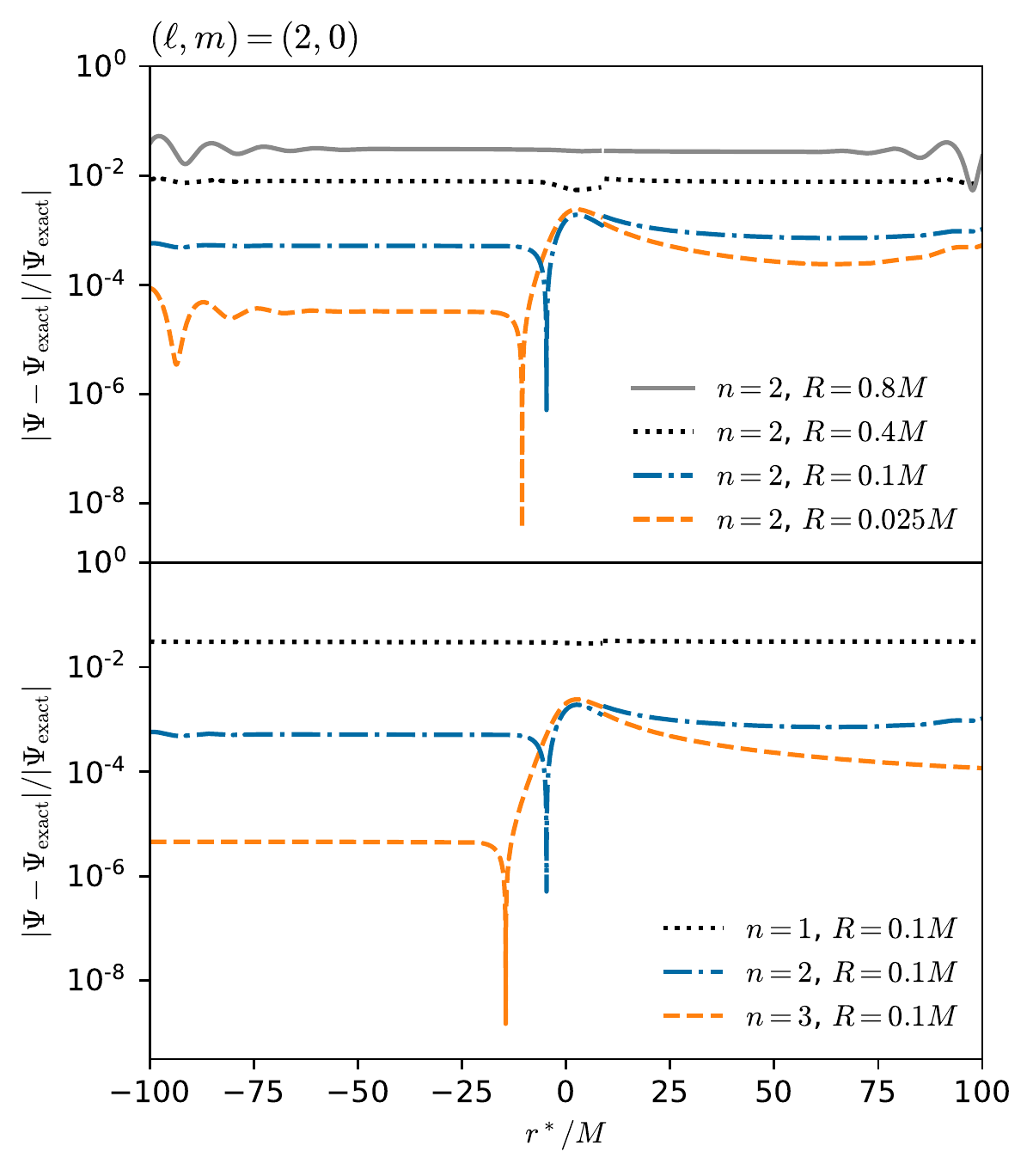}
  \caption{
    The relative finite-$R$ error in $\Psi_{20}$ as measured by comparison
    with the exact analytical solution $\Psi_{exact}$. In the upper panel we vary the
    tube radius $R$ at fixed model order $n=2$,
    and in the lower panel we vary $n$ at fixed $R=0.1M$. The relative
    difference is shown on a
    late-time $t=\text{const}$ slice. The numerical resolution is
    $h=0.005M$ in all cases.
  }
  \label{fig:FDS_relative_error_2}
\end{figure}

\begin{figure}
  \includegraphics[width=\columnwidth]{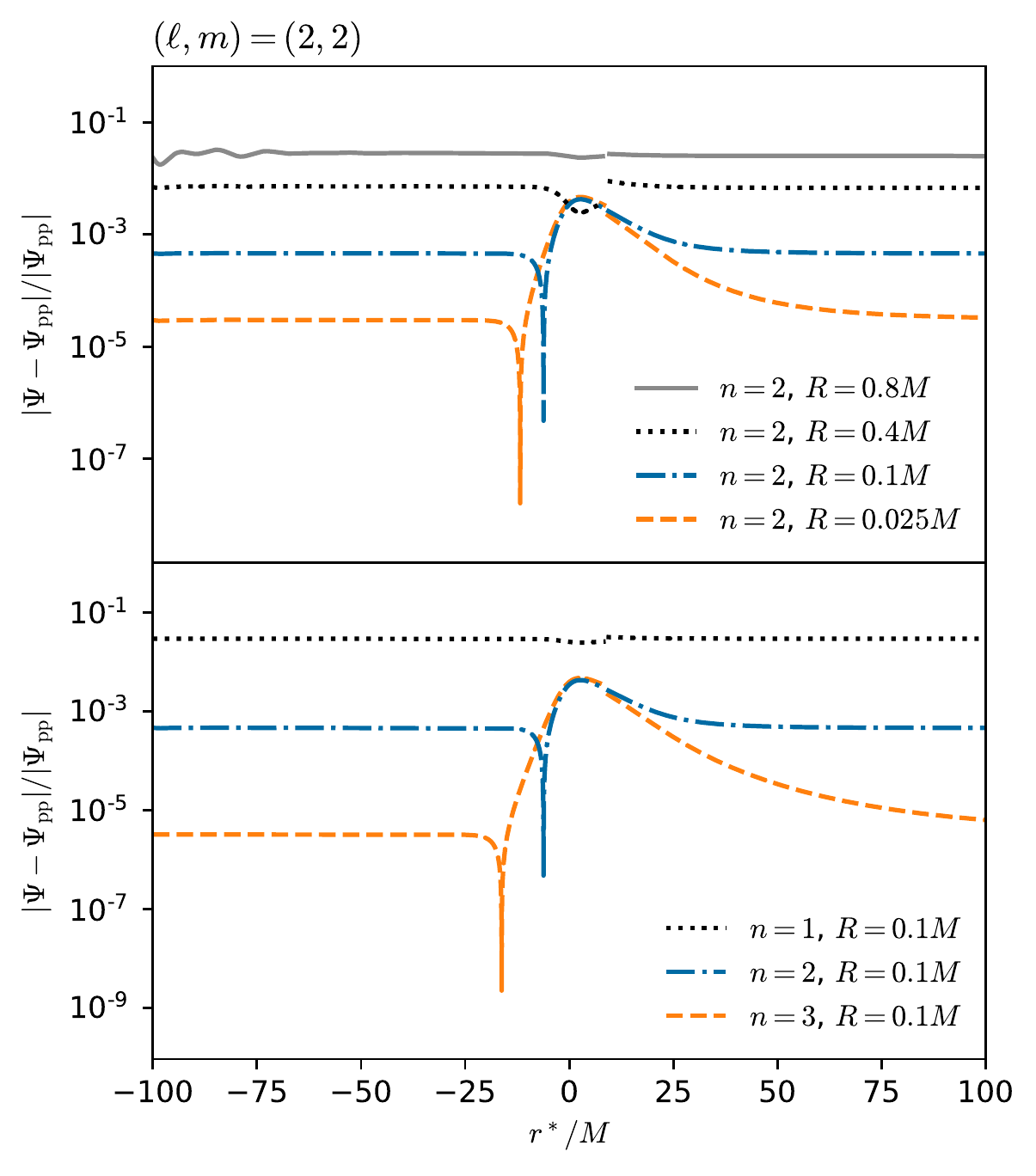}
  \caption{
    The relative finite-$R$ error in $\Psi_{22}$ as measured by comparison
    with the (accurate) numerical solution $\Psi_{pp}$ from our point-particle code.
    The format of this plot and all other details are as in
    Fig.~\ref{fig:FDS_relative_error_2}.
  }
  \label{fig:FDS_relative_error_pp_2}
\end{figure}

It is of greater interest, in the context of this work, to quantify and
understand the scaling of our solutions with the tube size
$R$ and model error $n$. Figures
\ref{fig:FDS_relative_error_2} and \ref{fig:FDS_relative_error_pp_2}
show how the local
finite-$R$ error in our numerical solutions varies as a
function of $R$ (at fixed $n$; top
panels) and as a function of $n$ (at fixed
$R$; bottom panels). In Fig.~\ref{fig:FDS_relative_error_2} we
measure the  finite-$R$ error by comparing with the exact
analytical solution for the static mode $(2,0)$ and in
Fig.~\ref{fig:FDS_relative_error_pp_2} we measure it by
comparing with numerical solutions obtained using our exposed point-particle
code. In both cases we display the relative
differences as functions of $r^*$ on a late-time
$t=\text{const}$ slice.

We see that, as expected, our solutions generally become more accurate as we
decrease $R$ or increase $n$.
We note that even with the simplest, linear ($n=1$)
analytical model, and with a tube radius as large as $R=0.1M$,
the worldtube-related error is only around $1\%$ almost
uniformly. There is a marked reduction in error at smaller
$R$ and larger $n$, except near the
worldtube (at $r^*\approx 9M$ in these figures, too narrow to be
resolved), where the error seems to saturate. As we demonstrate further below,
the saturation marks the point where finite-difference error becomes dominant
over $R$-related error, so that a further decrease in
$R$ (or increase in $n$) does not lead
to a further reduction in overall error. The effect is most pronounced near the
worldtube, since the finite-difference error is largest there (where field
gradients are largest), while worldtube error (we expect) remains roughly
spatially uniform. The effect is exacerbated by the fact that as we decrease
$R$ we expose more of the high-gradient region surrounding
the particle. To fully demonstrate convergence with $R$ or
$n$ near the tube would require a concurrent refinement of
resolution there.

\begin{figure}
  \includegraphics[width=\columnwidth]
  {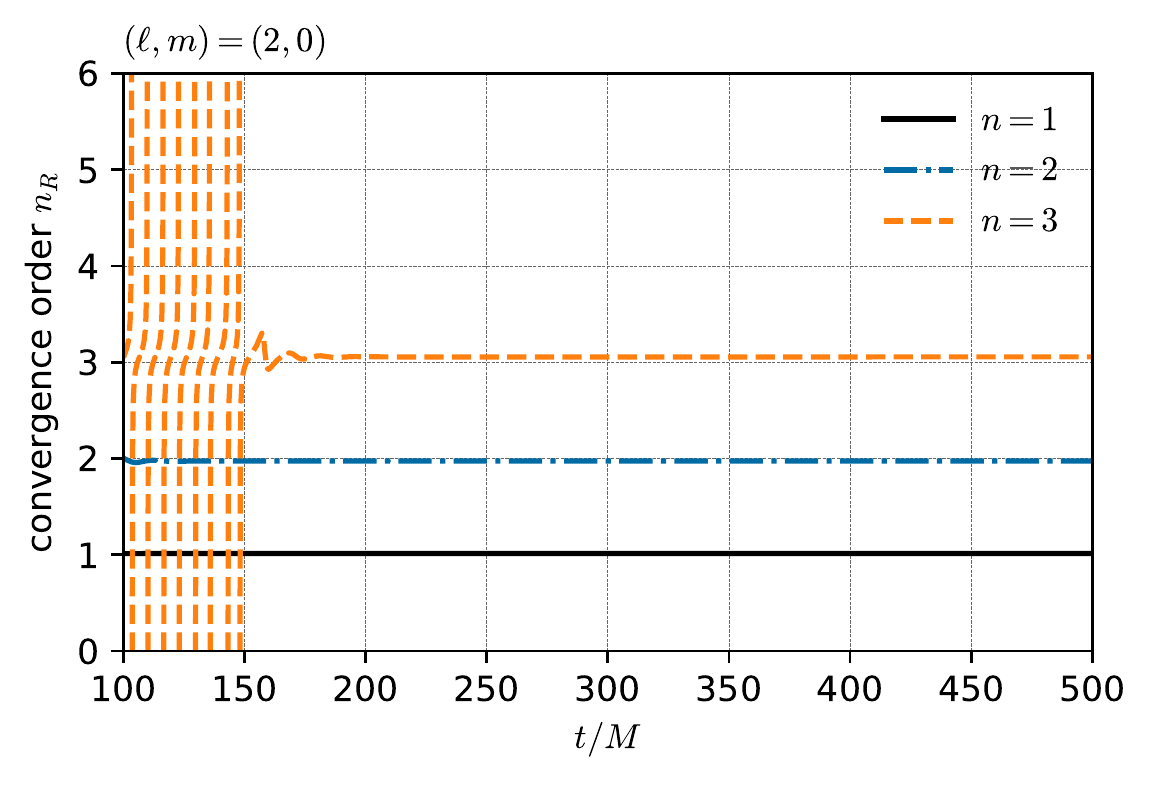}
  \caption{
    \label{fig:FDS_intr}
    Convergence of the numerical solution with respect to tube size $R$,
    at fixed model order $n$ and resolution $h(=0.005M)$.
    We plot the internal convergence index $n_R$, defined in Eq.~\eqref{eq:n_R},
    as a function of $t$ along a fixed radius of $r^*=8.93258M$.
    To obtain $n_R$ (for each model order $n$) we carry out three runs with
    $R=0.1M$, $0.05M$ and $0.025M$.
    After the decay of initial junk, the convergence order appears to be
    $n_R\approx n$, indicating that the dominant tube-related error is of $\mathcal{O}(R^n)$.
  }
\end{figure}

To quantity the rate of convergence with respect to $R$ (at
fixed $n$ and $h$), we construct the
index
\begin{equation}\label{eq:n_R}
  n_R=\log_2 \left|
  \frac{\Psi_R-\Psi_{\frac{R}{2}}}
  {\Psi_{\frac{R}{2}}-\Psi_{\frac{R}{4}}}
  \right| \, ,
\end{equation}
where $\Psi_{R'}$ represents the value of the field calculated with
a tube radius $R=R'$. This measures the ``internal''
convergence of the numerical solution as we decrease $R$
(as opposed to convergence to the exact solution, illustrated in
Figs.~\ref{fig:FDS_relative_error_2} and \ref{fig:FDS_relative_error_pp_2}).
Figure~\ref{fig:FDS_intr} shows $n_R$ as a function of
$t$ along an $r=\text{const}$. We observe
$n_R\approx n$, indicating that the dominant tube-related error is of
$\mathcal{O}(R^n)$---precisely as predicted in
Sec.~\ref{Subsection:error estimates}

So far we have been considering ``local'' measures of error, ones depending on
location and time. It is also informative to examine a global error norm, which
we now introduce and adopt for the rest of our analysis here and in Sec.\
\ref{section:Scheme2Results}. We denote by $||\Psi||_{L^1}$ the
$L^1$ norm of a numerical field $\Psi$
evaluated on a $t=\text{const}$ slice. The numerical data points for
this norm are sampled uniformly in $r^*$ in the domain
$[-100M,r^*_p-R]\cup [r^*_p+R, 100M]$.
When comparing norms corresponding to runs with
different $R$ values, the largest of the
$R$ values is used for all norms.

\begin{figure}
  \includegraphics[width=\columnwidth]
  {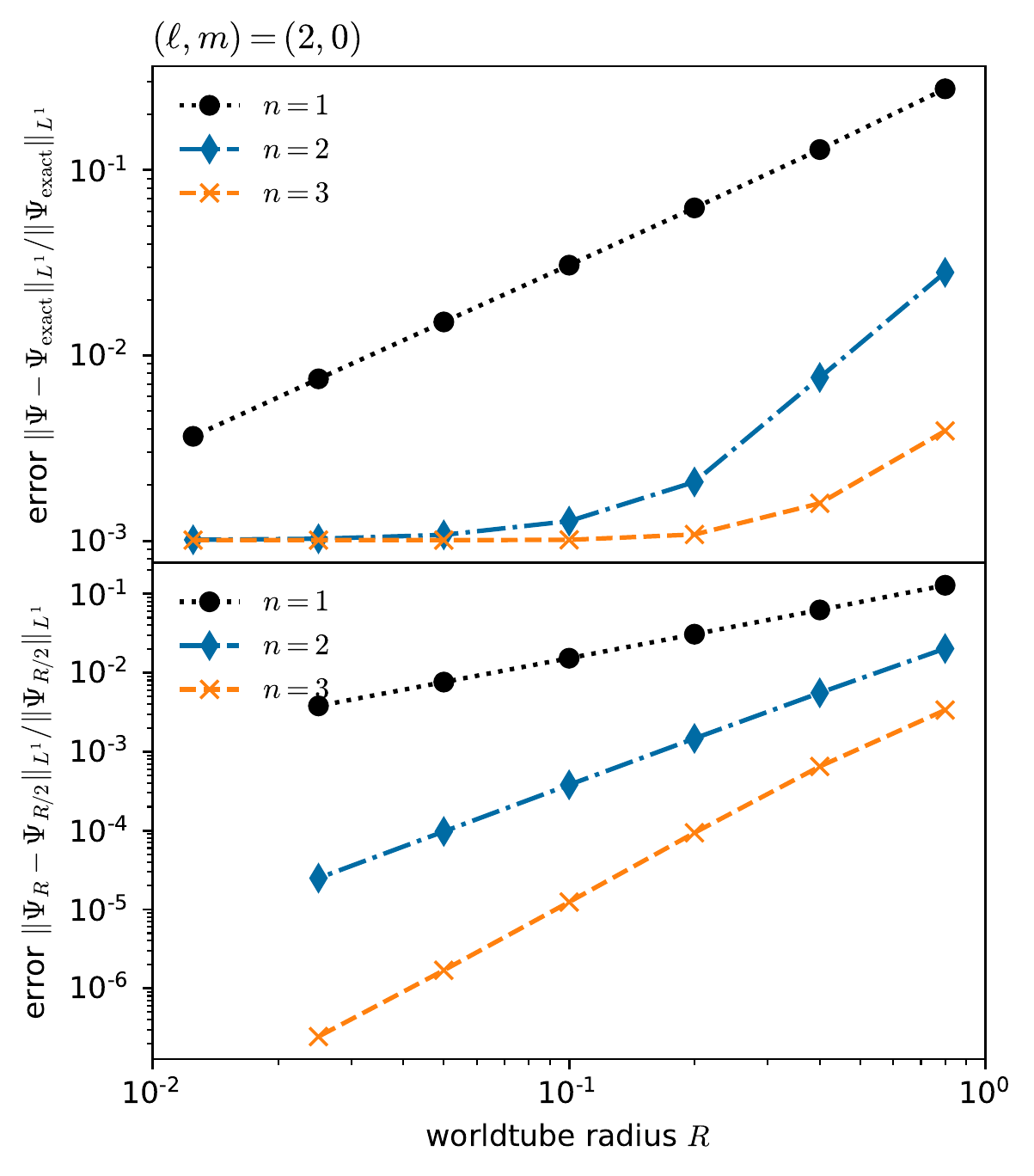}
  \caption{
    \label{fig:FDS_L1_error_2}
    Top panel: Relative $L^1$ error norm with respect to the exact
    analytical solution, as a function of worldtube radius $R$.
    Bottom panel: Internal relative error norm calculated by varying $R$.
    In both cases the finite-difference resolution is held fixed at $h=0.005$.
    The saturation of error in the upper panel is due to the
    finite-difference error becoming dominant at small $R$.}
\end{figure}

The top panel in Fig.~\ref{fig:FDS_L1_error_2} shows the value of the relative
error norm $||\Psi-\Psi_\mathrm{exact}||_{L^1}/||\Psi_\mathrm{exact}||_{L^1}$
as a function of $R$ for
the static mode $(2,0)$, with $\Psi_\mathrm{exact}$ being the
exact analytical solution. We see that the error norm decreases with increasing
model order $n$ and decreasing tube size
$R$, but for $n=2,3$ it seems to saturate
at small $R$. This behavior is consistent with what we saw
in Figs.~\ref{fig:FDS_relative_error_2} and~\ref{fig:FDS_relative_error_pp_2}:
when the
worldtube error magnitude falls below that of the discretization error, a
further reduction in tube size does not improve the accuracy of the solution.
This explanation is further supported by the data shown in the lower panel of
Fig.~\ref{fig:FDS_L1_error_2}, where we display the \emph{internal}
error norm $||\Psi_R-\Psi_{R/2}||_{L^1}/|\Psi_{R/2}||_{L^1}$:
Here we see a monotonic convergence with
$R$ at a constant rate even for $n=2,3$;
the field norm converges to a value that differs slightly from
$||\Psi_\mathrm{exact}||_{L^1}$ due to the dominating $h$-related
error.

\begin{figure}
  \includegraphics[width=\columnwidth]
  {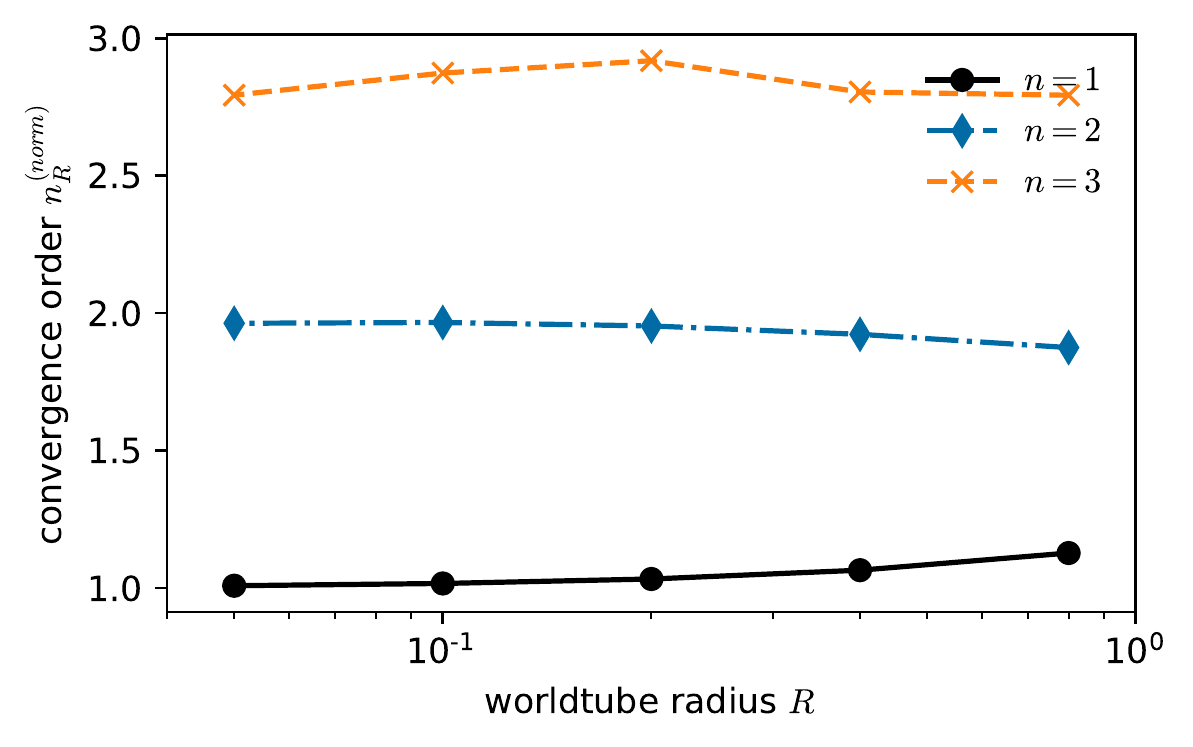}
  \caption{
    \label{fig:FDS_L1_conv_order}
    Convergence of the $L^1$ norm with respect to $R$, at fixed model
    order $n$ and resolution $h(=0.005M)$.
    We plot the internal convergence index $n_{R}^\mathrm{(norm)}$,
    defined in Eq.~\eqref{eq:n_R^norm}, and observe
    $n_{R}^\mathrm{(norm)}\approx n$. Thus the worldtube error scales
    like $\sim R^n$ also as measured by the global $L^1$ norm.
  }
\end{figure}

To quantify the rate of convergence of the global norm with
$R$, we introduce the convergence index
\begin{equation}\label{eq:n_R^norm}
  n_{R}^\mathrm{(norm)} :=\log_2
  \frac{||\Psi_R-\Psi_{\frac{R}{2}}||_{L^1}}
  {||\Psi_{\frac{R}{2}}-\Psi_{\frac{R}{4}}||_{L^1}}
  \, ,
\end{equation}
plotted in Fig.~\ref{fig:FDS_L1_conv_order}. We observe
$n_{R}^\mathrm{(norm)}\approx n$,
\ie{} the tube-related error is $\mathcal{O}(R^n)$ also
as measured by the
global $L^1$ norm. Similar results are obtained for other
$\ell,m$ modes.
\section{Scheme~II: Numerical Method}\label{section:Scheme2Method}
\subsection{Evolution Equations}
\label{Subsection:Scheme2EvolutionEquations}

By introducing the new variables
\begin{align}
  \label{eqn:definition_pi}
  \pi  & := \partial_t \Psi \,   ,   \\
  \label{eqn:definition_chi}
  \chi & := \partial_{r^*} \Psi \, ,
\end{align}
we reduce the evolution equation \eqref{eqn:overall} (in the vacuum region outside the worltube)
to the first-order system
\begin{subequations}
  \label{eqn:EOM_Scheme2_simple}
  \begin{align}
    \label{eqn:EOM_Psi_Scheme2}
    \partial_t \Psi & = \pi \,   ,                                                     \\
    \label{eqn:EOM_pi}
    \partial_t \pi  & = \partial_{r^*} \chi - V  \Psi   \,  ,                          \\
    \label{eqn:EOM_chi}
    \partial_t \chi & = \partial_{r^*} \pi + \kappa (\partial_{r^*} \Psi - \chi)  \, ,
  \end{align}
\end{subequations}
where $\kappa$ is a constraint damping parameter controlling how
strongly violations of the constraint~\eqref{eqn:definition_chi} are damped
(see~\cite{HolLinOwe04},
where the symbol $\gamma_2$ corresponds to our
$\kappa$).
In practice we chose $\kappa=1$ for all the results presented for
this scheme.
We further introduce the puncture and regular fields
$\pi^\mathcal{P}$, $\chi^\mathcal{P}$, $\pi^\mathcal{R}$,
$\chi^\mathcal{R}$
of the reduction variables $\pi$ and $\chi$,
as well as their corresponding
analytical approximations in the tube, $\pi^A = \pi^\mathcal{P} + \pi^\mathcal{R}$ and
$\chi^A = \chi^\mathcal{P} + \chi^\mathcal{R}$.

We now introduce auxiliary variables $\tilde\Psi(t)$,
$\tilde\Psi'(t)$ and $\tilde\Psi''(t)$
defined on the worldtube boundary $\partial\Gamma$, which are to act as
intermediaries in conveying information to and from the approximate
solution within the worldtube. These represent, respectively, the
field $\Psi$, and its first and second $r^*$
derivatives on
$\partial\Gamma$.  Similarly, we introduce $\tilde\pi$,
$\tilde\pi'$
and $\tilde\pi''$, as well as $\tilde\chi$,
$\tilde\chi'$ and
$\tilde\chi''$.    These fields are determined by solving a set of
ODEs along $\partial\Gamma$,
obtained from the restriction of  Eqs.~\eqref{eqn:EOM_Scheme2_simple} to the boundary:
\begin{subequations}
  \label{eqn:EOM_Scheme2_BoundaryODE}
  \begin{align}
    \label{eqn:EOM_BoundaryODE_Psi}
    \partial_t \tilde \Psi   & \bdryequal \tilde \pi\,,                                            \\
    \label{eqn:EOM_BoundaryODE_pi}
    \partial_t \tilde \pi    & \bdryequal  \tilde \chi' - V  \tilde \Psi\,,                        \\
    \label{eqn:EOM_BoundaryODE_chi}
    \partial_t \tilde \chi   & \bdryequal  \tilde \pi' + \kappa (\tilde \Psi' - \tilde \chi)  \, .
    \intertext{
      The symbol $\bdryequal$ denotes equality on $\partial\Gamma$.
      The ODEs for $(\tilde\Psi',\tilde\pi', \tilde\chi')$ arise from the radial derivative of Eqs.~\eqref{eqn:EOM_Scheme2_simple}, restricted to the boundary:}
    \label{eqn:EOM_BoundaryODE_dPsi}
    \partial_t \tilde \Psi'  & \bdryequal \tilde \pi'\,,                                           \\
    \label{eqn:EOM_BoundaryODE_dpi}
    \partial_t \tilde \pi'   & \bdryequal \tilde \chi''
    - \tilde \Psi \partial_{r^*} V
    - V \tilde \Psi'\,,
    \\
    \label{eqn:EOM_BoundaryODE_dchi}
    \partial_t\tilde \chi'
                             & \bdryequal \tilde \pi''
    + \kappa (\tilde \Psi'' - \tilde \chi')\,.
    \intertext{
      Finally, the ODEs for $(\tilde\Psi', \tilde\pi'', \tilde\chi')$ arise from the second spatial derivative of Eqs.~\eqref{eqn:EOM_Scheme2_simple}:
    }
    \label{eqn:EOM_BoundaryODE_ddPsi}
    \partial_t \tilde \Psi'' & \bdryequal \tilde \pi''\,,                                          \\
    \label{eqn:EOM_BoundaryODE_ddpi}
    \partial_t \tilde \pi''  & \bdryequal \partial_{r^*}^3 \chi^A
    - \tilde \Psi \partial_{r^*}^2 V
    - 2 \tilde \Psi' \partial_{r^*} V
    - V \tilde \Psi''\,,
    \\
    \label{eqn:EOM_BoundaryODE_ddchi}
    \partial_t\tilde \chi''  & \bdryequal \partial_{r^*}^3 \pi^A
    + \kappa (\partial_{r^*}^3 \Psi^A -  \tilde \chi'')  \, .
  \end{align}
\end{subequations}
To close the set of auxiliary ODEs,
Eqs.~\eqref{eqn:EOM_BoundaryODE_ddpi} and
\eqref{eqn:EOM_BoundaryODE_ddchi} couple to the matched
analytical approximations $\Psi^A$, $\pi^A$ and
$\chi^A$.
$\Psi^A$ is obtained by solving Eq.~\eqref{eq:approach2_matching}
using as $\partial_{r^*}^j \Psi^N$ the six boundary values
$\tilde\Psi, \tilde\Psi', \tilde\Psi''$ (three each at $\partial \Gamma_-$ and
at $\partial \Gamma_+$).
Analogously, $\pi^A$ and $\chi^A$ are obtained
using
$\tilde\pi$, $\tilde\pi'$, $\tilde\pi''$, and
$\tilde\chi$, $\tilde\chi'$,
$\tilde\chi''$ respectively.
The system of equations~\eqref{eqn:EOM_Scheme2_simple} are not yet coupled to
the exterior (bulk) solution; this coupling will be discussed and
incorporated below in Sec.~\ref{Subsection:BoundaryODECoupling}.

The ODE system above yields an approximate analytical solution whose
regular part is accurate through $\mathcal{O}(\Delta {r^*}^5)$.
In Sec.~\ref{section:Scheme2Results} we also investigate setups
with regular fields expanded to $\mathcal{O}(\Delta {r^*}^3)$ only.
For these tests we discard
$\tilde\Psi''$, $\tilde\pi''$ and $\tilde\chi''$
and the corresponding
evolution
equations~\eqref{eqn:EOM_BoundaryODE_ddPsi}--\eqref{eqn:EOM_BoundaryODE_ddchi},
and use a third-order analytical approximation to close the
system.
The auxiliary system could be extended to arbitrary derivative
orders in an obvious way, by taking sufficiently many derivatives
of the field equations.
Note that the scheme can be interpreted as evolving the
the regular parts only,
since the puncture field and its spatial and temporal derivatives are
known up to a given order through Eq.~\eqref{eqn:PunctureAnsatz}.

\subsection{Boundary Treatment}
\label{Subsection:Scheme2BoundaryTreatment}

Boundary conditions must be provided on the outer boundaries of the
computational domain and on the excision boundary $\partial\Gamma$.
Furthermore, if the computational domain is divided into smaller elements,
then boundary conditions are needed at the interfaces where neighboring
elements meet.
For the boundary implementation presented here we assume a numerical
scheme that is formulated on collocation points.
To derive the boundary implementation we perform a
characteristic decomposition~\cite{SarTig12}
of the system of PDEs. We write the system in the form
\begin{equation}
  \label{eqn:PDEDecomposition}
  \partial_t \mathbf{ u} = A^k \partial_k \mathbf{ u}
  + \mathbf{f}(\mathbf{ u}) \, ,
\end{equation}
where $\mathbf{ u} = (\Psi, \pi, \chi)$ is the vector of evolution variables,
$A^k$ are the principal part matrices and
$\mathbf{f}$ contains the
non-principal terms.
The characteristic vectors $\mathbf{v}$ are the left eigenvectors of
$A^k \hat s_k$,
where $\hat s$ is the respective outward pointing unit normal to
the element boundary; here, since we work in one spatial
dimension, $\hat s$ has only one component and the index is
suppressed in the
notation.
The characteristic speeds $\lambda$ are the
corresponding eigenvalues.
The characteristic variables are the inner products of $\mathbf{v}$
and
$\mathbf{u}$, \ie{} for our system,
\begin{align}
  \label{eqn:characteristic_variables}
  \begin{matrix*}[l]
    \hat u^\pm &= \pi \pm \hat s \chi + \kappa \Psi ~~& \lambda^\pm &= \pm 1\\
    \hat u^0   &= \Psi &\lambda^0 &= 0\, .
  \end{matrix*}
\end{align}
Characteristic variables with positive characteristic speed are referred to as
\emph{incoming} and boundary conditions must be provided for them,
\ie{} we have to provide boundary data for
$\hat u^+$.
Characteristic variables with $\lambda<0$ are
\emph{outgoing}
and no boundary condition is needed. Likewise we do not need
boundary conditions for the static characteristic variable
$\hat u^0$.

We modify
the time derivatives of the incoming characteristic variables at the boundary
points to impose boundary conditions.
At the interface between elements
we employ a penalty method~\cite{HilWeyBru15},
modifying the time derivative of the incoming characteristic
field $\hat u^+$ in the following way:
\begin{equation}
  \label{eqn:boundary_penalty_method}
  \partial_t \hat u^+ = D_t \hat u^+
  + p \lambda^+ (\hat u^+_\mathrm{neighbour} - \hat u^+) \, ,
\end{equation}
where $D_t$ is the time derivative constructed from the
evolution equations~\eqref{eqn:EOM_Scheme2_simple},
before the boundary modification has been added.
Furthermore, $\hat u^+_\mathrm{neighbour}$ denotes the value of
$\hat u^+$ on the
neighbouring element (i.e.\ evaluated with data from the neighboring element,
but still using the surface normal $\hat s$ pointing out of the
current element)
Finally, the penalty parameter $p$
is chosen to be $p = J/w$ as in~\cite{HilWeyBru15},
which guarantees stability of the method. Here
$J=\partial x/\partial r^*$ is the
Jacobian associated to transformations from the local
coordinates $x$ of the element
to the global coordinates $r^*$ of the computational domain
and $w$ is the integration
weight
of the point at the element boundary for the quadrature on the element grid.
The given expression for the
penalty parameter holds for integration weights satisfying a summation-by-parts
property~\cite{Str94} with respect to the derivative stencil of
the numerical method.
We now introduce
\begin{equation}
  b_\pm(r^*) =
  \begin{cases}
    1, & \mbox{if } r^* = r^*_\pm    \\
    0, & \mbox{if } r^* \neq r^*_\pm
  \end{cases} \, ,
\end{equation}
where $r^*_-$ and $r^*_+$ denote the position
of the element's left and
right boundary, with the corresponding boundary normals
being $\hat s=-1$ on $r^*_-$ and
$\hat s=1$ on $r^*_+$.
The modified equation of motion can be rewritten as
\begin{equation}
  \label{eqn:boundary_penalty_method_b}
  \begin{split}
    \partial_t \hat u^+ = D_t \hat u^+
    &+ \left [
      p \lambda^+ (\hat u^+_\mathrm{neighbour} - \hat u^+)
      \right ]_{r^*_-} b_- \\
    &+ \left [
      p \lambda^+ (\hat u^+_\mathrm{neighbour} - \hat u^+)
      \right ]_{r^*_+} b_+ \, ,
  \end{split}
\end{equation}
where the subscript on the square bracket denotes the point at which the
bracketed expression is evaluated.

An alternative way to impose boundary conditions
is the Bjørhus method~\cite{Bjo95}, where
the time derivatives are modified like
\begin{equation}
  \label{eqn:boundary_bjorhus_method}
  \begin{split}
    \partial_t \hat u^+ = D_t \hat u^+
    &+ \left [-
      \lambda^+ (\hat s \partial_{r^*} \hat u^+ - g)
      \right ]_{r^*_-} b_- \\
    &+ \left [-
      \lambda^+ (\hat s \partial_{r^*} \hat u^+ - g)
      \right ] _{r^*_+} b_+ \, .
  \end{split}
\end{equation}
The term $g$ models the expression
for $\hat s \partial_{r^*} \hat u^+$ that one desires to impose at the boundary.
At the excision and the outer boundary one can employ either the penalty or
the Bjørhus method, but for the results shown here
at the outer boundaries the latter is employed, whereas on the excision
boundary
the penalty method is used.

The modifications of the original equations~\eqref{eqn:EOM_Psi_Scheme2}
are obtained after transforming back from the characteristic variables to
the evolved variables.

On the excision boundary,
boundary conditions for the penalty method are obtained from the matched
analytical solutions,
\begin{equation}
  \label{eqn:boundary_condition_excision_pi_penalty}
  \hat u^+_\mathrm{neighbour} =  \pi^A + \hat s \chi^A + \kappa \Psi^A  \,  ,
\end{equation}
and similarly for the Bjørhus method:
\begin{equation}
  \label{eqn:boundary_condition_excision_pi_bjorhus}
  g = \hat s \partial_{r^*} (
  \pi^A  + \hat s \chi^A  + \kappa \Psi^A
  ) \,  .
\end{equation}
For the static modes we find it useful to choose
\begin{equation}
  \label{eqn:boundary_condition_outer_static_pi}
  g   =
  \frac{\partial_{r^*}^2 \Psi_\mathrm{exact}}{\hat s \partial_{r^*} \Psi_\mathrm{exact}}
  (\pi + \hat s \chi)
  + \kappa \hat s \chi \, .
\end{equation}
This choice ensures that in the static case, \ie{}
$\pi = 0$,
the field has the same derivative as the exact analytical
solution~\eqref{eqn:analytical}.
There are other choices consistent with the analytical solution, but we find
that
this particular choice has small numerical reflections at the boundary.

For the radiative modes we find that for large $r^*$ the
characteristic variables behave like
$\hat u^+ - \kappa \Psi \sim \exp[i m \Omega (r^* - t)] / {r^*}^2$.
Boundary conditions compatible with this functional form are given by:
\begin{equation}
  \label{eqn:boundary_condition_outer_positive_pi}
  g     =
  \left (-\frac{2}{r} + i m \Omega \right )
  (\pi + \hat s \chi)
  + \kappa s \chi \, .
\end{equation}
Near the horizon we impose
\begin{equation}
  \label{eqn:boundary_condition_outer_negative_pi}
  g     =
  -i m \Omega  (\pi + \hat s \chi)
  + \kappa \hat s \chi \,  ,
\end{equation}
corresponding to the behaviour
$\hat u^+ - \kappa \Psi \sim \exp(i m \Omega (-r^* - t)) $.

\subsection{Coupling to the Boundary Ordinary Differential Equations}
\label{Subsection:BoundaryODECoupling}

Besides coupling the auxiliary ODEs to the bulk
PDEs one also needs a prescription
for the opposite direction, \ie{}
some external input for the auxiliary system, Eqs.~\eqref{eqn:EOM_Scheme2_BoundaryODE}.
A stable system can be derived
realizing that the auxiliary system is equivalent to a spectral method employed
inside the excision worldtube. This means we can employ the same techniques as
before when imposing boundary conditions,
\ie{}, we will modify each block of
Eqs.~\eqref{eqn:EOM_Scheme2_BoundaryODE}
with extra terms in the spirit of Eq.~\eqref{eqn:boundary_penalty_method_b}
--if the penalty method is chosen-- or Eq.~\eqref{eqn:boundary_bjorhus_method},
if the Bjørhus method is used.
The necessary right-hand-side modifications are constructed by assuming a
fiducial spectral element spanning the worldtube $\Gamma$
on which the functions $b_\pm$ are evaluated.
On the fiducial element we collocate grid points employing
a Chebyshev-Gauss-Lobatto~\cite{Kop09} collocation scheme with
the number of grid points chosen to be consistent with the order of
expansion of the regular field, \ie{} one grid point more
than the order.

The boundary data in the modification terms for the penalty method,
Eq.~\eqref{eqn:boundary_penalty_method},
is given simply by the values of the bulk fields
\begin{equation}
  \label{eqn:boundary_ODE_penalty_neighbour}
  \hat {\tilde u}^+_\mathrm{neighbour} = \hat u^+|_{r^*_\pm} \, .
\end{equation}
Similarly in the case of the Bjørhus method, Eq.~\eqref{eqn:boundary_bjorhus_method},
one uses the derivatives of the bulk fields,
\begin{equation}
  \label{eqn:boundary_ODE_bjorhus_neighbour}
  \tilde g = \hat s (  \partial_{r^*} \pi|_{r^*_\pm}
  + \hat s  \partial_{r^*} \chi|_{r^*_\pm}
  + \kappa \chi|_{r^*_\pm}  ) \,  .
\end{equation}
Denoting by $C$ the square bracket expressions
in either~\eqref{eqn:boundary_penalty_method_b}
or~\eqref{eqn:boundary_bjorhus_method},
the corresponding modifications in the auxiliary system are given by
\begin{align}
  \label{eqn:boundary_ODE}
  \partial_t \hat {\tilde u}^+ & =
  D_t \hat {\tilde u}^+ + C_{r^*_-} b_- + C_{r^*_+} b_+ \, ,                \\
  \partial_t \partial_{r^*} \hat {\tilde u}^+
                               & = D_t \partial_{r^*} \hat {\tilde u}^+
  + C_{r^*_-} \partial_{r^*} b_-
  + C_{r^*_+} \partial_{r^*} b_+                                       \, , \\
  \partial_t \partial_{r^*}^2 \hat {\tilde u}^+
                               & = D_t \partial_{r^*}^2 \hat {\tilde u}^+
  + C_{r^*_-} \partial_{r^*}^2 b_-
  + C_{r^*_+} \partial_{r^*}^2 b_+  \, .
\end{align}
The system generalises in an obvious way to higher orders by taking higher
and higher derivatives in $r^*$, and it is understood that
$C_{r^*_\pm}$ are constants with respect to $r^*$.
When employing the penalty boundary
conditions~\eqref{eqn:boundary_penalty_method},
the penalty parameter contained in $C_{r^*_\pm}$ is determined from
the fiducial worldtube element grid.
As for the boundary modifications of the bulk PDEs, the
modifications of the ODE system~\eqref{eqn:EOM_Scheme2_BoundaryODE}
are obtained by
transforming back from the characteristic variables to
the evolved variables.

\subsection{Implementation Details}
\label{Subsection:Scheme2ImplementationDetails}

We evolve Eqs.~\eqref{eqn:EOM_Scheme2_simple}
and~\eqref{eqn:EOM_Scheme2_BoundaryODE} using the classic fourth-order
Runge–Kutta method on the spatial domain $[-300M,300M] \setminus \Gamma$,
where the worldtube $\Gamma$ is centered on the position
$r_p^*$ of the particle.
The numerical domain is divided into a set of spectral elements.
On each element field values are expanded into a series of basis polynomials
and on every element the series is truncated at the same order.
For the results presented here, Chebyshev polynomials are used as the
spectral basis and the grid is collocated on Chebyshev-Gauss-Lobatto points.
Details on how the method works can be found
in~\cite{Kop09}, which we mostly followed for our code.

For the evolution of the system to be stable, the time step has to satisfy
a CFL condition, where the relevant length interval is
given by the minimal interval $\Delta r^*_\mathrm{min}$ between
collocations points.
Henceforth we chose the time step to be
$\Delta t = c_\mathrm{CFL} \Delta r^*_\mathrm{min}$,
where it is important that the fiducial element, which was introduced for the
coupling to the auxiliary boundary system, is also taken into account in the
determination of $\Delta r^*_\mathrm{min}$.
The constant $c_\mathrm{CFL}$ is set to $0.5$ for
static setups ($m=0$ modes),
whereas for the evolution of radiative modes we chose $c_\mathrm{CFL}=0.25$,
which we have checked to be sufficiently small for the results	in our setups
to be converged with respect to $\Delta t$.
Furthermore, we note that the matching must be performed at every Runge-Kutta
substep, \ie{} in every right-hand-side evaluation.
Otherwise the scheme does
not converge with fourth order with respect to $\Delta t$.

For radiative modes it is important to treat the phase factor
$e^{-i m \Omega t}$, which is
implicit in $S_{\ell m}(t)$ in Eq.~\eqref{eqn:PunctureAnsatz}, with high
numerical precision. Because of the high precision of the spectral scheme,
we are sensitive to machine round-off in the argument of the phase factor.
If we were to use the total time $t$, the round-off error
would grow linearly
with $t$ and thus affect the precision of the phase factor
as well.
Instead we use an approach where after every time step $\Delta t$
we multiply the phase factor by $e^{-i m \Omega \Delta t}$ to update
it for the next time step, and normalise the
result such that the norm of the phase factor remains exactly 1.
The same updating procedure is employed for the
individual Runge-Kutta substeps, so that the phase factor corresponds to the
respective time of the substep.
Furthermore, we use quadruple precision in evaluating the evolution time.
We find this necessary to ensure that our convergence tests are not limited by
round-off error in the time at which different
configurations are compared.

\section{Scheme II: Tests and analysis}
\label{section:Scheme2Results}
\subsection{Test Setups}
\label{Subsubsection:Scheme2TestSetups}

We investigate the scheme with two different matching setups,
which differ by the polynomial order of the matched regular field
inside the worldtube.
We test one setup where the regular field is expanded to fifth
order and the auxiliary system is evolved using
the evolution equations as stated in Sec.~\ref{section:Scheme2Method}.
In the second setup the regular field is expanded to
third order only and the auxiliary system is evolved using
Eqs.~\eqref{eqn:EOM_BoundaryODE_Psi}--\eqref{eqn:EOM_BoundaryODE_dchi},
with $\tilde \Psi''$, $\tilde \pi''$, $\tilde \chi''$
replaced by their analytical approximations to close the system.
The expansion of the puncture field varies from second to fifth order and
hence the second setup can probe a situation where the truncation error
is dominated by the regular field, whereas the first setup always makes sure
that the error in the puncture field converges slower or as fast
as the regular field.
Combined these setups allow us to distinguish between the
convergence behaviour stemming from the truncation of the
puncture expansion and the one stemming from the expansion of the regular
field.

We use a numerical resolution of 15 grid points per spectral element
and up to 256 spectral elements covering $r^*$ values in the
interval $[-100, 300] \setminus [r_p^* - R, r_p^* + R]$.
Because the numerical domain changes when the worldtube radius is changed
we always compute norms on the reference $r^*$ interval
covering  values $[-100, 300] \setminus [r_p^* - 2 M, r_p^* + 2 M]$ to ensure comparability.
For the internal convergence tests we interpolate all solutions on a
common reference grid before subtracting the different solutions. This
accounts for the change of grid collocation when the resolution or
the worldtube radius is changed.

We first test the scheme for the static mode $\ell=2,m=0$, with the
scalar charge in
a fixed circular geodesic orbit with $r_p = 7M$ (as for Scheme I).
As initial data for the evolution, the static analytical
solution~\eqref{eqn:analytical} is used.
Because of the finite-resolution truncation error, the numerical solution does
not
settle to the precise analytical solution, but to a slightly different one.
The transition
from the analytical initial data to the numerical static solution
causes transient radiation propagating off the numerical domain, with a small
partial reflection at the outer domain boundaries.
After a coordinate time of $\sim 5000 M$ much of the transient
radiation has decayed and the difference with the analytical solution
saturates.
The numerical data at this time is taken as the representation of the numerical
static solution.

\begin{figure}
  \includegraphics[width=\columnwidth]
  {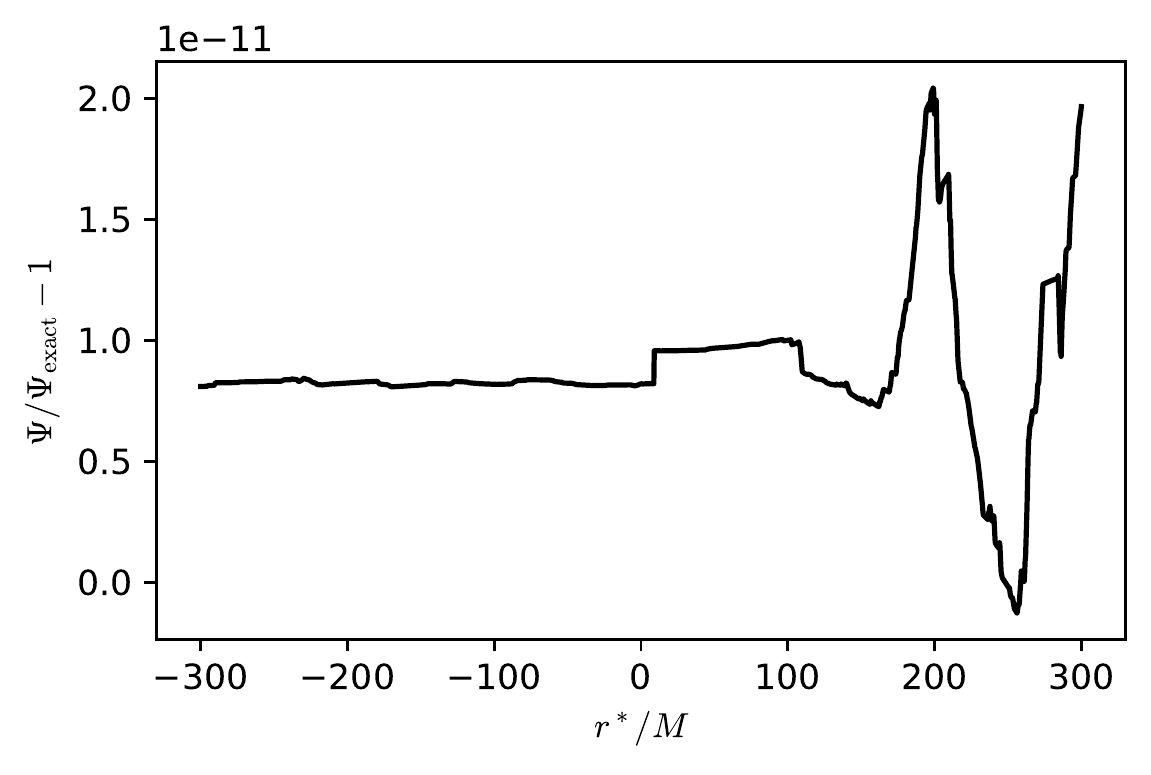}
  \caption{
    \label{fig:Scheme2_static_solution_relerror_Y20}
    Relative error of the $Y_{20}$ mode for 15 points per element using
    the 5th order expansions for the puncture and regular field, and
    a worldtube radius of $R = 0.1$.  }
\end{figure}

Figure~\ref{fig:Scheme2_static_solution_relerror_Y20} shows the relative
difference between the numerical and analytic solution. The small
differences are amplified at large $r^*$, because they are
normalised by the
solution which approaches zero in this limit.
The data shows that the numerical solution does not
settle down completely, but that there remains some residual
numerical noise propagating on the grid, which can be
attributed to the finite machine precision.
We have also tested the numerical scheme with different initial data and
find that it settles down to the same static solution. However, in that case
the transient radiation has a larger amplitude and takes longer to decay.

\begin{figure}
  \includegraphics[width=\columnwidth]
  {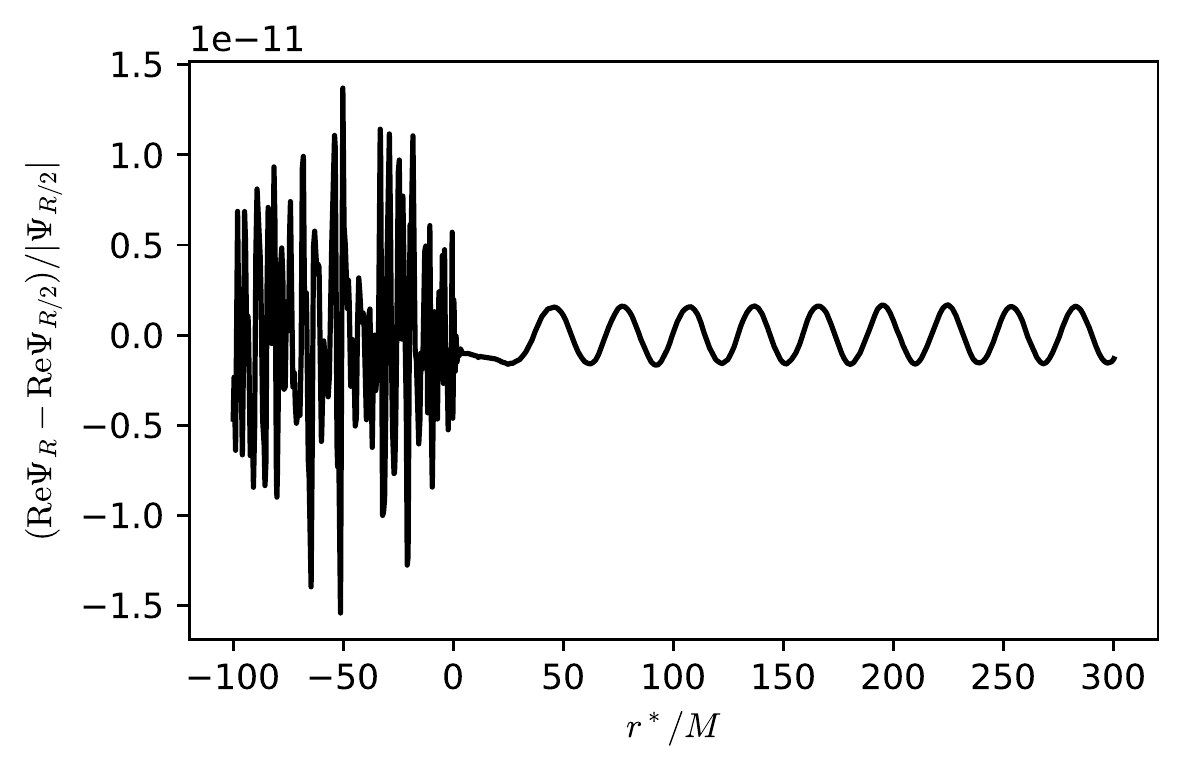}
  \caption{
    \label{fig:Scheme2_static_solution_relerror_Y44}
    Relative error of the $Y_{44}$ mode for 15 points per element using
    the 5th order expansions for the puncture and regular field and
    a worldtube radius of $R = 0.05$. The data is taken for the
    stationary end state at $t=11000 M$. }
\end{figure}

To test the convergence with $R$ in a non-static situation
we investigate the
$\ell=4, m=4$ mode,
with the charge
again being fixed at $r_p=7M$. For these modes there is no
solution known
in analytical form and hence we start with zero initial data,
\ie{}
$\Psi = 0, \dot \Psi = \pi = 0$.
Furthermore we estimate the error by taking the difference between
two runs that differ in the size of their worldtubes by a factor of two.
As in the static case, there is some transient radiation that is
radiated away until the system settles down to a stationary state.
The presence of this junk radiation is partially obscured by the periodic
changes in $\Psi$, but it can be observed when taking the
difference
of the modulus $|\Psi|$ of two simulations with different
worldtube size.
Figure~\ref{fig:Scheme2_static_solution_relerror_Y44} shows the estimated
relative error for the highest resolution used in our results.
It can be observed that
the error on the left side of the particle is dominated by numerical
noise, whereas on the right the difference between the two runs looks
smooth and is modulated by the periodic waveform of the signal.
The noise on the left side can be attributed to an insufficient resolution,
which is amplified by the relatively small modulus $|\Psi|$ in
that area; cf.\ Fig.~\ref{fig:FDS_waveform_inset}.

\subsection{Convergence Tests}
\label{Subsection:Scheme2Convergence}

\begin{figure}
  \includegraphics[width=\columnwidth]
  {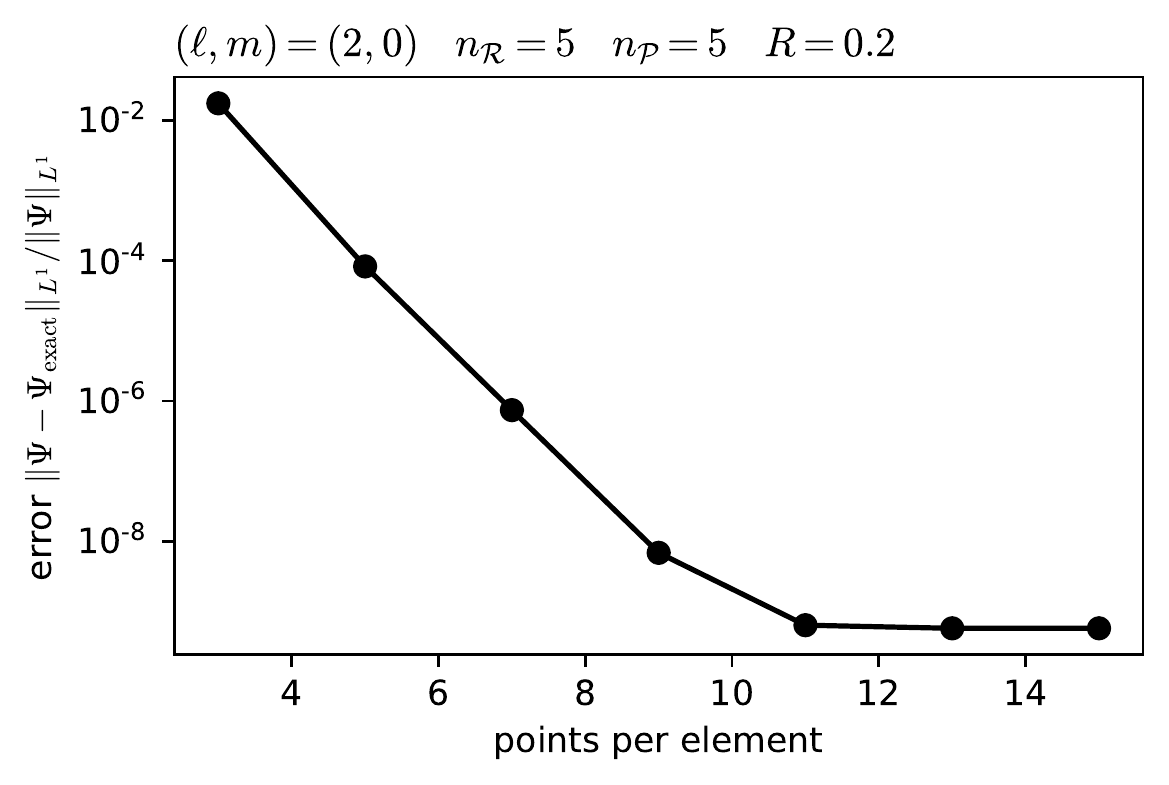}
  \caption{
    \label{fig:Scheme2_resolution_convergence}
    Error of the $Y_{20}$ mode as a function of the number of
    points per spectral element.
    The grid consists of 128 elements and
    a has worldtube radius of $R = 0.2$. The puncture and regular
    parts of fields are expanded to 5th order.
  }
\end{figure}

Figure~\ref{fig:Scheme2_resolution_convergence} shows the error as a function
of the number of collocation points per spectral element and
demonstrates the exponential
convergence of our spectral discretisation scheme.
The error levels off at high number of points, because error contributions
in the matched analytical solution inside the worldtube become the dominating
source of error.
Next the convergence with respect to the
size of the worldtube is tested. In these tests we report only error
quantities that are fully converged with respect to the number of collocation
points and the matched analytical solution is the only source of error.

\begin{figure}[tb]
  \includegraphics[width=\columnwidth]
  {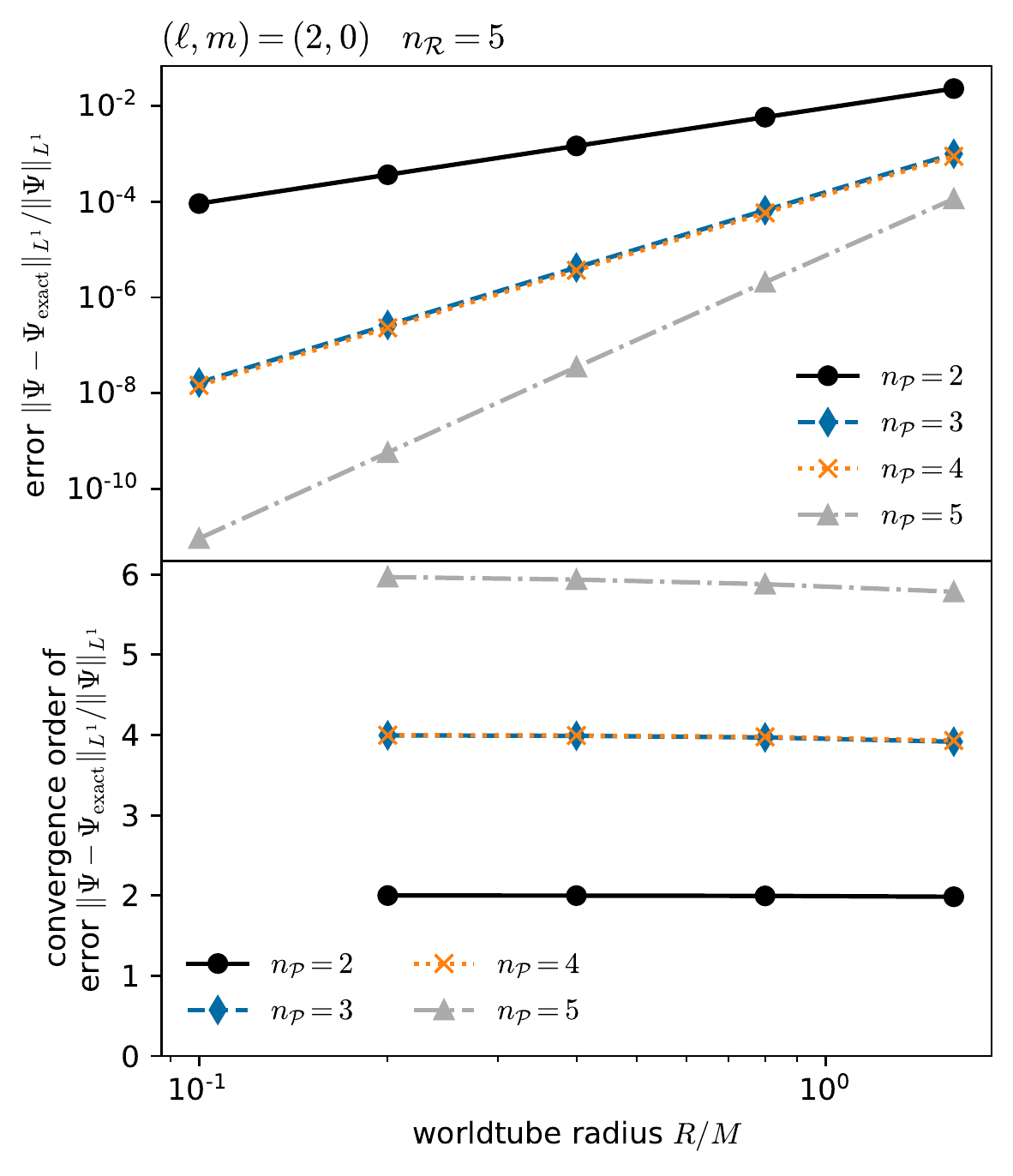}
  \caption{
    \label{fig:Scheme2_external_convergence_lm20}
    Top panel: $L^1$ norm of the difference to the analytical
    static solution with respect to the worldtube radius $R$
    for the $Y_{20}$ mode.
    The indicated order $n_\mathcal{P}$ denotes the order of expansion of
    the puncture field $\Psi^\mathcal{P}$. The regular fields are all expanded
    to order five.
    Bottom panel: Convergence order for the $L^1$ norm.  }
\end{figure}

Figure~\ref{fig:Scheme2_external_convergence_lm20} demonstrates the convergence
of the first setup, where the matched regular fields are all expanded
up to order five.
It shows the $L^1$ norm of the difference to the exact
analytical solution,
along with
the corresponding convergence order $n_R^\mathrm{(norm)}$
defined in Eq.~\eqref{eq:n_R^norm}.
For even expansion orders $n_\mathcal{P}$ of the puncture field, the
total scheme converges
with $n_\mathcal{P}$ as predicted from the analysis in
Sec.~\ref{Subsection:error estimates},
whereas for the odd orders the convergence
is one order higher.
It is not known to us what causes this irregular convergence pattern
with respect to $n_\mathcal{P}$, but this finding suggests that
the $\partial_n \Psi^N$ term in Eq.~\eqref{eq:kirchhoff_representation}
is suppressed for odd $n_\mathcal{P}$.

We can not only observe a decrease in the error with decreasing worldtube
size, but also with increasing order of the puncture expansion.
The error for 3rd and 4th order, however, are almost identical in their
magnitude. The reason for this particular behaviour is unknown to us.

\begin{figure}[tb]
  \includegraphics[width=\columnwidth]
  {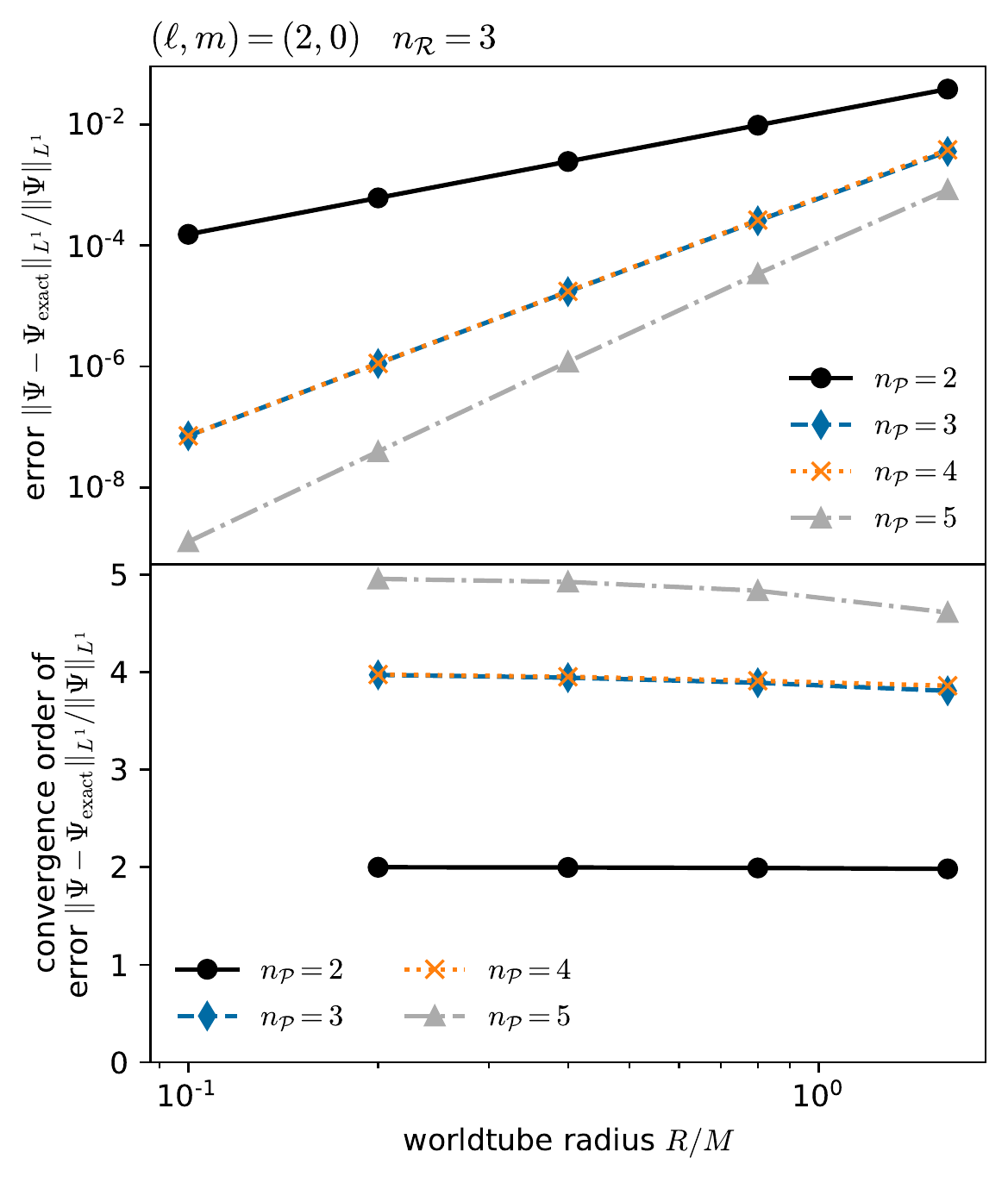}
  \caption{
    \label{fig:Scheme2_external_convergence_lm20_regular3}
    Top panel: $L^1$ norm of the difference to the analytical
    static solution with respect to the worldtube radius $R$
    for the $Y_{20}$ mode.
    The indicated order $n_\mathcal{P}$ denotes the order of expansion of
    the puncture field $\Psi^\mathcal{P}$. The regular fields are all expanded
    to order three.
    Bottom panel: Convergence order for the $L^1$ norm.  }
\end{figure}

Figure~\ref{fig:Scheme2_external_convergence_lm20_regular3} shows the
convergence behaviour of the second setup.
For the cases where the expansion order of the puncture is in the range
from two to four we find convergence
consistent with the findings of the first setup.
However, for a puncture expansion order of five the convergence order
is limited by the regular solution.
Since the regular field is truncated at third order in this setup,
the analysis of Sec.~\ref{Subsection:error estimates} would predict
a convergence order of only three and disregarding the $\partial_n \Psi^N$
terms
a convergence order of four, but it is observed that
the convergence is actually of fifth order. This apparent super-geometric
convergence could be explained by the fact that it is not only the field
$\Psi^\mathcal{R}$ that is expanded to third order, but also the
derivative $\chi^\mathcal{R}$.
Since the third order coefficient of $\chi^\mathcal{R}$ corresponds to the
fourth
order coefficient of $\Psi^\mathcal{R}$ this could lead to a scheme where
the
effective expansion order of the regular field is one order higher than
naively expected.

\begin{figure}[tb]
  \includegraphics[width=\columnwidth]
  {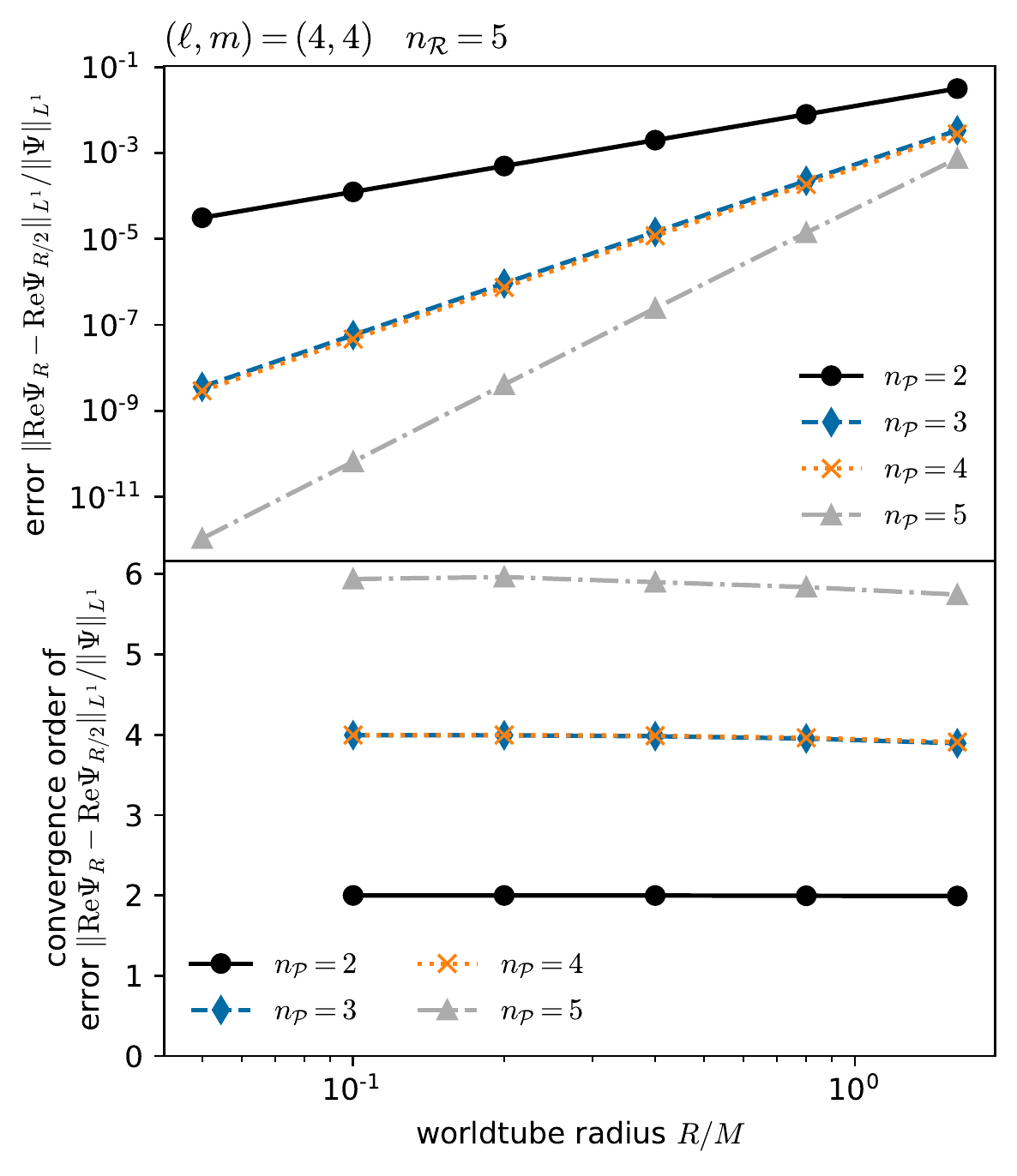}
  \caption{
    \label{fig:Scheme2_external_convergence_L1_lm44_psi}
    Top panel: $L^1$ norm of the difference between solutions with
    worldtube radius $R$ and $R/2$
    for the $Y_{44}$ mode.
    The indicated order $n_\mathcal{P}$ denotes the order of expansion of
    the puncture field $\Psi^\mathcal{P}$. The regular fields are all expanded
    to order five.
    Bottom panel: Convergence order for the $L^1$ norm. }
\end{figure}

\begin{figure}[tb]
  \includegraphics[width=\columnwidth]
  {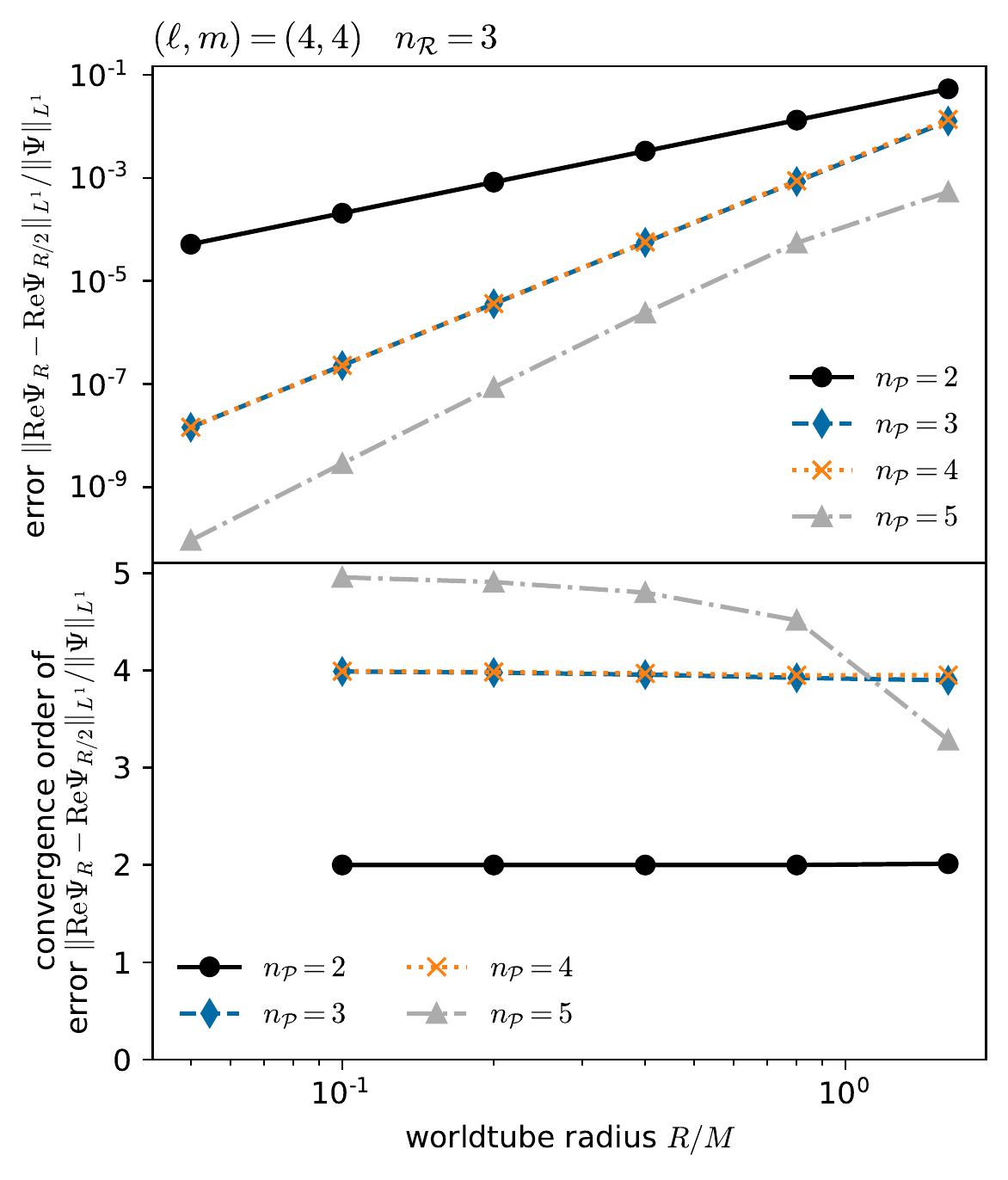}
  \caption{
    \label{fig:Scheme2_external_convergence_L1_lm44_psi_regular3}
    Top panel: $L^1$ norm of the difference between solutions with
    worldtube radius $R$ and $R/2$
    for the $Y_{44}$ mode.
    The indicated order $n_\mathcal{P}$ denotes the order of expansion of
    the puncture field $\Psi^\mathcal{P}$. The regular fields are all expanded
    to order three.
    Bottom panel: Convergence order for the $L^1$ norm. }
\end{figure}

The radiative modes exhibit an error convergence behaviour
that is qualitatively identical to the static modes. For completeness we
show the corresponding convergence behaviour in
Figs.~\ref{fig:Scheme2_external_convergence_L1_lm44_psi}
and~\ref{fig:Scheme2_external_convergence_L1_lm44_psi_regular3}.

\section{Conclusion}\label{Section: Conclusion}

Numerical simulations of binary black holes are increasingly costly
with more extreme mass ratio $q$, because the CFL
instability
forces a reduction of the evolution time step $\Delta t\lesssim
  m_2\propto q$.	This
article explores a new technique to circumvent
the time step limitations that arise from solving the field equations fully
numerically in
the region of spacetime near the small mass $m_2$.  We place
a
worldtube of radius $R\gg m_2$ around $m_2$.
Inside this worldtube,
we replace the fully numerical solution by a perturbative solution
while retaining a fully numerical solution outside the worldtube.
Thus, the smallest length scales remaining on the numerical grid are
of order $R$, and the CFL limit is relaxed to
$\Delta t\sim R$.

The present article explores the toy problem of computing the scalar
field generated by a point-charge orbiting a Schwarzschild black hole
on a circular orbit.  The solutions of this problem are illustrated
in Fig.~\ref{fig:FDS_waveform_inset}.
We explore two different algorithms to match the
perturbed solution inside the worldtube with the fully numerical
solution outside, a matching scheme that fits over an extended
region, and a boundary scheme that imposes conditions precisely at one
radius.  We furthermore explore two different numerical
implementations for the numerical exterior solution, one based on
finite differences on a characteristic grid, the other on spectral
methods on spatial hypersurfaces.
The most accurate solutions are obtained with the spectral scheme,
which allows significantly smaller discretisation errors than the
finite-difference scheme.

We achieved stable evolutions in both schemes.	We also
established convergence of the results in various quantities: in the
grid spacing of the numerical domain, in the radius $R$ of
the
worldtube, as well as in the order of the approximate perturbative
solution inside the worldtube.
Of primary concern is convergence with worldtube radius
$R$.  For the
finite-difference scheme, Fig.~\ref{fig:FDS_L1_error_2} demonstrates
convergence $\propto R^n$ for a perturbative solution of order
$n$.   For the spectral
scheme, Figs.~\ref{fig:Scheme2_external_convergence_lm20} and
\ref{fig:Scheme2_external_convergence_L1_lm44_psi} demonstrate up to sixth-order convergence in
$R$.

For a worldtube radius of $R=0.8M$, we achieve relative errors
$\lesssim
  10^{-6}$ for a fifth-order perturbative solution, and
$\approx 10^{-4}$
for a third-order internal solution.  This level of accuracy at such
large worldtube radius is encouraging for our ultimate goal, the
application of the worldtube scheme to full GR.

The work in this paper has focused on the implementation of the
worldtube architecture and exploration of matching approaches, thus
demonstrating in principle that the proposed worldtube excision method
works.

Future steps include 3+1D simulations of the scalar-charge
problem, extending over the 1+1D simulations presented here.  Such
simulations will first proceed with the charge on a fixed circular
orbit, and subsequently they will include scalar self-force effects
resulting in an inspiral of the scalar charge.	These steps will
provide valuable additional insights needed to address our ultimate
goal, the full binary black hole problem in GR at intermediate
mass ratios.

We expect much of our methodology to carry over to those more complex cases,
but with some foreseeable differences and challenges. In the 3+1D scalar toy
problem, in local Fermi coordinates $(t,x^i)$ centered on the
scalar charge, the scalar field has the form $\Phi = \Phi^\mathcal{P}+ \Phi^\mathcal{R}$
with~\cite{Poisson:2011nh}
\begin{equation}
  \Phi^\mathcal{P}=\frac{e}{s} - \frac{e}{6s}\mathcal{E}_{ij}(t)x^ix^j +\mathcal{O}(s^2).
\end{equation}
Here $s=\sqrt{\delta_{ij}x^ix^j}$ is the distance from the particle, and
$\mathcal{E}_{ij} = R_{titj}$ (evaluated on the particle) is the tidal field due to
the large black hole. This local solution takes the place of our 1+1D local
solution~\eqref{eq:psiA def} [with Eq.~\eqref{eqn:PunctureAnsatz}]. All the unknowns in the 3+1D local solution
are contained in $\Phi^\mathcal{R}$, just as in our 1+1D calculation, and
after transforming the local solution to whatever external coordinates are
convenient, our matching approaches should apply with little or no fundamental
change.

More significant differences appear in the full gravity problem, where the
local solution will instead be the metric of a tidally perturbed black hole. In
local coordinates centered on the small black hole, a typical component has the
form
\begin{equation}\label{eq:tidally perturbed BH}
  g_{tt} = - (1-m_2/s) + e_1(m_2/s)\mathcal{E}_{ij}x^ix^j +\mathcal{O}(s^3/\mathcal{R}^3)
\end{equation}
for some function $e_1(m_2/s)$~\cite{Poisson:2005pi,Poisson:2018qqd}, where
$m_2$ is the black hole's mass. Here $\mathcal{E}_{ij}$
is now an \emph{effective} tidal field that can only be determined
through a matching procedure; due to nonlinearities, it is influenced by the
small black hole's own effect on the external geometry~\cite{Pound:2017psq}.
The $m_2/s$ term in $g_{tt}$ plays the role of
the $e/s$ term in $\Phi^\mathcal{P}$, but as mentioned
in Sec.~\ref{Section: Matching Methods}, beyond that term the metric does not have a
neat decomposition into singular and regular pieces. Due to the
$m_2/s$ dependence in $e_1$, the tidal term
is singular at $s=0$. A regular piece can still be
constructed by setting explicit appearances of $m_2$ to 0,
yielding a metric with components of the form $g^\mathcal{R}_{tt} = - 1 + \mathcal{E}_{ij}x^ix^j+\mathcal{O}(s^3/\mathcal{R}^3)$, which is
a smooth vacuum solution analogous to $\Phi^\mathcal{R}$. But the
remainder $g^\mathcal{P}_{\alpha\beta}=g_{\alpha\beta}-g^\mathcal{R}_{\alpha\beta}$ does not satisfy a simple field equation. Any
such field equation will involve complicated nonlinear dependence on
$g^\mathcal{R}_{\alpha\beta}$, such that $g^\mathcal{P}_{\alpha\beta}$ cannot be determined
without simultaneously determining $g^\mathcal{R}_{\alpha\beta}$. This should not
pose a problem for us because, by design, our matching approaches allow us to
match the total metric to the external solution rather than solving field
equations for a residual field as in a traditional puncture scheme discussed in Sec.~\ref{Subsection:error estimates}.

A more pronounced practical complication in the gravity problem is that we will
not know in advance the transformation from the local coordinates of
Eq.~\eqref{eq:tidally perturbed BH} to the coordinates used in the external numerics.
The transformation must be determined as part of the matching procedure.
Likewise, the motion of the small black hole must be determined dynamically
through matching. We leave these challenges for future work.

\begin{acknowledgments}

MD acknowledges support from an STFC studentship, Project Reference No.~2283146.
AP gratefully acknowledges the support of a Royal Society University Research Fellowship and a Research Fellows Enhancement Award.
LB acknowledges support from STFC through Grant No.~ST/R00045X/1.
We thank Ian Hawke, Barry Wardell and Nikolas Wittek for helpful discussions.

\end{acknowledgments}
\bibliography{worldtube_excision_1D.bbl}

\begin{thebibliography}{44}%
\makeatletter
\providecommand \@ifxundefined [1]{%
 \@ifx{#1\undefined}
}%
\providecommand \@ifnum [1]{%
 \ifnum #1\expandafter \@firstoftwo
 \else \expandafter \@secondoftwo
 \fi
}%
\providecommand \@ifx [1]{%
 \ifx #1\expandafter \@firstoftwo
 \else \expandafter \@secondoftwo
 \fi
}%
\providecommand \natexlab [1]{#1}%
\providecommand \enquote  [1]{``#1''}%
\providecommand \bibnamefont  [1]{#1}%
\providecommand \bibfnamefont [1]{#1}%
\providecommand \citenamefont [1]{#1}%
\providecommand \href@noop [0]{\@secondoftwo}%
\providecommand \href [0]{\begingroup \@sanitize@url \@href}%
\providecommand \@href[1]{\@@startlink{#1}\@@href}%
\providecommand \@@href[1]{\endgroup#1\@@endlink}%
\providecommand \@sanitize@url [0]{\catcode `\\12\catcode `\$12\catcode
  `\&12\catcode `\#12\catcode `\^12\catcode `\_12\catcode `\%12\relax}%
\providecommand \@@startlink[1]{}%
\providecommand \@@endlink[0]{}%
\providecommand \url  [0]{\begingroup\@sanitize@url \@url }%
\providecommand \@url [1]{\endgroup\@href {#1}{\urlprefix }}%
\providecommand \urlprefix  [0]{URL }%
\providecommand \Eprint [0]{\href }%
\providecommand \doibase [0]{https://doi.org/}%
\providecommand \selectlanguage [0]{\@gobble}%
\providecommand \bibinfo  [0]{\@secondoftwo}%
\providecommand \bibfield  [0]{\@secondoftwo}%
\providecommand \translation [1]{[#1]}%
\providecommand \BibitemOpen [0]{}%
\providecommand \bibitemStop [0]{}%
\providecommand \bibitemNoStop [0]{.\EOS\space}%
\providecommand \EOS [0]{\spacefactor3000\relax}%
\providecommand \BibitemShut  [1]{\csname bibitem#1\endcsname}%
\let\auto@bib@innerbib\@empty
\bibitem [{\citenamefont {Abbott}\ \emph {et~al.}(2019)\citenamefont {Abbott}
  \emph {et~al.}}]{LIGOScientific:2018mvr}%
  \BibitemOpen
  \bibfield  {author} {\bibinfo {author} {\bibfnamefont {B.~P.}\ \bibnamefont
  {Abbott}} \emph {et~al.} (\bibinfo {collaboration} {LIGO Scientific,
  Virgo}),\ }\bibfield  {title} {\bibinfo {title} {{GWTC-1: A
  Gravitational-Wave Transient Catalog of Compact Binary Mergers Observed by
  LIGO and Virgo during the First and Second Observing Runs}},\ }\href
  {https://doi.org/10.1103/PhysRevX.9.031040} {\bibfield  {journal} {\bibinfo
  {journal} {Phys. Rev. X}\ }\textbf {\bibinfo {volume} {9}},\ \bibinfo {pages}
  {031040} (\bibinfo {year} {2019})},\ \Eprint
  {https://arxiv.org/abs/1811.12907} {arXiv:1811.12907 [astro-ph.HE]}
  \BibitemShut {NoStop}%
\bibitem [{\citenamefont {{Abbott}}\ \emph {et~al.}(2021)\citenamefont
  {{Abbott}}, \citenamefont {{Abbott}}, \citenamefont {{Abraham}},
  \citenamefont {{Acernese}}, \citenamefont {{Ackley}}, \citenamefont
  {{Adams}}, \citenamefont {{Adams}}, \citenamefont {{Adhikari}} \emph
  {et~al.}}]{Abbott:2020niy}%
  \BibitemOpen
  \bibfield  {author} {\bibinfo {author} {\bibfnamefont {R.}~\bibnamefont
  {{Abbott}}}, \bibinfo {author} {\bibfnamefont {T.~D.}\ \bibnamefont
  {{Abbott}}}, \bibinfo {author} {\bibfnamefont {S.}~\bibnamefont {{Abraham}}},
  \bibinfo {author} {\bibfnamefont {F.}~\bibnamefont {{Acernese}}}, \bibinfo
  {author} {\bibfnamefont {K.}~\bibnamefont {{Ackley}}}, \bibinfo {author}
  {\bibfnamefont {A.}~\bibnamefont {{Adams}}}, \bibinfo {author} {\bibfnamefont
  {C.}~\bibnamefont {{Adams}}}, \bibinfo {author} {\bibfnamefont {R.~X.}\
  \bibnamefont {{Adhikari}}}, \emph {et~al.} (\bibinfo {collaboration} {LIGO
  Scientific, Virgo}),\ }\bibfield  {title} {\bibinfo {title} {{GWTC-2: Compact
  Binary Coalescences Observed by LIGO and Virgo During the First Half of the
  Third Observing Run}},\ }\href {https://doi.org/10.1103/PhysRevX.11.021053}
  {\bibfield  {journal} {\bibinfo  {journal} {Phys. Rev. X}\ }\textbf {\bibinfo
  {volume} {11}},\ \bibinfo {pages} {021053} (\bibinfo {year} {2021})},\
  \Eprint {https://arxiv.org/abs/2010.14527} {arXiv:2010.14527 [gr-qc]}
  \BibitemShut {NoStop}%
\bibitem [{\citenamefont {Abbott}\ \emph
  {et~al.}(2020{\natexlab{a}})\citenamefont {Abbott} \emph
  {et~al.}}]{Abbott:2020khf}%
  \BibitemOpen
  \bibfield  {author} {\bibinfo {author} {\bibfnamefont {R.}~\bibnamefont
  {Abbott}} \emph {et~al.} (\bibinfo {collaboration} {LIGO Scientific,
  Virgo}),\ }\bibfield  {title} {\bibinfo {title} {{GW190814: Gravitational
  Waves from the Coalescence of a 23 Solar Mass Black Hole with a 2.6 Solar
  Mass Compact Object}},\ }\href {https://doi.org/10.3847/2041-8213/ab960f}
  {\bibfield  {journal} {\bibinfo  {journal} {Astrophys. J. Lett.}\ }\textbf
  {\bibinfo {volume} {896}},\ \bibinfo {pages} {L44} (\bibinfo {year}
  {2020}{\natexlab{a}})},\ \Eprint {https://arxiv.org/abs/2006.12611}
  {arXiv:2006.12611 [astro-ph.HE]} \BibitemShut {NoStop}%
\bibitem [{\citenamefont {Abbott}\ \emph
  {et~al.}(2020{\natexlab{b}})\citenamefont {Abbott} \emph
  {et~al.}}]{KAGRA:2020npa}%
  \BibitemOpen
  \bibfield  {author} {\bibinfo {author} {\bibfnamefont {B.~P.}\ \bibnamefont
  {Abbott}} \emph {et~al.} (\bibinfo {collaboration} {KAGRA, LIGO Scientific,
  Virgo}),\ }\bibfield  {title} {\bibinfo {title} {{Prospects for observing and
  localizing gravitational-wave transients with Advanced LIGO, Advanced Virgo
  and KAGRA}},\ }\href {https://doi.org/10.1007/s41114-020-00026-9} {\bibfield
  {journal} {\bibinfo  {journal} {Living Rev. Rel.}\ }\textbf {\bibinfo
  {volume} {23}},\ \bibinfo {pages} {3} (\bibinfo {year}
  {2020}{\natexlab{b}})}\BibitemShut {NoStop}%
\bibitem [{\citenamefont {Hall}\ and\ \citenamefont
  {Evans}(2019)}]{Hall:2019xmm}%
  \BibitemOpen
  \bibfield  {author} {\bibinfo {author} {\bibfnamefont {E.~D.}\ \bibnamefont
  {Hall}}\ and\ \bibinfo {author} {\bibfnamefont {M.}~\bibnamefont {Evans}},\
  }\bibfield  {title} {\bibinfo {title} {{Metrics for next-generation
  gravitational-wave detectors}},\ }\href
  {https://doi.org/10.1088/1361-6382/ab41d6} {\bibfield  {journal} {\bibinfo
  {journal} {Class. Quant. Grav.}\ }\textbf {\bibinfo {volume} {36}},\ \bibinfo
  {pages} {225002} (\bibinfo {year} {2019})},\ \Eprint
  {https://arxiv.org/abs/1902.09485} {arXiv:1902.09485 [astro-ph.IM]}
  \BibitemShut {NoStop}%
\bibitem [{\citenamefont {Amaro-Seoane}\ \emph {et~al.}(2017)\citenamefont
  {Amaro-Seoane} \emph {et~al.}}]{LISA:2017pwj}%
  \BibitemOpen
  \bibfield  {author} {\bibinfo {author} {\bibfnamefont {P.}~\bibnamefont
  {Amaro-Seoane}} \emph {et~al.} (\bibinfo {collaboration} {LISA}),\
  }\href@noop {} {\bibinfo {title} {{Laser Interferometer Space Antenna}}}
  (\bibinfo {year} {2017}),\ \Eprint {https://arxiv.org/abs/1702.00786}
  {arXiv:1702.00786 [astro-ph.IM]} \BibitemShut {NoStop}%
\bibitem [{\citenamefont {Jani}\ \emph {et~al.}(2019)\citenamefont {Jani},
  \citenamefont {Shoemaker},\ and\ \citenamefont {Cutler}}]{Jani:2019ffg}%
  \BibitemOpen
  \bibfield  {author} {\bibinfo {author} {\bibfnamefont {K.}~\bibnamefont
  {Jani}}, \bibinfo {author} {\bibfnamefont {D.}~\bibnamefont {Shoemaker}},\
  and\ \bibinfo {author} {\bibfnamefont {C.}~\bibnamefont {Cutler}},\
  }\bibfield  {title} {\bibinfo {title} {{Detectability of Intermediate-Mass
  Black Holes in Multiband Gravitational Wave Astronomy}},\ }\href
  {https://doi.org/10.1038/s41550-019-0932-7} {\bibfield  {journal} {\bibinfo
  {journal} {Nature Astron.}\ }\textbf {\bibinfo {volume} {4}},\ \bibinfo
  {pages} {260} (\bibinfo {year} {2019})},\ \Eprint
  {https://arxiv.org/abs/1908.04985} {arXiv:1908.04985 [gr-qc]} \BibitemShut
  {NoStop}%
\bibitem [{\citenamefont {Salcido}\ \emph {et~al.}(2016)\citenamefont
  {Salcido}, \citenamefont {Bower}, \citenamefont {Theuns}, \citenamefont
  {McAlpine}, \citenamefont {Schaller}, \citenamefont {Crain}, \citenamefont
  {Schaye},\ and\ \citenamefont {Regan}}]{Salcido:2016oor}%
  \BibitemOpen
  \bibfield  {author} {\bibinfo {author} {\bibfnamefont {J.}~\bibnamefont
  {Salcido}}, \bibinfo {author} {\bibfnamefont {R.~G.}\ \bibnamefont {Bower}},
  \bibinfo {author} {\bibfnamefont {T.}~\bibnamefont {Theuns}}, \bibinfo
  {author} {\bibfnamefont {S.}~\bibnamefont {McAlpine}}, \bibinfo {author}
  {\bibfnamefont {M.}~\bibnamefont {Schaller}}, \bibinfo {author}
  {\bibfnamefont {R.~A.}\ \bibnamefont {Crain}}, \bibinfo {author}
  {\bibfnamefont {J.}~\bibnamefont {Schaye}},\ and\ \bibinfo {author}
  {\bibfnamefont {J.}~\bibnamefont {Regan}},\ }\bibfield  {title} {\bibinfo
  {title} {{Music from the heavens \textendash{} gravitational waves from
  supermassive black hole mergers in the EAGLE simulations}},\ }\href
  {https://doi.org/10.1093/mnras/stw2048} {\bibfield  {journal} {\bibinfo
  {journal} {Mon. Not. Roy. Astron. Soc.}\ }\textbf {\bibinfo {volume} {463}},\
  \bibinfo {pages} {870} (\bibinfo {year} {2016})},\ \Eprint
  {https://arxiv.org/abs/1601.06156} {arXiv:1601.06156 [astro-ph.GA]}
  \BibitemShut {NoStop}%
\bibitem [{\citenamefont {Volonteri}\ \emph {et~al.}(2020)\citenamefont
  {Volonteri} \emph {et~al.}}]{Volonteri:2020wkx}%
  \BibitemOpen
  \bibfield  {author} {\bibinfo {author} {\bibfnamefont {M.}~\bibnamefont
  {Volonteri}} \emph {et~al.},\ }\bibfield  {title} {\bibinfo {title} {{Black
  hole mergers from dwarf to massive galaxies with the NewHorizon and
  Horizon-AGN simulations}},\ }\href {https://doi.org/10.1093/mnras/staa2384}
  {\bibfield  {journal} {\bibinfo  {journal} {Mon. Not. Roy. Astron. Soc.}\
  }\textbf {\bibinfo {volume} {498}},\ \bibinfo {pages} {2219} (\bibinfo {year}
  {2020})},\ \Eprint {https://arxiv.org/abs/2005.04902} {arXiv:2005.04902
  [astro-ph.GA]} \BibitemShut {NoStop}%
\bibitem [{\citenamefont {Baumgarte}\ and\ \citenamefont
  {Shapiro}(2010)}]{BauSha10}%
  \BibitemOpen
  \bibfield  {author} {\bibinfo {author} {\bibfnamefont {T.~W.}\ \bibnamefont
  {Baumgarte}}\ and\ \bibinfo {author} {\bibfnamefont {S.~L.}\ \bibnamefont
  {Shapiro}},\ }\href@noop {} {\emph {\bibinfo {title} {Numerical Relativity:
  Solving {E}instein's Equations on the Computer}}}\ (\bibinfo  {publisher}
  {Cambridge University Press},\ \bibinfo {address} {Cambridge},\ \bibinfo
  {year} {2010})\BibitemShut {NoStop}%
\bibitem [{\citenamefont {Boyle}\ \emph {et~al.}(2019)\citenamefont {Boyle}
  \emph {et~al.}}]{Boyle:2019kee}%
  \BibitemOpen
  \bibfield  {author} {\bibinfo {author} {\bibfnamefont {M.}~\bibnamefont
  {Boyle}} \emph {et~al.},\ }\bibfield  {title} {\bibinfo {title} {{The SXS
  Collaboration catalog of binary black hole simulations}},\ }\href
  {https://doi.org/10.1088/1361-6382/ab34e2} {\bibfield  {journal} {\bibinfo
  {journal} {Class. Quant. Grav.}\ }\textbf {\bibinfo {volume} {36}},\ \bibinfo
  {pages} {195006} (\bibinfo {year} {2019})},\ \Eprint
  {https://arxiv.org/abs/1904.04831} {arXiv:1904.04831 [gr-qc]} \BibitemShut
  {NoStop}%
\bibitem [{\citenamefont {Lousto}\ and\ \citenamefont
  {Healy}(2020)}]{Lousto:2020tnb}%
  \BibitemOpen
  \bibfield  {author} {\bibinfo {author} {\bibfnamefont {C.~O.}\ \bibnamefont
  {Lousto}}\ and\ \bibinfo {author} {\bibfnamefont {J.}~\bibnamefont {Healy}},\
  }\bibfield  {title} {\bibinfo {title} {{Exploring the Small Mass Ratio Binary
  Black Hole Merger via Zeno\textquoteright{}s Dichotomy Approach}},\ }\href
  {https://doi.org/10.1103/PhysRevLett.125.191102} {\bibfield  {journal}
  {\bibinfo  {journal} {Phys. Rev. Lett.}\ }\textbf {\bibinfo {volume} {125}},\
  \bibinfo {pages} {191102} (\bibinfo {year} {2020})},\ \Eprint
  {https://arxiv.org/abs/2006.04818} {arXiv:2006.04818 [gr-qc]} \BibitemShut
  {NoStop}%
\bibitem [{\citenamefont {Rosato}\ \emph {et~al.}(2021)\citenamefont {Rosato},
  \citenamefont {Healy},\ and\ \citenamefont {Lousto}}]{Rosato:2021jsq}%
  \BibitemOpen
  \bibfield  {author} {\bibinfo {author} {\bibfnamefont {N.}~\bibnamefont
  {Rosato}}, \bibinfo {author} {\bibfnamefont {J.}~\bibnamefont {Healy}},\ and\
  \bibinfo {author} {\bibfnamefont {C.~O.}\ \bibnamefont {Lousto}},\ }\bibfield
   {title} {\bibinfo {title} {{Adapted gauge to small mass ratio binary black
  hole evolutions}},\ }\href {https://doi.org/10.1103/PhysRevD.103.104068}
  {\bibfield  {journal} {\bibinfo  {journal} {Phys. Rev. D}\ }\textbf {\bibinfo
  {volume} {103}},\ \bibinfo {pages} {104068} (\bibinfo {year} {2021})},\
  \Eprint {https://arxiv.org/abs/2103.09326} {arXiv:2103.09326 [gr-qc]}
  \BibitemShut {NoStop}%
\bibitem [{\citenamefont {Barack}\ and\ \citenamefont
  {Pound}(2019)}]{Barack:2018yvs}%
  \BibitemOpen
  \bibfield  {author} {\bibinfo {author} {\bibfnamefont {L.}~\bibnamefont
  {Barack}}\ and\ \bibinfo {author} {\bibfnamefont {A.}~\bibnamefont {Pound}},\
  }\bibfield  {title} {\bibinfo {title} {{Self-force and radiation reaction in
  general relativity}},\ }\href {https://doi.org/10.1088/1361-6633/aae552}
  {\bibfield  {journal} {\bibinfo  {journal} {Rept. Prog. Phys.}\ }\textbf
  {\bibinfo {volume} {82}},\ \bibinfo {pages} {016904} (\bibinfo {year}
  {2019})},\ \Eprint {https://arxiv.org/abs/1805.10385} {arXiv:1805.10385
  [gr-qc]} \BibitemShut {NoStop}%
\bibitem [{\citenamefont {Pound}\ and\ \citenamefont
  {Wardell}(2021)}]{Pound:2021qin}%
  \BibitemOpen
  \bibfield  {author} {\bibinfo {author} {\bibfnamefont {A.}~\bibnamefont
  {Pound}}\ and\ \bibinfo {author} {\bibfnamefont {B.}~\bibnamefont
  {Wardell}},\ }\href@noop {} {\bibinfo {title} {{Black hole perturbation
  theory and gravitational self-force}}} (\bibinfo {year} {2021}),\ \Eprint
  {https://arxiv.org/abs/2101.04592} {arXiv:2101.04592 [gr-qc]} \BibitemShut
  {NoStop}%
\bibitem [{\citenamefont {van~de Meent}(2018)}]{vandeMeent:2017bcc}%
  \BibitemOpen
  \bibfield  {author} {\bibinfo {author} {\bibfnamefont {M.}~\bibnamefont
  {van~de Meent}},\ }\bibfield  {title} {\bibinfo {title} {{Gravitational
  self-force on generic bound geodesics in Kerr spacetime}},\ }\href
  {https://doi.org/10.1103/PhysRevD.97.104033} {\bibfield  {journal} {\bibinfo
  {journal} {Phys. Rev. D}\ }\textbf {\bibinfo {volume} {97}},\ \bibinfo
  {pages} {104033} (\bibinfo {year} {2018})},\ \Eprint
  {https://arxiv.org/abs/1711.09607} {arXiv:1711.09607 [gr-qc]} \BibitemShut
  {NoStop}%
\bibitem [{\citenamefont {Chua}\ \emph {et~al.}(2021)\citenamefont {Chua},
  \citenamefont {Katz}, \citenamefont {Warburton},\ and\ \citenamefont
  {Hughes}}]{Chua:2020stf}%
  \BibitemOpen
  \bibfield  {author} {\bibinfo {author} {\bibfnamefont {A.~J.~K.}\
  \bibnamefont {Chua}}, \bibinfo {author} {\bibfnamefont {M.~L.}\ \bibnamefont
  {Katz}}, \bibinfo {author} {\bibfnamefont {N.}~\bibnamefont {Warburton}},\
  and\ \bibinfo {author} {\bibfnamefont {S.~A.}\ \bibnamefont {Hughes}},\
  }\bibfield  {title} {\bibinfo {title} {{Rapid generation of fully
  relativistic extreme-mass-ratio-inspiral waveform templates for LISA data
  analysis}},\ }\href {https://doi.org/10.1103/PhysRevLett.126.051102}
  {\bibfield  {journal} {\bibinfo  {journal} {Phys. Rev. Lett.}\ }\textbf
  {\bibinfo {volume} {126}},\ \bibinfo {pages} {051102} (\bibinfo {year}
  {2021})},\ \Eprint {https://arxiv.org/abs/2008.06071} {arXiv:2008.06071
  [gr-qc]} \BibitemShut {NoStop}%
\bibitem [{\citenamefont {Hughes}\ \emph {et~al.}(2021)\citenamefont {Hughes},
  \citenamefont {Warburton}, \citenamefont {Khanna}, \citenamefont {Chua},\
  and\ \citenamefont {Katz}}]{Hughes:2021exa}%
  \BibitemOpen
  \bibfield  {author} {\bibinfo {author} {\bibfnamefont {S.~A.}\ \bibnamefont
  {Hughes}}, \bibinfo {author} {\bibfnamefont {N.}~\bibnamefont {Warburton}},
  \bibinfo {author} {\bibfnamefont {G.}~\bibnamefont {Khanna}}, \bibinfo
  {author} {\bibfnamefont {A.~J.~K.}\ \bibnamefont {Chua}},\ and\ \bibinfo
  {author} {\bibfnamefont {M.~L.}\ \bibnamefont {Katz}},\ }\bibfield  {title}
  {\bibinfo {title} {{Adiabatic waveforms for extreme mass-ratio inspirals via
  multivoice decomposition in time and frequency}},\ }\href
  {https://doi.org/10.1103/PhysRevD.103.104014} {\bibfield  {journal} {\bibinfo
   {journal} {Phys. Rev. D}\ }\textbf {\bibinfo {volume} {103}},\ \bibinfo
  {pages} {104014} (\bibinfo {year} {2021})},\ \Eprint
  {https://arxiv.org/abs/2102.02713} {arXiv:2102.02713 [gr-qc]} \BibitemShut
  {NoStop}%
\bibitem [{\citenamefont {Pound}\ \emph {et~al.}(2020)\citenamefont {Pound},
  \citenamefont {Wardell}, \citenamefont {Warburton},\ and\ \citenamefont
  {Miller}}]{Pound:2019lzj}%
  \BibitemOpen
  \bibfield  {author} {\bibinfo {author} {\bibfnamefont {A.}~\bibnamefont
  {Pound}}, \bibinfo {author} {\bibfnamefont {B.}~\bibnamefont {Wardell}},
  \bibinfo {author} {\bibfnamefont {N.}~\bibnamefont {Warburton}},\ and\
  \bibinfo {author} {\bibfnamefont {J.}~\bibnamefont {Miller}},\ }\bibfield
  {title} {\bibinfo {title} {{Second-Order Self-Force Calculation of
  Gravitational Binding Energy in Compact Binaries}},\ }\href
  {https://doi.org/10.1103/PhysRevLett.124.021101} {\bibfield  {journal}
  {\bibinfo  {journal} {Phys. Rev. Lett.}\ }\textbf {\bibinfo {volume} {124}},\
  \bibinfo {pages} {021101} (\bibinfo {year} {2020})},\ \Eprint
  {https://arxiv.org/abs/1908.07419} {arXiv:1908.07419 [gr-qc]} \BibitemShut
  {NoStop}%
\bibitem [{\citenamefont {Warburton}\ \emph {et~al.}(2021)\citenamefont
  {Warburton}, \citenamefont {Pound}, \citenamefont {Wardell}, \citenamefont
  {Miller},\ and\ \citenamefont {Durkan}}]{Warburton:2021kwk}%
  \BibitemOpen
  \bibfield  {author} {\bibinfo {author} {\bibfnamefont {N.}~\bibnamefont
  {Warburton}}, \bibinfo {author} {\bibfnamefont {A.}~\bibnamefont {Pound}},
  \bibinfo {author} {\bibfnamefont {B.}~\bibnamefont {Wardell}}, \bibinfo
  {author} {\bibfnamefont {J.}~\bibnamefont {Miller}},\ and\ \bibinfo {author}
  {\bibfnamefont {L.}~\bibnamefont {Durkan}},\ }\href@noop {} {\bibinfo {title}
  {{Gravitational-wave energy flux for compact binaries through second order in
  the mass ratio}}} (\bibinfo {year} {2021}),\ \Eprint
  {https://arxiv.org/abs/2107.01298} {arXiv:2107.01298 [gr-qc]} \BibitemShut
  {NoStop}%
\bibitem [{\citenamefont {Wardell}\ \emph {et~al.}(2021)\citenamefont
  {Wardell}, \citenamefont {Pound}, \citenamefont {Warburton}, \citenamefont
  {Durkan},\ and\ \citenamefont {Miller}}]{WardellMG}%
  \BibitemOpen
  \bibfield  {author} {\bibinfo {author} {\bibfnamefont {B.}~\bibnamefont
  {Wardell}}, \bibinfo {author} {\bibfnamefont {A.}~\bibnamefont {Pound}},
  \bibinfo {author} {\bibfnamefont {N.}~\bibnamefont {Warburton}}, \bibinfo
  {author} {\bibfnamefont {L.}~\bibnamefont {Durkan}},\ and\ \bibinfo {author}
  {\bibfnamefont {J.}~\bibnamefont {Miller}},\ }\bibfield  {title} {\bibinfo
  {title} {Progress towards waveforms for extreme mass ratio inspirals}}
  (\bibinfo {year} {2021}),\ \bibinfo {note} {talk given at the 16th Marcel
  Grossmann Meeting. Recording available at
  \url{https://indico.icranet.org/event/1/sessions/82/\#20210705}}\BibitemShut
  {NoStop}%
\bibitem [{\citenamefont {van~de Meent}\ and\ \citenamefont
  {Pfeiffer}(2020)}]{vandeMeent:2020xgc}%
  \BibitemOpen
  \bibfield  {author} {\bibinfo {author} {\bibfnamefont {M.}~\bibnamefont
  {van~de Meent}}\ and\ \bibinfo {author} {\bibfnamefont {H.~P.}\ \bibnamefont
  {Pfeiffer}},\ }\bibfield  {title} {\bibinfo {title} {{Intermediate mass-ratio
  black hole binaries: Applicability of small mass-ratio perturbation
  theory}},\ }\href {https://doi.org/10.1103/PhysRevLett.125.181101} {\bibfield
   {journal} {\bibinfo  {journal} {Phys. Rev. Lett.}\ }\textbf {\bibinfo
  {volume} {125}},\ \bibinfo {pages} {181101} (\bibinfo {year} {2020})},\
  \Eprint {https://arxiv.org/abs/2006.12036} {arXiv:2006.12036 [gr-qc]}
  \BibitemShut {NoStop}%
\bibitem [{\citenamefont {Abbott}\ \emph
  {et~al.}(2020{\natexlab{c}})\citenamefont {Abbott} \emph
  {et~al.}}]{Abbott:2020tfl}%
  \BibitemOpen
  \bibfield  {author} {\bibinfo {author} {\bibfnamefont {R.}~\bibnamefont
  {Abbott}} \emph {et~al.} (\bibinfo {collaboration} {LIGO Scientific,
  Virgo}),\ }\bibfield  {title} {\bibinfo {title} {{GW190521: A Binary Black
  Hole Merger with a Total Mass of $150 M_{\odot}$}},\ }\href
  {https://doi.org/10.1103/PhysRevLett.125.101102} {\bibfield  {journal}
  {\bibinfo  {journal} {Phys. Rev. Lett.}\ }\textbf {\bibinfo {volume} {125}},\
  \bibinfo {pages} {101102} (\bibinfo {year} {2020}{\natexlab{c}})},\ \Eprint
  {https://arxiv.org/abs/2009.01075} {arXiv:2009.01075 [gr-qc]} \BibitemShut
  {NoStop}%
\bibitem [{\citenamefont {Schutz}(2017)}]{Schutz2017}%
  \BibitemOpen
  \bibfield  {author} {\bibinfo {author} {\bibfnamefont {B.}~\bibnamefont
  {Schutz}},\ }\bibfield  {title} {\bibinfo {title} {Discussion on emri/imri
  using numerical relativity.}} (\bibinfo {year} {2017}),\ \bibinfo {note}
  {20th Capra Meeting}\BibitemShut {NoStop}%
\bibitem [{\citenamefont {Poisson}(2005)}]{Poisson:2005pi}%
  \BibitemOpen
  \bibfield  {author} {\bibinfo {author} {\bibfnamefont {E.}~\bibnamefont
  {Poisson}},\ }\bibfield  {title} {\bibinfo {title} {{Metric of a tidally
  distorted, nonrotating black hole}},\ }\href
  {https://doi.org/10.1103/PhysRevLett.94.161103} {\bibfield  {journal}
  {\bibinfo  {journal} {Phys. Rev. Lett.}\ }\textbf {\bibinfo {volume} {94}},\
  \bibinfo {pages} {161103} (\bibinfo {year} {2005})},\ \Eprint
  {https://arxiv.org/abs/gr-qc/0501032} {arXiv:gr-qc/0501032} \BibitemShut
  {NoStop}%
\bibitem [{\citenamefont {Poisson}\ and\ \citenamefont
  {Vlasov}(2010)}]{Poisson:2009qj}%
  \BibitemOpen
  \bibfield  {author} {\bibinfo {author} {\bibfnamefont {E.}~\bibnamefont
  {Poisson}}\ and\ \bibinfo {author} {\bibfnamefont {I.}~\bibnamefont
  {Vlasov}},\ }\bibfield  {title} {\bibinfo {title} {{Geometry and dynamics of
  a tidally deformed black hole}},\ }\href
  {https://doi.org/10.1103/PhysRevD.81.024029} {\bibfield  {journal} {\bibinfo
  {journal} {Phys. Rev. D}\ }\textbf {\bibinfo {volume} {81}},\ \bibinfo
  {pages} {024029} (\bibinfo {year} {2010})},\ \Eprint
  {https://arxiv.org/abs/0910.4311} {arXiv:0910.4311 [gr-qc]} \BibitemShut
  {NoStop}%
\bibitem [{\citenamefont {Taylor}\ and\ \citenamefont
  {Poisson}(2008)}]{Taylor:2008xy}%
  \BibitemOpen
  \bibfield  {author} {\bibinfo {author} {\bibfnamefont {S.}~\bibnamefont
  {Taylor}}\ and\ \bibinfo {author} {\bibfnamefont {E.}~\bibnamefont
  {Poisson}},\ }\bibfield  {title} {\bibinfo {title} {{Nonrotating black hole
  in a post-Newtonian tidal environment}},\ }\href
  {https://doi.org/10.1103/PhysRevD.78.084016} {\bibfield  {journal} {\bibinfo
  {journal} {Phys. Rev. D}\ }\textbf {\bibinfo {volume} {78}},\ \bibinfo
  {pages} {084016} (\bibinfo {year} {2008})},\ \Eprint
  {https://arxiv.org/abs/0806.3052} {arXiv:0806.3052 [gr-qc]} \BibitemShut
  {NoStop}%
\bibitem [{\citenamefont {Poisson}\ and\ \citenamefont
  {Corrigan}(2018)}]{Poisson:2018qqd}%
  \BibitemOpen
  \bibfield  {author} {\bibinfo {author} {\bibfnamefont {E.}~\bibnamefont
  {Poisson}}\ and\ \bibinfo {author} {\bibfnamefont {E.}~\bibnamefont
  {Corrigan}},\ }\bibfield  {title} {\bibinfo {title} {{Nonrotating black hole
  in a post-Newtonian tidal environment II}},\ }\href
  {https://doi.org/10.1103/PhysRevD.97.124048} {\bibfield  {journal} {\bibinfo
  {journal} {Phys. Rev. D}\ }\textbf {\bibinfo {volume} {97}},\ \bibinfo
  {pages} {124048} (\bibinfo {year} {2018})},\ \Eprint
  {https://arxiv.org/abs/1804.01848} {arXiv:1804.01848 [gr-qc]} \BibitemShut
  {NoStop}%
\bibitem [{\citenamefont {Le~Tiec}\ \emph {et~al.}(2021)\citenamefont
  {Le~Tiec}, \citenamefont {Casals},\ and\ \citenamefont
  {Franzin}}]{LeTiec:2020bos}%
  \BibitemOpen
  \bibfield  {author} {\bibinfo {author} {\bibfnamefont {A.}~\bibnamefont
  {Le~Tiec}}, \bibinfo {author} {\bibfnamefont {M.}~\bibnamefont {Casals}},\
  and\ \bibinfo {author} {\bibfnamefont {E.}~\bibnamefont {Franzin}},\
  }\bibfield  {title} {\bibinfo {title} {{Tidal Love Numbers of Kerr Black
  Holes}},\ }\href {https://doi.org/10.1103/PhysRevD.103.084021} {\bibfield
  {journal} {\bibinfo  {journal} {Phys. Rev. D}\ }\textbf {\bibinfo {volume}
  {103}},\ \bibinfo {pages} {084021} (\bibinfo {year} {2021})},\ \Eprint
  {https://arxiv.org/abs/2010.15795} {arXiv:2010.15795 [gr-qc]} \BibitemShut
  {NoStop}%
\bibitem [{\citenamefont {Damour}\ and\ \citenamefont
  {Nagar}(2009)}]{Damour:2009vw}%
  \BibitemOpen
  \bibfield  {author} {\bibinfo {author} {\bibfnamefont {T.}~\bibnamefont
  {Damour}}\ and\ \bibinfo {author} {\bibfnamefont {A.}~\bibnamefont {Nagar}},\
  }\bibfield  {title} {\bibinfo {title} {{Relativistic tidal properties of
  neutron stars}},\ }\href {https://doi.org/10.1103/PhysRevD.80.084035}
  {\bibfield  {journal} {\bibinfo  {journal} {Phys. Rev. D}\ }\textbf {\bibinfo
  {volume} {80}},\ \bibinfo {pages} {084035} (\bibinfo {year} {2009})},\
  \Eprint {https://arxiv.org/abs/0906.0096} {arXiv:0906.0096 [gr-qc]}
  \BibitemShut {NoStop}%
\bibitem [{\citenamefont {Raposo}\ and\ \citenamefont
  {Pani}(2020)}]{Raposo:2020yjy}%
  \BibitemOpen
  \bibfield  {author} {\bibinfo {author} {\bibfnamefont {G.}~\bibnamefont
  {Raposo}}\ and\ \bibinfo {author} {\bibfnamefont {P.}~\bibnamefont {Pani}},\
  }\bibfield  {title} {\bibinfo {title} {{Axisymmetric deformations of neutron
  stars and gravitational-wave astronomy}},\ }\href
  {https://doi.org/10.1103/PhysRevD.102.044045} {\bibfield  {journal} {\bibinfo
   {journal} {Phys. Rev. D}\ }\textbf {\bibinfo {volume} {102}},\ \bibinfo
  {pages} {044045} (\bibinfo {year} {2020})},\ \Eprint
  {https://arxiv.org/abs/2002.02555} {arXiv:2002.02555 [gr-qc]} \BibitemShut
  {NoStop}%
\bibitem [{\citenamefont {Yagi}\ and\ \citenamefont
  {Yunes}(2016)}]{Yagi:2016ejg}%
  \BibitemOpen
  \bibfield  {author} {\bibinfo {author} {\bibfnamefont {K.}~\bibnamefont
  {Yagi}}\ and\ \bibinfo {author} {\bibfnamefont {N.}~\bibnamefont {Yunes}},\
  }\bibfield  {title} {\bibinfo {title} {{I-Love-Q Relations: From Compact
  Stars to Black Holes}},\ }\href
  {https://doi.org/10.1088/0264-9381/33/9/095005} {\bibfield  {journal}
  {\bibinfo  {journal} {Class. Quant. Grav.}\ }\textbf {\bibinfo {volume}
  {33}},\ \bibinfo {pages} {095005} (\bibinfo {year} {2016})},\ \Eprint
  {https://arxiv.org/abs/1601.02171} {arXiv:1601.02171 [gr-qc]} \BibitemShut
  {NoStop}%
\bibitem [{\citenamefont {Hinderer}\ \emph {et~al.}(2010)\citenamefont
  {Hinderer}, \citenamefont {Lackey}, \citenamefont {Lang},\ and\ \citenamefont
  {Read}}]{Hinderer:2009ca}%
  \BibitemOpen
  \bibfield  {author} {\bibinfo {author} {\bibfnamefont {T.}~\bibnamefont
  {Hinderer}}, \bibinfo {author} {\bibfnamefont {B.~D.}\ \bibnamefont
  {Lackey}}, \bibinfo {author} {\bibfnamefont {R.~N.}\ \bibnamefont {Lang}},\
  and\ \bibinfo {author} {\bibfnamefont {J.~S.}\ \bibnamefont {Read}},\
  }\bibfield  {title} {\bibinfo {title} {{Tidal deformability of neutron stars
  with realistic equations of state and their gravitational wave signatures in
  binary inspiral}},\ }\href {https://doi.org/10.1103/PhysRevD.81.123016}
  {\bibfield  {journal} {\bibinfo  {journal} {Phys. Rev. D}\ }\textbf {\bibinfo
  {volume} {81}},\ \bibinfo {pages} {123016} (\bibinfo {year} {2010})},\
  \Eprint {https://arxiv.org/abs/0911.3535} {arXiv:0911.3535 [astro-ph.HE]}
  \BibitemShut {NoStop}%
\bibitem [{\citenamefont {Poisson}\ \emph {et~al.}(2011)\citenamefont
  {Poisson}, \citenamefont {Pound},\ and\ \citenamefont
  {Vega}}]{Poisson:2011nh}%
  \BibitemOpen
  \bibfield  {author} {\bibinfo {author} {\bibfnamefont {E.}~\bibnamefont
  {Poisson}}, \bibinfo {author} {\bibfnamefont {A.}~\bibnamefont {Pound}},\
  and\ \bibinfo {author} {\bibfnamefont {I.}~\bibnamefont {Vega}},\ }\bibfield
  {title} {\bibinfo {title} {{The Motion of point particles in curved
  spacetime}},\ }\href {https://doi.org/10.12942/lrr-2011-7} {\bibfield
  {journal} {\bibinfo  {journal} {Living Rev. Rel.}\ }\textbf {\bibinfo
  {volume} {14}},\ \bibinfo {pages} {7} (\bibinfo {year} {2011})},\ \Eprint
  {https://arxiv.org/abs/1102.0529} {arXiv:1102.0529 [gr-qc]} \BibitemShut
  {NoStop}%
\bibitem [{\citenamefont {Pound}(2010)}]{Pound:2009sm}%
  \BibitemOpen
  \bibfield  {author} {\bibinfo {author} {\bibfnamefont {A.}~\bibnamefont
  {Pound}},\ }\bibfield  {title} {\bibinfo {title} {{Self-consistent
  gravitational self-force}},\ }\href
  {https://doi.org/10.1103/PhysRevD.81.024023} {\bibfield  {journal} {\bibinfo
  {journal} {Phys. Rev. D}\ }\textbf {\bibinfo {volume} {81}},\ \bibinfo
  {pages} {024023} (\bibinfo {year} {2010})},\ \Eprint
  {https://arxiv.org/abs/0907.5197} {arXiv:0907.5197 [gr-qc]} \BibitemShut
  {NoStop}%
\bibitem [{\citenamefont {Sciama}\ \emph {et~al.}(1969)\citenamefont {Sciama},
  \citenamefont {Waylen},\ and\ \citenamefont {Gilman}}]{Sciama:1969vtz}%
  \BibitemOpen
  \bibfield  {author} {\bibinfo {author} {\bibfnamefont {D.~W.}\ \bibnamefont
  {Sciama}}, \bibinfo {author} {\bibfnamefont {P.~C.}\ \bibnamefont {Waylen}},\
  and\ \bibinfo {author} {\bibfnamefont {R.~C.}\ \bibnamefont {Gilman}},\
  }\bibfield  {title} {\bibinfo {title} {{Generally covariant integral
  formulation of einstein's field equation}},\ }\href
  {https://doi.org/10.1103/PhysRev.187.1762} {\bibfield  {journal} {\bibinfo
  {journal} {Phys. Rev.}\ }\textbf {\bibinfo {volume} {187}},\ \bibinfo {pages}
  {1762} (\bibinfo {year} {1969})}\BibitemShut {NoStop}%
\bibitem [{\citenamefont {Lousto}\ and\ \citenamefont
  {Price}(1997)}]{Lousto:1997wf}%
  \BibitemOpen
  \bibfield  {author} {\bibinfo {author} {\bibfnamefont {C.~O.}\ \bibnamefont
  {Lousto}}\ and\ \bibinfo {author} {\bibfnamefont {R.~H.}\ \bibnamefont
  {Price}},\ }\bibfield  {title} {\bibinfo {title} {{Understanding initial data
  for black hole collisions}},\ }\href
  {https://doi.org/10.1103/PhysRevD.56.6439} {\bibfield  {journal} {\bibinfo
  {journal} {Phys. Rev. D}\ }\textbf {\bibinfo {volume} {56}},\ \bibinfo
  {pages} {6439} (\bibinfo {year} {1997})},\ \Eprint
  {https://arxiv.org/abs/gr-qc/9705071} {arXiv:gr-qc/9705071} \BibitemShut
  {NoStop}%
\bibitem [{\citenamefont {Holst}\ \emph {et~al.}(2004)\citenamefont {Holst},
  \citenamefont {Lindblom}, \citenamefont {Owen}, \citenamefont {Pfeiffer},
  \citenamefont {Scheel},\ and\ \citenamefont {Kidder}}]{HolLinOwe04}%
  \BibitemOpen
  \bibfield  {author} {\bibinfo {author} {\bibfnamefont {M.}~\bibnamefont
  {Holst}}, \bibinfo {author} {\bibfnamefont {L.}~\bibnamefont {Lindblom}},
  \bibinfo {author} {\bibfnamefont {R.}~\bibnamefont {Owen}}, \bibinfo {author}
  {\bibfnamefont {H.}~\bibnamefont {Pfeiffer}}, \bibinfo {author}
  {\bibfnamefont {M.}~\bibnamefont {Scheel}},\ and\ \bibinfo {author}
  {\bibfnamefont {L.}~\bibnamefont {Kidder}},\ }\bibfield  {title} {\bibinfo
  {title} {Optimal constraint projection for hyperbolic evolution systems},\
  }\href@noop {} {\bibfield  {journal} {\bibinfo  {journal} {Phys. Rev. D}\
  }\textbf {\bibinfo {volume} {70}},\ \bibinfo {pages} {084017} (\bibinfo
  {year} {2004})},\ \Eprint {https://arxiv.org/abs/gr-qc/0407011}
  {gr-qc/0407011} \BibitemShut {NoStop}%
\bibitem [{\citenamefont {Sarbach}\ and\ \citenamefont
  {Tiglio}(2012)}]{SarTig12}%
  \BibitemOpen
  \bibfield  {author} {\bibinfo {author} {\bibfnamefont {O.}~\bibnamefont
  {Sarbach}}\ and\ \bibinfo {author} {\bibfnamefont {M.}~\bibnamefont
  {Tiglio}},\ }\bibfield  {title} {\bibinfo {title} {Continuum and discrete
  initial-boundary value problems and einstein's field equations},\ }\href
  {http://www.livingreviews.org/lrr-2012-9} {\bibfield  {journal} {\bibinfo
  {journal} {Living Reviews in Relativity}\ }\textbf {\bibinfo {volume} {15}}
  (\bibinfo {year} {2012})},\ \Eprint {https://arxiv.org/abs/1203.6443}
  {arXiv:1203.6443 [gr-qc]} \BibitemShut {NoStop}%
\bibitem [{\citenamefont {Hilditch}\ \emph {et~al.}(2016)\citenamefont
  {Hilditch}, \citenamefont {Weyhausen},\ and\ \citenamefont
  {Br{\"u}gmann}}]{HilWeyBru15}%
  \BibitemOpen
  \bibfield  {author} {\bibinfo {author} {\bibfnamefont {D.}~\bibnamefont
  {Hilditch}}, \bibinfo {author} {\bibfnamefont {A.}~\bibnamefont
  {Weyhausen}},\ and\ \bibinfo {author} {\bibfnamefont {B.}~\bibnamefont
  {Br{\"u}gmann}},\ }\bibfield  {title} {\bibinfo {title} {{Pseudospectral
  method for gravitational wave collapse}},\ }\href
  {https://doi.org/10.1103/PhysRevD.93.063006} {\bibfield  {journal} {\bibinfo
  {journal} {Phys. Rev.}\ }\textbf {\bibinfo {volume} {D93}},\ \bibinfo {pages}
  {063006} (\bibinfo {year} {2016})},\ \Eprint
  {https://arxiv.org/abs/1504.04732} {arXiv:1504.04732 [gr-qc]} \BibitemShut
  {NoStop}%
\bibitem [{\citenamefont {Strand}(1994)}]{Str94}%
  \BibitemOpen
  \bibfield  {author} {\bibinfo {author} {\bibfnamefont {B.}~\bibnamefont
  {Strand}},\ }\bibfield  {title} {\bibinfo {title} {Summation by parts for
  finite differencing approximations for d/dx},\ }\href
  {https://doi.org/10.1006/jcph.1994.1005} {\bibfield  {journal} {\bibinfo
  {journal} {J. Comput. Phys.}\ }\textbf {\bibinfo {volume} {110}},\ \bibinfo
  {pages} {47} (\bibinfo {year} {1994})}\BibitemShut {NoStop}%
\bibitem [{\citenamefont {Bj{\o}rhus}(1995)}]{Bjo95}%
  \BibitemOpen
  \bibfield  {author} {\bibinfo {author} {\bibfnamefont {M.}~\bibnamefont
  {Bj{\o}rhus}},\ }\bibfield  {title} {\bibinfo {title} {{The ODE Formulation
  of Hyperbolic PDEs Discretized by the Spectral Collocation Method}},\
  }\href@noop {} {\bibfield  {journal} {\bibinfo  {journal} {SIAM J. Sci.
  Comput.}\ }\textbf {\bibinfo {volume} {16}},\ \bibinfo {pages} {542}
  (\bibinfo {year} {1995})}\BibitemShut {NoStop}%
\bibitem [{\citenamefont {Kopriva}(2009)}]{Kop09}%
  \BibitemOpen
  \bibfield  {author} {\bibinfo {author} {\bibfnamefont {D.~A.}\ \bibnamefont
  {Kopriva}},\ }\href@noop {} {\emph {\bibinfo {title} {Implementing Spectral
  Methods for Partial Differential Equations}}}\ (\bibinfo  {publisher}
  {Springer},\ \bibinfo {address} {{New York}},\ \bibinfo {year}
  {2009})\BibitemShut {NoStop}%
\bibitem [{\citenamefont {Pound}(2017)}]{Pound:2017psq}%
  \BibitemOpen
  \bibfield  {author} {\bibinfo {author} {\bibfnamefont {A.}~\bibnamefont
  {Pound}},\ }\bibfield  {title} {\bibinfo {title} {{Nonlinear gravitational
  self-force: second-order equation of motion}},\ }\href
  {https://doi.org/10.1103/PhysRevD.95.104056} {\bibfield  {journal} {\bibinfo
  {journal} {Phys. Rev. D}\ }\textbf {\bibinfo {volume} {95}},\ \bibinfo
  {pages} {104056} (\bibinfo {year} {2017})},\ \Eprint
  {https://arxiv.org/abs/1703.02836} {arXiv:1703.02836 [gr-qc]} \BibitemShut
  {NoStop}%
\end{thebibliography}%
\end{document}